\documentclass{aa}

\usepackage{graphicx}
\usepackage{tikz}
\usepackage{xcolor}

\usepackage{pgf}	
\usepackage{txfonts}
\usepackage{natbib}
\usepackage{epsf}
\usepackage{rotating}
\usepackage{graphics}
\usepackage{verbatim}
\usepackage{longtable}
\usepackage{hyperref} 
\hypersetup{colorlinks,allcolors=blue}
\usepackage{amssymb}
\usepackage{amsmath}

\usepackage{todonotes}
\usepackage{color}
\hypersetup{colorlinks=true,citecolor=blue}
\usepackage{tabularx}
\usepackage[normalem]{ulem} 
\usepackage{subfigure}
\usepackage{lscape}

\def \1s{$1\,\sigma$}

\def \t0{T$_0$}

\DeclareUnicodeCharacter{2212}{ --}

\begin{document}

\title{Null transit detections of 68 radial-velocity exoplanets observed by TESS}

\author{
F.~V.~Lovos\inst{1, 2} \and
R.~F.~D\'iaz\inst{2, 3} \and
L.~A.~Nieto\inst{3, 4}
}

 \offprints{F~.V~Lovos (flavialovos@unc.edu.ar)}

\institute{Universidad Nacional de C\'ordoba, Observatorio Astron\'omico, Laprida 854, X5000BGR, C\'ordoba, Argentina.
\and
Consejo Nacional de Investigaciones Cient\'ificas y T\'ecnicas (CONICET), Godoy Cruz 2290, CABA, CPC 1425FQB, Argentina.
\and
International Center for Advanced Studies (ICAS) and ICIFI (CONICET), ECyT-UNSAM, Campus Miguelete, 25 de Mayo y Francia, (1650) Buenos Aires, Argentina.
\and
Gerencia de Tecnología de la información y de las Comunicaciones (GTIC), Subgerencia Vinculación y Desarrollo de Nuevas Tecnologías de la Información, DCAP-CNEA. Centro Atómico Constituyentes, Av. Gral. Paz 1499, (1650) Buenos Aires, Argentina.}

\date{Received April 12, 2022; accepted July 11, 2022}     

\abstract{
In recent years, the number of exoplanets has grown considerably. The most successful techniques in these detections are the radial velocity (RV)  and planetary transits techniques, the latter of which has been significantly advanced by the {Kepler}, K2 and, more recently, the Transiting Exoplanet Survey Satellite (TESS) missions. The detection of exoplanets by means of  both transits and RVs is of importance because this allows the characterization of their bulk densities and internal compositions. The TESS survey offers a unique possibility to search for transits of extrasolar planets detected using RVs. In this work, we present the results of our search for transits of RV-detected planets using the photometry of the TESS space mission. We focus on systems with super-Earth- and Neptune-sized planets on orbits with periods of shorter than 30 days. This cut is intended to keep objects with a relatively high transit probability, and is also consistent with the duration of TESS observations on a single sector. Given the summed geometric transit probabilities, the expected number of transiting planets is $3.4 \pm 1.8$. The sample contains two known transiting planets. We report null results for the remaining 66 out of 68 planets studied, and we exclude in all cases planets larger than 2.4 R$_{\oplus}$ under the assumption of central transits. The remaining two planets orbit HD~136352 and were recently announced.
}

\authorrunning{Lovos et al.}
\titlerunning{Null transit detections of RV planets}

\maketitle

\section{Introduction \label{sect.intro}}

Our knowledge of exoplanet systems has grown over the past 15 years thanks to space-based missions \citep{lissauer2014b,borucki2016}  that saw the number of confirmed exoplanets increase to more than 5000\footnote{https://exoplanetarchive.ipac.caltech.edu/} (e.g., CoRoT-22~b, \citet{moutou2011}; Kepler-10~b, \citet{batalha2010}; Kepler-11~b, \citet{lissauer2011}).
The Transiting Exoplanet Survey Satellite (TESS) is monitoring the nearest and brightest stars, looking for transiting exoplanets of all sizes, but in particular with the ability to detect planets of sizes as small as Earth or Neptune.
Unlike Kepler, the TESS mission was designed to survey over $85\%$ of the sky, searching for planets around stars that are typically 30-100 times brighter than those surveyed by Kepler. These stars are therefore more accessible to follow-up observations, and in particular to ground-based spectroscopy \citep{ricker2015}. The TESS satellite has been operating since 2018 in an Earth-centered orbit with a period of 13.7 days. It observed different sectors in the southern ecliptic hemisphere in the first year and in the northern ecliptic hemisphere for the second year. Each sector comprises two TESS orbits for a duration of 27 days.
Its primary mission finished in july 2020, providing around 110 TESS confirmed planets, and  more than 5700 TESS candidate planets waiting for confirmation. Since then, an extended mission has been in operation, re-observing the southern hemisphere with a slightly different strategy. TESS provides a unique opportunity to detect and improve our understanding of small,  short-period planets. 

Of particular interest are those planets that were first detected by radial velocity (RV) surveys.
The combination of precise photometric and spectroscopic measurements is particularly important in 
providing a detailed description of planetary systems. As is commonly known, the planetary radius 
derived from transit photometry and the mass derived from RV measurements give us planet densities and allow planetary interior structures to be modeled (e.g., \citet{valencia2006,Fortney_2007,seager2007}).   
In addition, the detection of transiting planets around bright and nearby stars allows atmospheric characterization by transit transmission spectrophotometric measurements (e.g., 55~Cnc~e \citep{demory2011},  HD~97658~b \citep{dragomir2013}, HD~189733~b \citep{brogi2018}, HD~209458~b \citep{sanchez2019}, among many others). During a transit event, the planet and its atmosphere block light from the parent star, producing a partial transmission of the stellar light through the atmosphere of the planet. The transit depth therefore depends on the wavelength, allowing the exploration of atmospheric compositions and environments \citep{sea-dem2010}. These observations require planets to orbit very bright stars, with magnitudes of below $V={8}$. If further planets were detected in transit around similarly bright stars, they would become important assets to understand planetary atmospheres and interiors.

In this context, the main goal of this work is to search for transit signals of known RV planets using TESS data (see \citet{Pepper2020} for the first RV planet revealed to be transiting by TESS data, HD 118203 b), focusing on low-mass planets with sizes smaller than Neptune. Additionally, to achieve high 
geometric transit probabilities, we restricted the search to planets with periods shorter than 30 days, which is close to the minimum observation time for TESS of 27 days (for one sector). Here, we report 66 null detections and two planets that were detected independently, HD~136352~b and c, and which were reported by \citet{Kane2020}.

This article is organized as follows. In Sect.~\ref{sect.sample} we present our sample of systems. A description of the TESS photometry that we used and the reduction process are given in Sect.~\ref{sect.data}. In 
Sect.~\ref{sect.transitsearch} we present the details of a transit search that was performed by applying the Box
Least Square (BLS) algorithm.  
In Sect.~\ref{sect.model} we model the possible transits of each surveyed planetary system based on interior models and physical and orbital parameters from the literature.
In Sect.~\ref{sect.results} we discuss some interesting cases from the BLS results.
In Sect.~\ref{sect.summ} we present some brief conclusions.

\section{Sample selection \label{sect.sample} }

We selected a sample of low-mass planets from the catalog of known planets detected by RVs, that is with masses of less than 20 M$_{\oplus}$ (Neptune- or Earth-sized planets). We specifically selected planets with no prior transit detections. We further restricted our sample to planets with periods shorter than 30 days in order to improve the geometric probability of transit occurrence, and to coincide with the minimum observing time of TESS. With these criteria, we selected 94 planets in 58 systems, of which only 43 were included in the TESS observing list of selected targets for short cadence exposure (2 minutes) and have all the relevant data products. The data considered here include all sectors of TESS photometry observed in the primary mission (Cycle 1 and 2) plus part of Cycle 3. The final sample consists of 68 RV planets around 43 stars. The sum of their geometric transit probabilities is 3.36. The expected number of transiting planets in the sample is therefore $3.4 \pm 1.8$. At this point the work of \citet{Dalba2019} was brought to our attention, they predict the detection of three transiting planets among RV-detected companions, which is consistent with our geometric transit probabilities. The physical parameters of the host stars are listed in Table~\ref{tab1:star}.

The orbital and physical parameters for each planet were compiled from the literature with the aid of the NASA exoplanet archive\footnote{https://exoplanetarchive.ipac.caltech.edu/}. The planets are listed together with their main properties in Table \ref{apptab2:planet}, where we report the orbital periods (Col. 2) and minimum masses (Col. 3). Column 4 presents the reference for each object. Based on the values from the literature and their reported uncertainties, we computed the times of inferior conjunction (T$_C$) assuming the parameters are normally distributed with mean and variance corresponding to the reported values. These times are reported in Col. 5 of Table \ref{apptab2:planet}, and were used for the transit search. The remaining columns list the
 planetary radii from models and the limiting orbital inclinations inferred from the lack of transits (see Sects.~\ref{sect.model} and \ref{sect.results}).

\begin{figure}[!ht]
\begin{minipage}{\linewidth}
\begin{center}
\includegraphics[angle=0,trim = 1mm 00mm 15mm 00mm, clip, width=\linewidth]{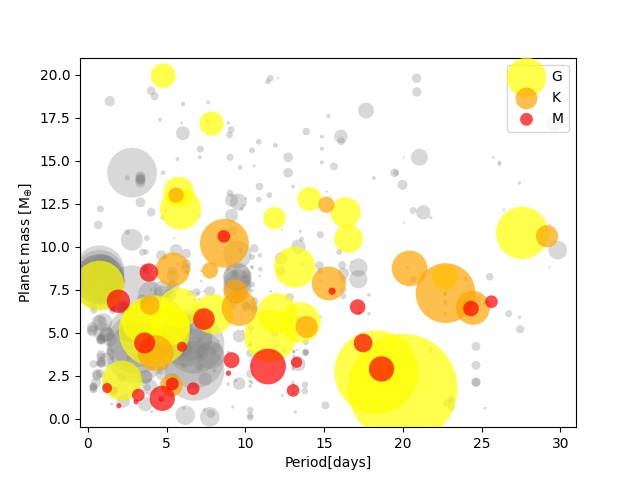}
\end{center}
\caption{\small{Planet minimum mass vs. orbital period. The color scale corresponds to the spectral types of the host stars, and the sizes are proportional to $1.9^{ \Delta\,T_{\rm mag}}$, i.e., larger sizes correspond to brighter stars. All planets outside the sample selection are represented in gray.}}
\label{fig:PM}
 \end{minipage}
\end{figure}

 In Fig.~\ref{fig:PM} we plot the planetary masses and orbital periods of the sample planets, together with those of all the known planets in the same region of parameter space. The colors indicate the  spectral types of the host stars, and sizes correspond to their relative brightness. We can see stars with spectral types between G and M in a wide range of magnitudes. 
 
 In general, planets with longer periods are more massive, probably reflecting observational biases. We see that most of the stars in the sample are M-dwarf stars with low-mass planets on short-period orbits, again probably an effect of the limitations of the RV technique. Thanks to the small sizes of these late stars, the transits are relatively deep compared to larger stars of earlier spectral types, which facilitates the detection of the transits of their companion planets.

\begin{table*}																	
\begin{center}																	
\caption{Stars with planets in the sample.}																	
\label{tab1:star}																	
\begin{tabular}{lcccccccc}																	
\hline																	
\hline																	
Star	&	T$_{\rm eff}$	&	Spectral Type	&	R$_{\star}$	&	\emph{V}	&	M$_{\star}$	&	References	&	\emph{T$_{\rm mag}$}	&	Tess sectors	\\
    	&	[K]	&	            	&	[R$_{\odot}$]	&		&	[M$_{\odot}$]	&	        	&		&		\\
\hline																	
\object{61 Vir}	&	5577	&	G5 V	&	0.963	&	4.69	&	0.942	&	 1 	&	4.09	&	10	\\
\object{BD-06 1339}	&	4324	&	K7/M0 V	&	0.602	&	9.7	&	0.7	&	 2 	&	8.29	&	6	\\
\object{BD-08 2823}	&	4746	&	K3 V	&	0.710	&	9.86	&	0.74	&	 3 	&	8.86	&	8, 35	\\
\object{DMPP-1}	&	6196	&	F8 V	&	1.260	&	7.98	&	1.21	&	 4 	&	7.47	&	6, 33	\\
\object{GJ 1061}	&	2953	&	M5.5 V	&	0.156	&	12.7	&	0.12	&	 5 	&	9.47	&	3, 4, 30, 31	\\
\object{GJ 1132}	&	3270	&	M4.5 V	&	0.207	&	13.68	&	0.181	&	 6 	&	12.14	&	9, 10, 36	\\
\object{GJ 15 A}	&	3607	&	M1 V	&	0.380	&	7.22	&	0.38	&	 7 	&	6.23	&	17	\\
\object{GJ 163}	&	3500	&	M 3.5 V	&	0.425	&	11.79	&	0.38	&	 8 	&	9.47	&	3--5, 30, 31	\\
\object{GJ 180}	&	3562	&	M3	&	0.420	&	10.914	&	0.39	&	 8 	&	8.82	&	5, 32	\\
\object{GJ 273}	&	3382	&	M 3.5 V	&	0.293	&	9.84	&	0.29	&	 9 	&	7.31	&	7	\\
\object{GJ 3138}	&	3717	&	M0 V	&	0.50	&	10.83	&	0.681	&	 9 	&	10.19	&	3, 30	\\
\object{GJ 3293}	&	3466	&	M2.5 	&	0.404	&	11.945	&	0.42	&	 9 	&	9.82	&	5, 31, 32	\\
\object{GJ 3323}	&	3159	&	M4	&	0.119	&	12.57	&	0.164	&	 9 	&	9.43	&	5, 32	\\
\object{GJ 3473}	&	3347	&	M4	&	0.364	&	13.74	&	0.36	&	 10, 11 	&	11.20	&	7, 34	\\
\object{GJ 357}	&	3505	&	M2.5 V	&	0.337	&	10.91	&	0.342	&	  12 	&	8.74	&	8, 35	\\
\object{GJ 433}	&	3472	&	M1.5 V	&	0.500	&	9.81	&	0.48	&	 13, 14 	&	7.81	&	10, 36	\\
\object{GJ 676} A	&	3734	&	M0 	&	0.690	&	9.58	&	0.73	&	 11, 14 	&	7.90	&	12	\\
\object{GJ 682}	&	3172	&	M4 	&	0.340	&	10.94	&	0.31	&	 8 	&	8.25	&	12	\\
\object{GJ 876}	&	3227	&	M2.5 V	&	0.400	&	10.16	&	0.34	&	 8 	&	7.57	&	2, 29	\\
\object{GJ 887}	&	3688	&	M1 V	&	0.471	&	7.39	&	0.489	&	 15 	&	5.48	&	2, 28	\\
\object{HD 10180}	&	5911	&	G1 V	&	1.109	&	7.33	&	1.06	&	 16, 17 	&	6.75	&	1, 2, 28, 29	\\
\object{HD 109271}	&	5783	&	G5 V	&	1.258	&	8.04	&	1.047	&	 2 	&	7.44	&	10	\\
\object{HD 136352}	&	5664	&	G4 V	&	1.010	&	5.65	&	0.81	&	  18 	&	5.05	&	12	\\
\object{HD 1461}	&	5765	&	G5 V	&	1.110	&	6.479	&	1.03	&	 8 	&	5.86	&	3	\\
\object{HD 158259}	&	5834	&	G0	&	1.290	&	6.479	&	1.15	&	 8 	&	5.90	&	17, 20, 24--26	\\
\object{HD 181433}	&	4900	&	K2 IV/ V	&	0.830	&	8.4	&	0.77	&	 8 	&	7.49	&	27	\\
\object{HD 20003}	&	5494	&	G8 V	&	0.922	&	8.39	&	0.875	&	  18 	&	7.71	&	1, 2, 6, 13, 27, 29	\\
\object{HD 20781}	&	5256	&	K0 V	&	0.834	&	8.48	&	0.7	&	  18 	&	7.72	&	4, 31	\\
\object{HD 20794}	&	5401	&	G8 V	&	0.920	&	4.26	&	0.85	&	 8 	&	3.58	&	3, 4, 30, 31	\\
\object{HD 213885}	&	5978	&	G0 V	&	1.068	&	7.95	&	1.068	&	 19 	&	7.38	&	1, 27, 28	\\
\object{HD 215497}	&	5113	&	K3 V	&	0.850	&	8.96	&	0.86	&	  14 	&	8.11	&	1, 28	\\
\object{HD 21693}	&	5430	&	G8 V	&	0.914	&	7.95	&	0.8	&	  18 	&	7.27	&	2--4, 29--31	\\
\object{HD 219134}	&	4966	&	K3 V	&	0.778	&	5.569	&	0.78	&	 20 	&	4.63	&	17, 24	\\
\object{HD 31527}	&	5898	&	G2 V	&	1.074	&	7.49	&	0.96	&	  18 	&	6.93	&	5, 32	\\
\object{HD 40307}	&	4827	&	K3 V	&	0.720	&	7.17	&	0.74	&	 8 	&	6.25	&	1--8, 10--13, 27--36\\
\object{HD 45184}	&	5869	&	G1.5 V	&	1.080	&	6.38	&	1.03	&	  18 	&	5.79	&	6, 33	\\
\object{HD 51608}	&	5358	&	G7 V	&	0.916	&	8.17	&	0.8	&	  18 	&	7.46	&	2, 3, 5--9, 12, 13, 29, 32, 33, 35, 36\\
\object{HD 69830}	&	5361	&	K0 V	&	0.850	&	5.96	&	0.9	&	 8 	&	5.26	&	7, 34	\\
\object{HD 7924}	&	5177	&	K0 V	&	0.780	&	7.17	&	0.65	&	  14 	&	6.39	&	18, 19	\\
\object{HIP 54373}	&	3848	&	K5 	&	0.500	&	10.38	&	0.57	&	 8 	&	8.67	&	9, 36	\\
\object{HIP 57274}	&	4510	&		&	0.780	&	8.97	&	0.29	&	 8 	&	7.90	&	22	\\
\object{YZ Cet}	&	3056	&	M4 V	&	0.168	&	12.04	&	0.13	&	 21 	&	12.29	&	3, 30	\\
\object{$\tau$~Cet}	&	5283	&	G8.5	&	0.830	&	3.496	&	0.8	&	 8 	&	2.75	&	3, 30	\\
\object{55 Cnc}	&	5235	&	G8 V	&	0.930	&	5.95	&	0.96	&	 8 	&	5.20	&	21	\\

\hline																									
\hline																									
																
\end{tabular}																	
\tablebib{(1) \citet{Vogt2010}; (2) \citet{LoCurto2013}; (3) \citet{Hebrard2010}; (4) \citet{Staab2020}; (5) \citet{Dreizler2020}; (6) \citep{BertaThompson2015}; (7) \citet{Pinamonti2018}; (8) \citet{Turnbull2015}; (9) \citet{AstuDefru2017a}; (10) \citet{Kemmer2020}; (11) \citet{hawley96}; (12) \citet{Luque2019}; (13) \citet{tuomi2014}; (14) \citet{Stass2017}; (15) \citet{Jeffers2020}; (16) \citet{Kane2014}; (17) \citet{Lovis2011}; (18) \citet{Udry2019}; (19) \citet{Espinoza2020}; (20) \citet{motalebi2015}; (21) \citet{AstuDefru2017b}.}																	
\end{center}																	
\end{table*}

\section{TESS photometry and data reduction \label{sect.data}}
 
We used the high-precision photometry from TESS. In the first and second cycles, TESS data have a minimum cadence of approximately 2 minutes for small areas around selected targets \citep{Stass2019} called target pixel files (TPFs); starting from the third observing cycle 20s cadence data were added  for
selected targets. In particular,  in this work we used the calibrated light curves with two-minute cadence acquired from simple aperture photometry (SAP) of TPFs which are processed and corrected for common instrumental systematic errors and background contamination by the Science Processing Operations Center (SPOC)
pipeline. These flux time-series data are known as pre-search data conditioning SAP (PDCSAP$_{-}$FLUX) light curves\footnote{https://heasarc.gsfc.nasa.gov/docs/tess/data-products.html} \citep{jenkins2016}. To obtain these data files, we employed the \texttt{Python} package \texttt{Lightkurve}\footnote{ Lightkurve is a Python package for Kepler and TESS data analysis \citep{2018ascl.soft12013L}.}, which is also a useful tool for time-series analysis. 

The PDCSAP light curves still show fluctuations mostly due to residual instrumental effects and/or from stellar variability. We therefore detrended the light curves using a Savitzky-Golay filter \citep{SavGol-1964}. This removes low-frequency trends by fitting a low-degree polynomial within a sliding interval, the width of which is selected to avoid affecting the transit curves. For this work, we chose a third-degree polynomial, after evaluating a series of possible degrees. Considering that the timescales of the transits are much shorter than those of the photometric variability, we were able to remove the fluctuations with negligible effect on the subsequent data analysis. We visually inspected the corrected light curves of a number of known transiting planets to verify that the transit shapes were not affected. After that, we normalized the flux time-series data from different sectors\footnote{https://tess.mit.edu/observations/} by their median value and concatenated them. 
It is important to mention that although the data processing pipeline and the filtering described above help to correct most of the noise and systematic errors, some residual effects still remain in the data, affecting several cadences. Among them, the most important ones are random pointing variations due to momentum dump events from the spacecraft and contamination by diffuse light from the Earth and the Moon (see the TESS Instrument Handbook\footnote{https://archive.stsci.edu/missions/tess/doc/TESS$_{-}$Instrument$_{-}$Hand\-book
$_{-}$v0.1.pdf}). For every target, we studied the TESS data release notes (DRN) \footnote{Data release notes of TESS observed sectors: https://archive.stsci.edu/tess/tess$_{-}$drn.html} of all relevant sectors and recorded the cadences affected to take them into account in each analysis.

To account for correlated noise, we used a simple method proposed by \citet{pont2006}. Similar analyses were performed by \citet{cowan2012}, \citet{wong2016}, and \citet{wong2020}. The goal is to check how the scatter behaves as the light curve is binned with the different time bin sizes. 

\begin{figure*}
\centering
\includegraphics[width=0.34\linewidth]{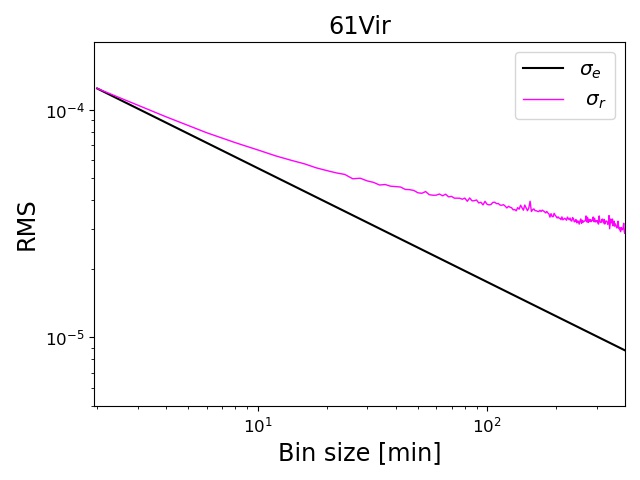}
\includegraphics[width=0.34\linewidth]{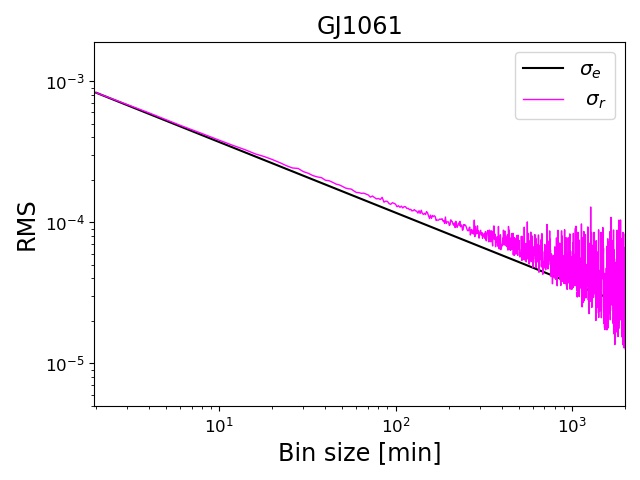}
\includegraphics[width=0.34\linewidth]{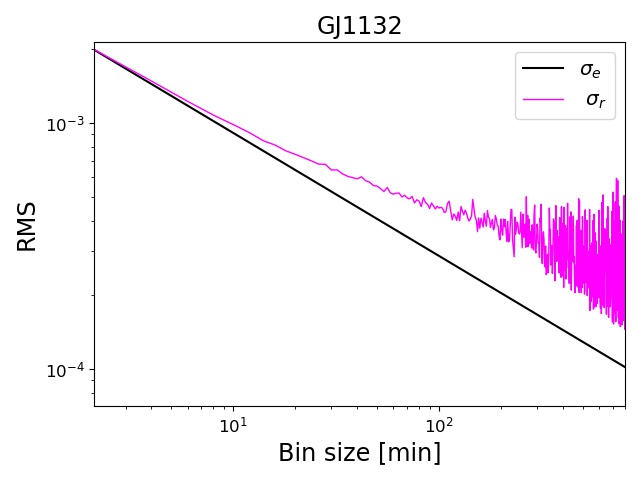}

\caption{\small{Root mean square for normalized light curves from three host stars of the sample, binned at various intervals (pink curve). The black line indicates the expected standard deviation, assuming pure additive Gaussian white noise, which scales as $\frac{1}{\sqrt{n}}$. The deviation of the pink curve from the black line indicates the presence of correlated red noise at the corresponding timescales. Left panel: 61 Vir has a $T_{mag}$ = 4.08. Middle panel: GJ 1061 has a $T_{mag}$ = 9.47. Right panel: GJ 1132 has a $T_{mag}$ = 12.14}}
\label{fig1:noise}

\end{figure*}

In this work, we compared the root mean square (RMS) of the time-binned light curves, $\sigma_{\rm r}$, with the RMS we expect for pure additive Gaussian white noise, that is $\sigma_{\rm e}$ = $\sigma_{1}/\sqrt{n}$, where $\sigma_{1}$ is the standard deviation of the unbinned data and $n$ is the number of points in each bin.
In Fig.~\ref{fig1:noise} we present the observed and expected RMS as a function of time bin-size for three stars in the sample with different magnitudes. 
The minimum bin size corresponds to the cadence of the telescope (2 min) and the maximum binsize usually depends on the number of data points available for each target.
We can see that both RMS values coincide for a bin size of 2 min 
and that the noise level is lower for brighter objects. On the other hand, the observed noise does not follow the expected relation as the data are binned. 

These plots allow estimation of the contribution of red noise for different timescales. The contribution of correlated noise is present in all cases, even for the smallest bins (e.g., in 20 minutes the red noise contribution is 10\%–25\% of the total noise).
In particular, for a timescale of around 2 hr (i.e., 120 min), which corresponds to the typical transit duration, red noise constitutes $\sim$50\% of the observed scatter. Comparing our results with Fig.\,1 in \citet{wong2020} we find a greater red noise contribution. This is explained by the relatively strong filtering of instrumental systematic noise applied by these latter author compared to that applied here. For example, they discard sharp segments and short-term flux variations, among others, all of which we elected to keep.

\section{Transit search \label{sect.transitsearch}}

In this section, we explore the complete photometric data looking for transits of known planets reported in Table~\ref{apptab2:planet} by means of a BLS search \citep{kovacs2002}. We searched for transits in the period range around the known planet periods (local-BLS), but we also performed a global search for transits of putative additional planets (global-BLS).

The frequency grid for the BLS\footnote{This research made use of Astropy (http://www.astropy.org) a community-developed core Python package for Astronomy \citep{astropy:2013, astropy:2018}} was defined by the maximum temporal extent of contiguous observed sectors, but we used resolution elements corresponding to the full span of the observations. For example, GJ~1061 was observed by TESS in sectors 3, 4, 30, and 31. Therefore, we explored periods up to 27 x 2 days, but we use a resolution element $\mathrm{d}\nu = 0.01/\Delta t$, where $\Delta t$ is the time difference between the first observation in Sector 3 and the last observation in Sector 31, and the factor in the numerator implies an oversampling factor of 100. 

For the local-BLS spectrum, we sampled frequencies in a range of $\pm 10 \sigma$ from the planet period, where $\sigma$ is the period uncertainty as reported in the literature. We used a resolution element as needed to sufficiently sample the frequency range around the planetary period. 

 Finally, as each periodogram has a rising trend of values toward longer periods because of the intrinsic effects of the distribution, we normalized them following \cite{ofir2014}. To remove this tendency and get the correct significance of every peak, we normalized the power spectra by a smoothed version with a moving median filter by choosing a suitable window size.

Six stars of the sample have known transiting planets (GJ~1132, GJ~3473, GJ~357, HD~158259, HD~213885 and HD~219134). In these cases, we obtained the parameters of the known transiting planet(s) using a BLS periodogram. We used these parameters to mask the data to eliminate known transits, and computed a second global-BLS periodogram. In the case of HD~219134, we performed this procedure twice, because it has two previously known transiting planets.

The period of the highest peak in the BLS periodogram and the corresponding signal-to-noise-ratio (SNR) are listed in Table \ref{apptab3:bls}. The first three columns of the table correspond to the global-BLS exploration of each star, and the last three columns correspond to the local-BLS exploration results of each planet. 

\begin{figure*}
\centering
\includegraphics[width=0.5\linewidth]{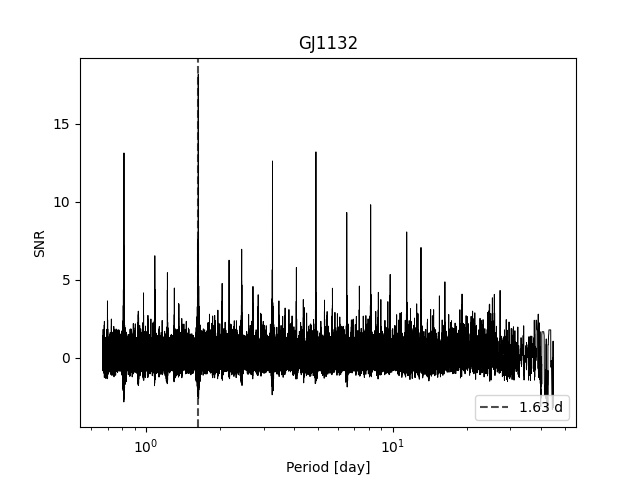}
\includegraphics[width=0.5\linewidth]{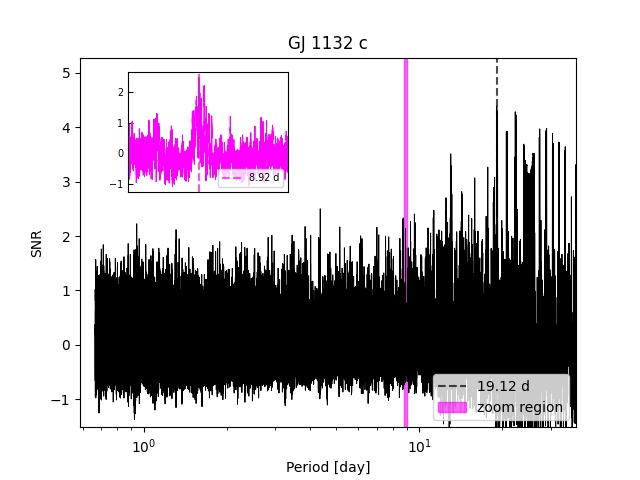}

\caption{\small{BLS power periodogram. The global-BLS is in black. The inset on the upper left corner shows the period range of pink shadowed zone. Dashed vertical lines point out maximum peaks in both range of exploration. Left panel: Global-BLS analysis for star GJ~1132 without masking the data before to run, the highest peak is detected at the period of the known transiting planet. Right panel: Global- and Local-BLS spectra computed using masked data; the latter explores the range of periods corresponding to the RV-detected planet GJ~1132~c.}}
\label{fig2:bls}

\end{figure*}

In Fig. \ref{fig2:bls} we show the BLSs of target GJ~1132 as an example. The left panel corresponds to BLS of unmasked data, clearly  showing the period associated with planet b \citep{BertaThompson2015}. The right panel presents the BLS periodogram of masked data. The global-BLS periodogram is represented in black, with a vertical dashed line at the position of the highest peak. The pink shadowed area corresponds to the period range of the local-BLS for the RV-detected planet GJ~1132~c \citep{bonfils2018}, and a zoomed-in version of this area is shown in the upper-left box inset.

Taking the SNR metric at face value would indicate that many systems have significant signals. However, visual inspection of the BLS periodograms indicates that the actual noise level is much higher and that the real limit for significant signal detection is probably much larger.
Following the empirical thresholds for transit detections proposed by \cite{dressing2015} and  \cite{livingston2018}, in  Sect.~\ref{sect.results} we discuss the peaks appearing with SNR $\geq$ 6 in the BLS spectra (see Table \ref{apptab3:bls}). Visual inspection of the light curves shows that in most cases these peaks are associated with fluctuations of flux due to instrumental systematic errors and other effects.

\section{Transit model \label{sect.model}}

In this section, we describe the models that are used to visualize the potential planetary transits. To model the transits we used the code \texttt{Batman} \citep{Kreidberg2015}. The transit model is parameterized by the planet period (P), the semimajor axis (\emph{a}/$R_\star$), the orbital inclination\footnote{We assumed edge-on orbits, that is \emph{i} $= 90^{\circ}$.}, the orbital eccentricity (\emph{e}), the longitude of pericenter (\emph{$\omega$}), and radius ratio ($R_P$/$R_\star$). Most of the parameters were obtained from the literature as described in Sect.~\ref{sect.sample}. However, the planetary radii were obtained from the internal structure model given by \cite{Fortney_2007,Fortney_Err2007}. This model predicts the radii from masses ranging from 0.01 to 1000 M$_{\oplus}$ for different combinations of ice, rock, and iron. Based on the results of the modeling, the authors provided analytical relationships between masses and radii for (1) planets of pure ice and ice and rock mixtures (``Ocean planets''; $R_\text{wat}$), and
(2) planets composed of pure rock, rock and iron mixtures, and pure iron (``Terrestrial planets''; $R_\text{ter}$):

\begin{equation*}
\begin{split}
R_\text{wat} =&(\,0.0912~imf + 0.1603\,) (\,log M\,)^2 +\\     &(\,0.3330~imf + 0.7387\,) (\,log M\,) +
(\,0.4639~imf + 0.1193\,)\\
R_\text{ter} =& (\,0.0592~rmf + 0.0975\,) (\,log M\,)^2 +\\ &(\,0.2337~rmf + 0.4938\,) (\,log M\,) +
(\,0.3102~rmf + 0.7932\,)
\end{split}
\end{equation*}

\noindent with radii in R$_{\oplus}$, and masses in M$_{\oplus}$. The $imf$ and $rmf$ coefficients indicate the composition, and stand for the "ice mass fraction" for ocean planets (1.0 for pure ice and 0.0 for pure rock) and the "rock mass fraction" for terrestrial planets (1.0 for pure rock and 0.0 for pure iron), respectively. In all cases, the ice layer is always above the rock layer, which in turn overlays iron. From the measured minimum masses, and assuming perfectly edge-on orbits ($i = 90^\circ$), we computed radii for pure ice, pure rock, and pure iron compositions. The resulting radii are reported in columns 6, 7, and 8 of Table~\ref{apptab2:planet}.

We modeled the light curves using a quadratic limb darkening (LD) law. The quadratic coefficients $a$ and $b$ for the TESS bandpass were computed from the tables in \cite{claret2018} by interpolating the values of T$_{eff}$ and log\,(g) of the host star. 

Given the large pixel size of TESS (21$''$), the contribution to the flux inside the photometric aperture from nearby stars could be significant. However, PDC$_{-}$SAP light curves are corrected for flux contamination. Also, these are relatively bright stars, and so the contamination that may be present is assumedly low. We therefore decided not to include contamination in our transit model.

\subsection*{Detection limits \label{sect.detection}}

We estimated upper limits for transit depths and planetary radii for each planet. Results are presented in Table~\ref{apptab4:upplim}. To do this, we phase folded the light curves using the mid-transit times and periods from the literature (see Table~\ref{apptab2:planet}). Assuming edge-on orbits, we estimated the respective transit windows (i.e., transit duration). 

We then computed the standard deviation of the measured relative flux within the transit window. We considered this as a conservative upper detection limit for the depth of transits (Col. 2, upper limit $(R_{\rm p}/R_{\star})^2$). The corresponding planetary radius R$_{\rm p}$ is presented in column 3\footnote{We could not compute the upper limits for planets DMPP-1 c and d, GJ~180~c, GJ~682~b, and GJ~887 b and c because we have incomplete orbital parameters.}. This gives us a limit above which a transiting planet would be detected. In all cases, the computed upper limit is below 2.4 R$_{\oplus}$, and so we excluded transits of planets bigger than 2.4 R$_{\oplus}$ for all systems; the median excluded radius is 1.4 R$_{\oplus}$, and the minimum is 0.48 R$_{\oplus}$.

Finally, we compared the upper planetary radius with the radius obtained from the models (showing in Table~\ref{apptab2:planet}). Of the 68 sampled planets, 28 have upper planetary radii smaller than R$_{iron}$, the radius of a planet made of iron. For these systems, a transit can be fully discarded under the model assumptions. For 51 planets, the upper limit is smaller than R$_{rock}$.

\section{Results \label{sect.results}}

To test our methodology, we applied it to 55~Cnc~e, a well-studied transiting super-Earth planet orbiting a nearby bright star. This planet was discovered by \citet{mcarthur2004} who reported an 
incorrect orbital period of 2.8 days. Some years later, \citet{dawsonfabrycky2010} argued that the correct period of the planet was about 0.74 days, which was later confirmed by transit observations \citep{winn2011}. Over the years, combining RV measurements and transit observations \citep[among others]{winn2011, demory2011, bou2018} has considerably improved the determination of the parameters of  55~Cnc~e, such as mass ($\sim$8~M$_{\oplus}$) and radius ($\sim$1.9~R$_{\oplus}$).

\begin{figure*}
\centering
\includegraphics[angle=0,trim = 02mm 0mm 15mm 00mm, clip,width=0.32\linewidth]{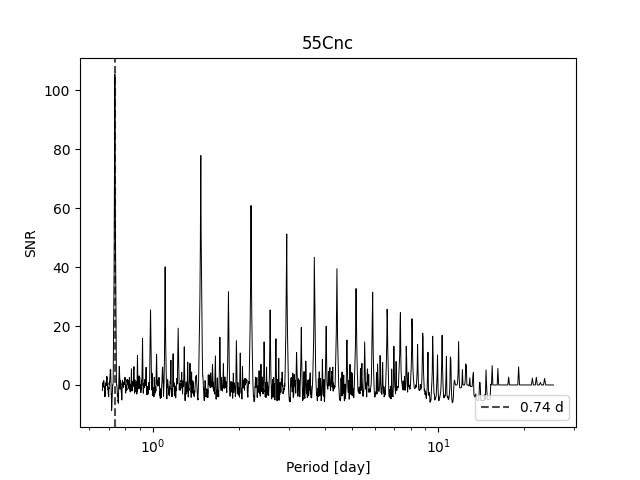}
\includegraphics[angle=0,trim = 00mm 0mm 00mm 00mm, clip,width=0.35\linewidth]{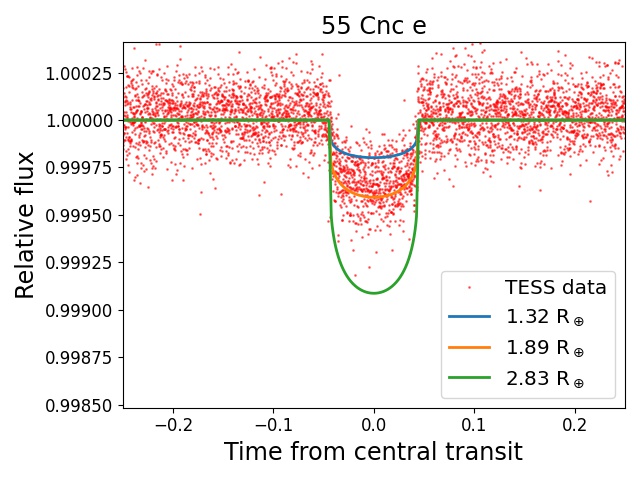}
\includegraphics[angle=0,trim = 00mm 0mm 00mm 00mm, clip,width=0.35\linewidth]{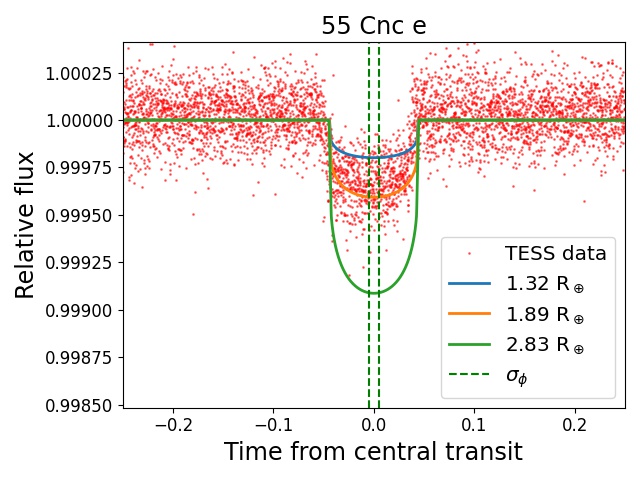}
\caption{\small{55~Cnc~e.~Left: BLS periodogram, dashed vertical black line indicates the maximum peak.~Middle: Transit model plus light curve folded with BLS results.~Right: Transit model plus light curve folded with literature parameters. Solid lines of different colors correspond to models for different radii (green for R$_{ice}$, orange for R$_{rock}$ and blue for R$_{iron}$), green dashed vertical lines indicate  $\pm \sigma_{\phi}$, the orbital phase uncertainty.}}
\label{fig3:55Cnc}

\end{figure*}

Figure~\ref{fig3:55Cnc} (left panel) shows the TESS BLS power spectrum, which exhibits a clear peak with a SNR of 105 at the period corresponding to the known transiting planet and its harmonics. In the middle panel, we plot the phase-folded curve using the period and phase obtained with the BLS, where we have superimposed the synthetic models computed using literature parameters and estimated radii, as described above. Finally, the rightmost panel of the figure presents the light curve phase folded using the period and transit phase from the literature (Table~\ref{apptab2:planet}.) with vertical dashed lines that indicate the phase uncertainty ($\sigma_{\phi}$).
These figures show the good accuracy both in BLS results and in the calculation of T$_{C}$ from the literature parameters. In addition, we see that the estimated radii are in agreement with the measured value of 1.9 R$_{\oplus}$ We can also check this in Table~\ref{apptab4:upplim}, where the upper limit to the radius is smaller than the radius of a planet of pure rock. This shows the reliability of the methodology applied.

\subsection{Null transit detections\label{sect.null}}

In this section we discuss the noteworthy cases from the BLS results, defined as those with maximum peak with SNR~$\geq$~6 (see Table~\ref{apptab3:bls}). These discussions are based on visual inspection of the light curves and the TESS DRN\footnote{TESS data release notes: https://archive.stsci.edu/tess/tess$_{-}$drn.html} of the relevant sectors.
Six systems of the sample already have at least one known transiting planet. These are GJ~1132, GJ~3473, GJ~357, HD~158259, HD~213885, and HD~2191345. In these cases, we computed the BLS spectrum masking the cadences that correspond to the known transits. 

\subsubsection{61~Vir}

61~Virginis is a system with three RV planets with periods of 4.2, 38.0, and 124.0 days \citep{Vogt2010}. Only the innermost 5.1-M$_{\oplus}$ planet fulfills the criterion (period $<$ 30 days) to be included in our sample. This target was observed by TESS in a total baseline of 25.27 days.

\begin{figure}[!]
\begin{minipage}{\linewidth}
\begin{center}

\includegraphics[angle=0,trim = 0mm 00mm 0mm 00mm, clip, width=\linewidth]{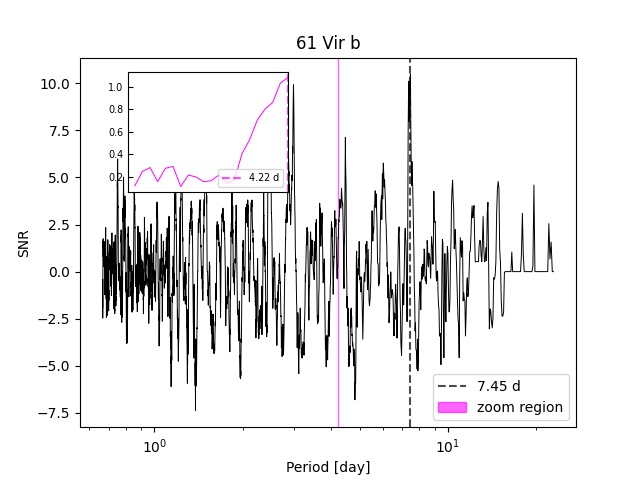}
\includegraphics[angle=0,trim = 0mm 00mm 0mm 00mm, clip, width=\linewidth]{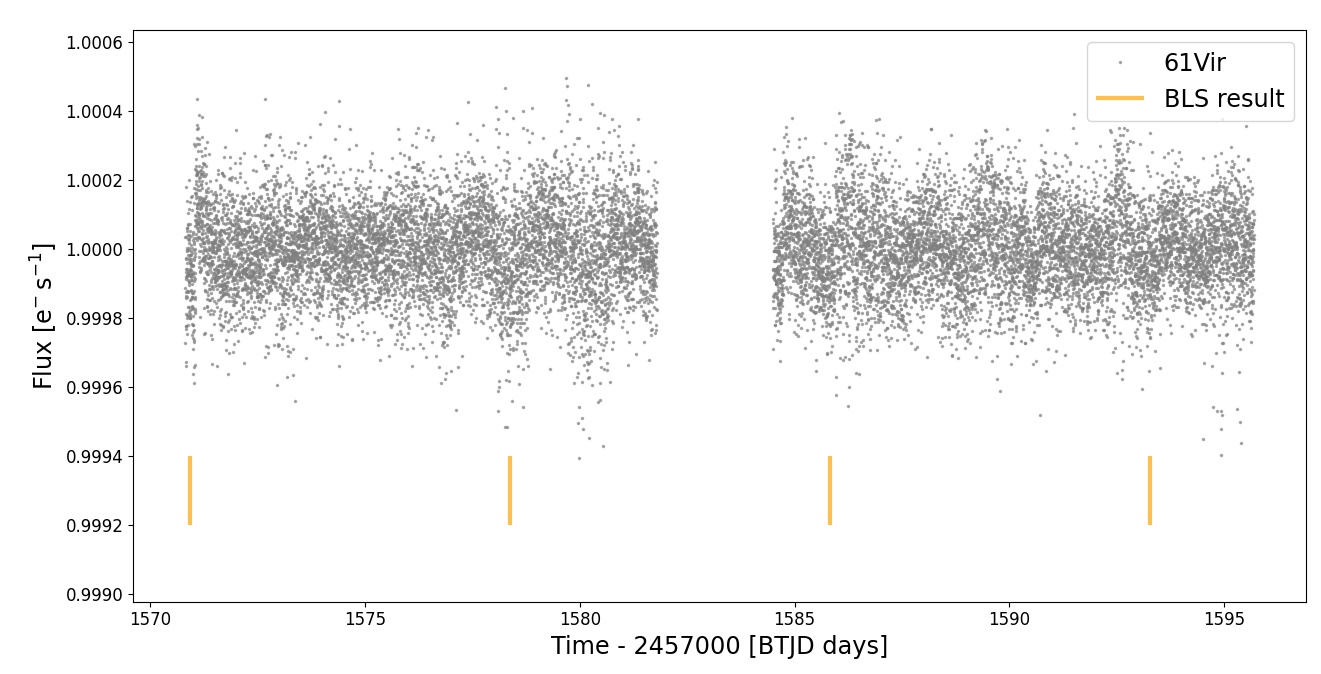}

\end{center}
\caption{\small{61~Vir.~Top: Global-BLS periodogram plus inset with a local-BLS spectrum. Dashed vertical black line indicates the maximum peak and period respective, and pink zone points out the region of the local-BLS analysis.~Bottom: Light curve from TESS photometry. Orange lines show the signals detected by global-BLS.}}
\label{fig:61vir}
 \end{minipage}
\end{figure}

The BLS power spectrum is presented in Fig.~\ref{fig:61vir}. A peak with SNR $\sim$ 10.5 is seen at a period of 7.5 days. However, this is not due to a transiting planet. We can see in the light curve (bottom panel of Fig.~\ref{fig:61vir}) that  one of the putative transits falls just at the beginning of the orbit in sector 10.
This part of the time series comes immediately before cadences associated with instrumental systematic errors that causes the measured SNR. Additionally, the second transit is aligned with a pointing instability, according to the corresponding data release notes. 

\begin{figure}[!]
\begin{minipage}{\linewidth}
\begin{center}
\includegraphics[angle=0,trim = 0mm 00mm 0mm 00mm, clip, width=\linewidth]{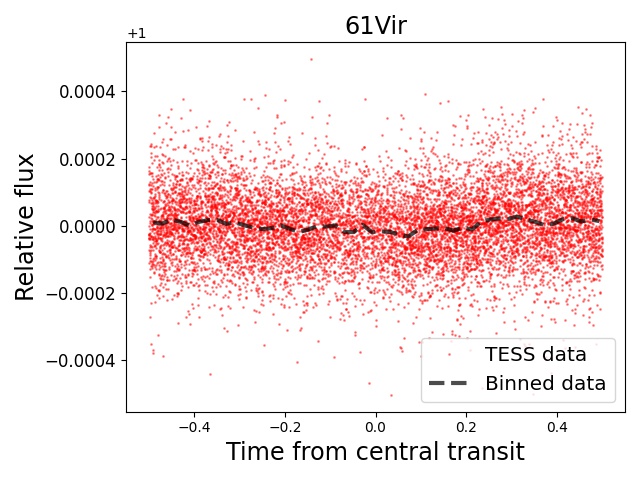}
\end{center}
\caption{\small{61~Vir. Light curve folded at a period of 1.49 days, detected from the BLS-spectrum obtained discarding cadences at $<$ 1571.05 days. Red points are the TESS data and the dashed line corresponds to the binned flux.}}
\label{fig:61virnew}
 \end{minipage}
\end{figure}

To verify that this detection is spurious, we computed the BLS spectrum without including the cadences at the beginning of the orbit ($<$ 1571.05 days). We obtained the highest peak at a period of 1.49 days with SNR $\sim$ 9. We phase folded the light curve to this period and binned the folded data (see Fig.~\ref{fig:61virnew}). No transit feature is detected in the folded curve, and so we discarded a possible planetary transit there.

\begin{figure}[!]
\begin{minipage}{\linewidth}
\begin{center}
\includegraphics[angle=0,trim = 0mm 00mm 0mm 00mm, clip, width=\linewidth]{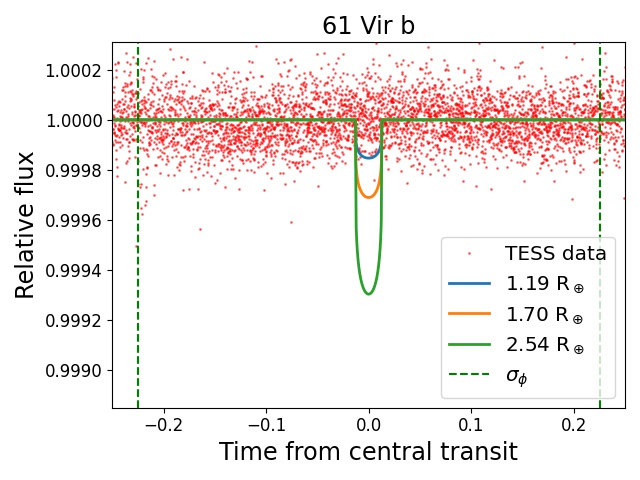}
\end{center}
\caption{\small{61~Vir~b. Transit model plus light curve folded with literature parameters. Red points are the TESS data binned. Solid lines of different colors correspond to models for different radii (green for R$_{ice}$, orange for R$_{rock}$ and blue for R$_{iron}$), and the dashed green lines indicate  $\pm ~\sigma_\phi$, the orbital phase uncertainty.}}
\label{fig:61virb}
 \end{minipage}
\end{figure}

On the other hand, in Fig.~\ref{fig:61virb}, we show the folded light curve using the literature ephemerides for planet b, together with the models of different planetary radii. A central planetary transit is clearly absent from the light curve for planets with radii above that of a planet of pure iron. This agrees with the computed local-BLS spectrum, which does not show any peak in the region around the period of planet b.

\subsubsection{GJ~15~A }

\begin{figure}[!]
\begin{minipage}{\linewidth}
\begin{center}

\includegraphics[angle=0,trim = 0mm 00mm 0mm 00mm, clip, width=\linewidth]{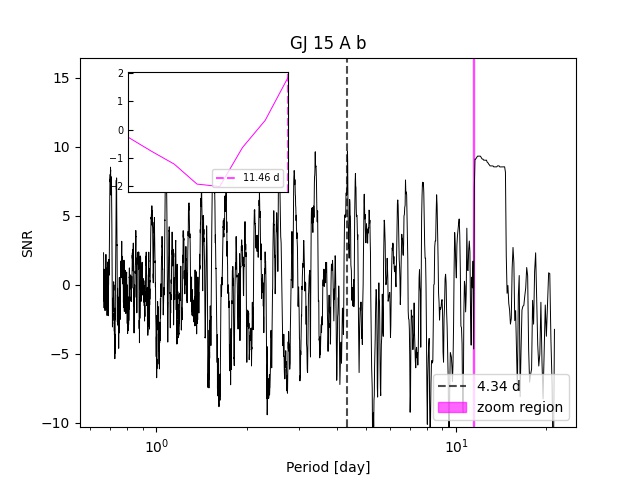}
\includegraphics[angle=0,trim = 0mm 00mm 0mm 00mm, clip, width=\linewidth]{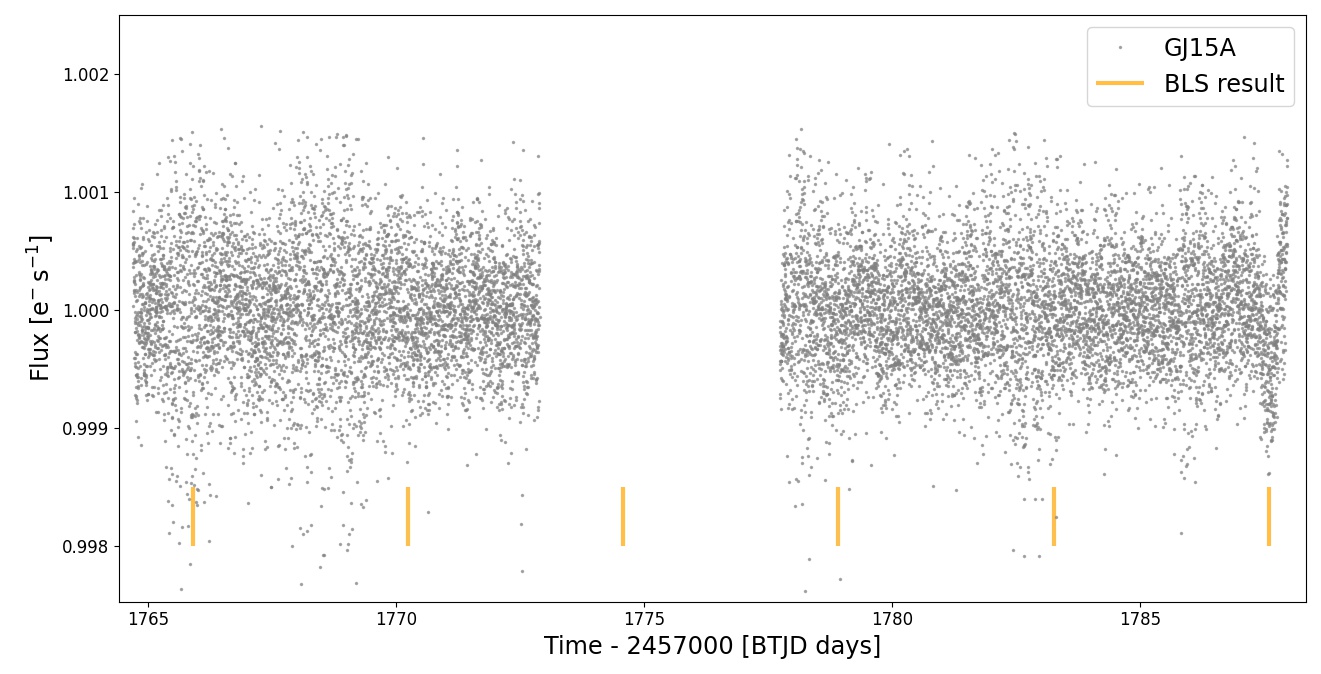}

\end{center}
\caption{\small{Same as Fig.~\ref{fig:61vir} but for GJ~15~A.}}

\label{fig:GJ15A}
 \end{minipage}
\end{figure}

GJ\,15\,A~b is a planet with a $M\sin i$ of 3.03 M$_{\oplus}$ orbiting a M-dwarf star with a period of 11.44 days \citep{howard2014,Pinamonti2018}. This star was observed by TESS in Sector 17. The global- and local-BLS spectra are presented in the top panel of Fig.~\ref{fig:GJ15A}, and the results are listed in Table \ref{apptab3:bls}.

\begin{figure}[!]
\begin{minipage}{\linewidth}
\begin{center}
\includegraphics[angle=0,trim = 0mm 00mm 0mm 00mm, clip, width=\linewidth]{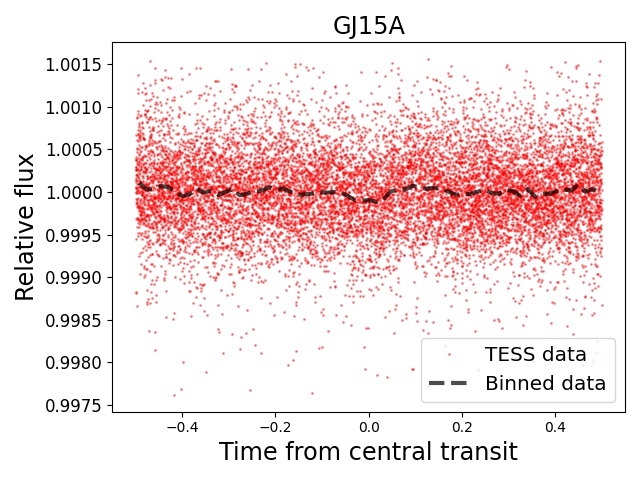}
\end{center}
\caption{\small{GJ~15~A. Folded light curve with a period of 1.3 days obtained from the BLS spectrum after discarding final cadences of sector 17. Red points are the TESS data and the dashed line corresponds to the binned flux.}}
\label{fig:GJ15Anew}
 \end{minipage}
\end{figure}

The global-BLS exhibits a peak at 4.34 days with a SNR $\sim$ 9.7. In the bottom panel of Fig.~\ref{fig:GJ15A} we show the light curve with vertical lines indicating the positions of the transits. Visual inspection of the light curve shows that one putative transit is located at the end of sector 17, in a region with increased instrumental variability that was not adequately corrected. We reran the BLS algorithm discarding those cadences and obtained a new peak close to 1.3 days with a SNR $\sim$ 9. The folded light curve with this new period is presented in Fig.~\ref{fig:GJ15Anew}. We can see a flux drop of $\sim$ 0.01\% in depth (100 ppm), which, for the stellar radius listed in Table~\ref{tab1:star}, would correspond to a planet with a radius of 0.41 $R_\oplus$. However, for a circular orbit, the transit of this planet should last around 63 minutes, while the measured duration is of about 3 hours. This signal is therefore probably not of planetary origin. On the other hand, the duration measured would be consistent with depth and period if we were to consider an eccentric orbit with $e$ $>$ 0.738. 

\begin{figure}[!]
\begin{minipage}{\linewidth}
\begin{center}
\includegraphics[angle=0,trim = 0mm 00mm 0mm 00mm, clip, width=\linewidth]{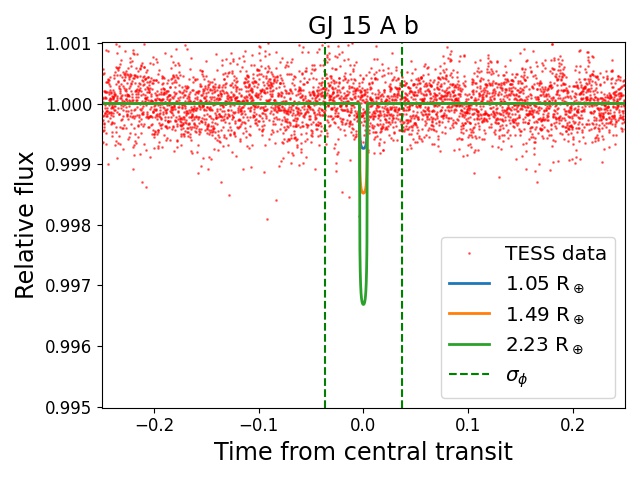}
\end{center}
\caption{\small{Same as Fig.~\ref{fig:61virb} but for GJ~15~A~b.}}

\label{fig:GJ15Ab}
 \end{minipage}
\end{figure}

Also, there is no significant peak in the local-BLS spectrum around the period of GJ~15~A b. In Fig.~\ref{fig:GJ15Ab} we present the phase-folded TESS light curve using the period from the literature. Model transit curves computed as described in Sect.~\ref{sect.model} are superimposed. As in the previous case, we conclude that no transit is detected.

\subsubsection{GJ~887} 

GJ~887 is a multiplanet system with two super-Earth planets discovered by \cite{Jeffers2020} using radial velocity measurements. They have orbital periods of 9.262 days and 21.789 days, for planet b and c, respectively. This target was observed in the years 2018 (sector 2) and 2020 (sector 28) by TESS (see Table~\ref{tab1:star}). \cite{Jeffers2020} also analyzed TESS photometry without finding evidence of transits, but these authors only used data from sector 2.

\begin{figure}[!]
\begin{minipage}{\linewidth}
\begin{center}
\includegraphics[angle=0,trim = 0mm 00mm 0mm 00mm, clip, width=\linewidth]{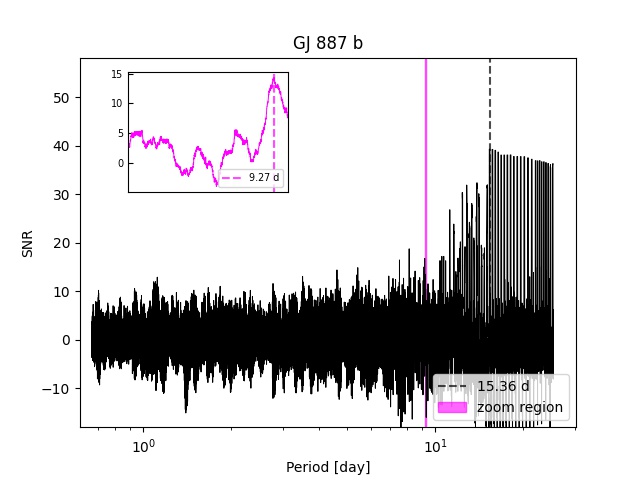}
\end{center}
\caption{\small{GJ~887. Global-BLS periodogram plus a small box with a zoom region showing the local-BLS for planet b. The dashed vertical black line indicates the maximum peak and respective period, and the pink zone points out the zoomed-in region shown in the small box.}}
\label{fig:GJ887}
 \end{minipage}
\end{figure}

\begin{figure*}

%
\centering
\includegraphics[width=0.5\linewidth]{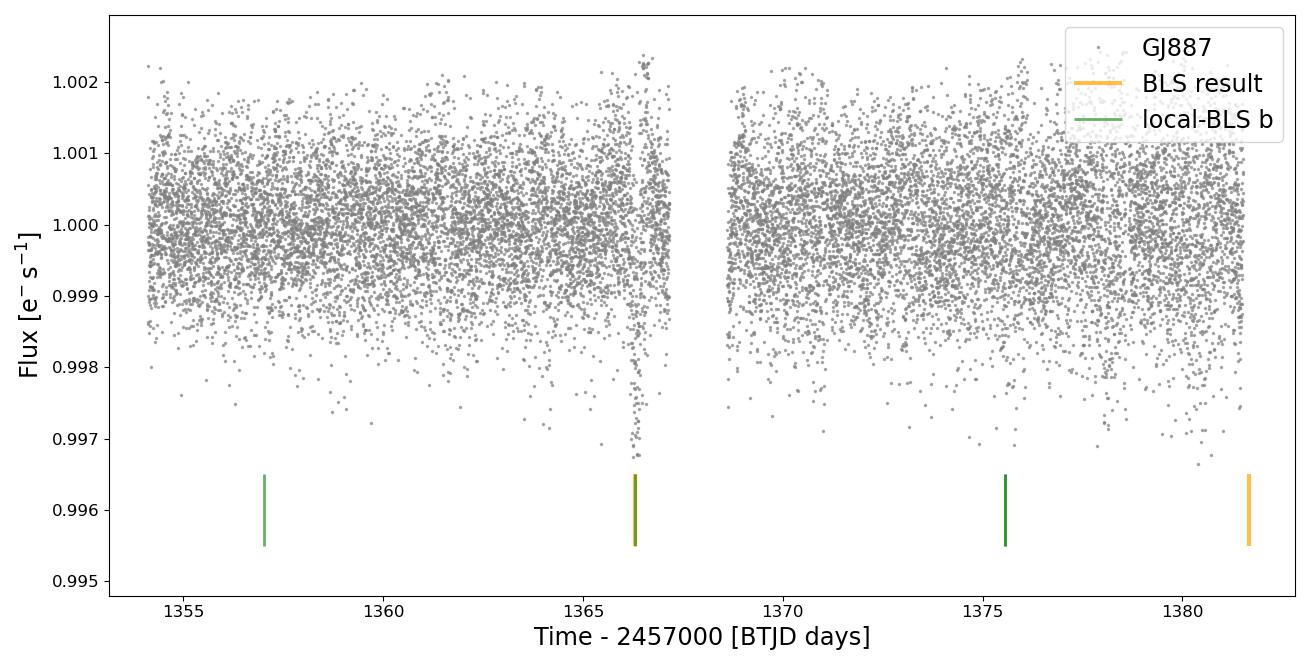}
\includegraphics[width=0.5\linewidth]{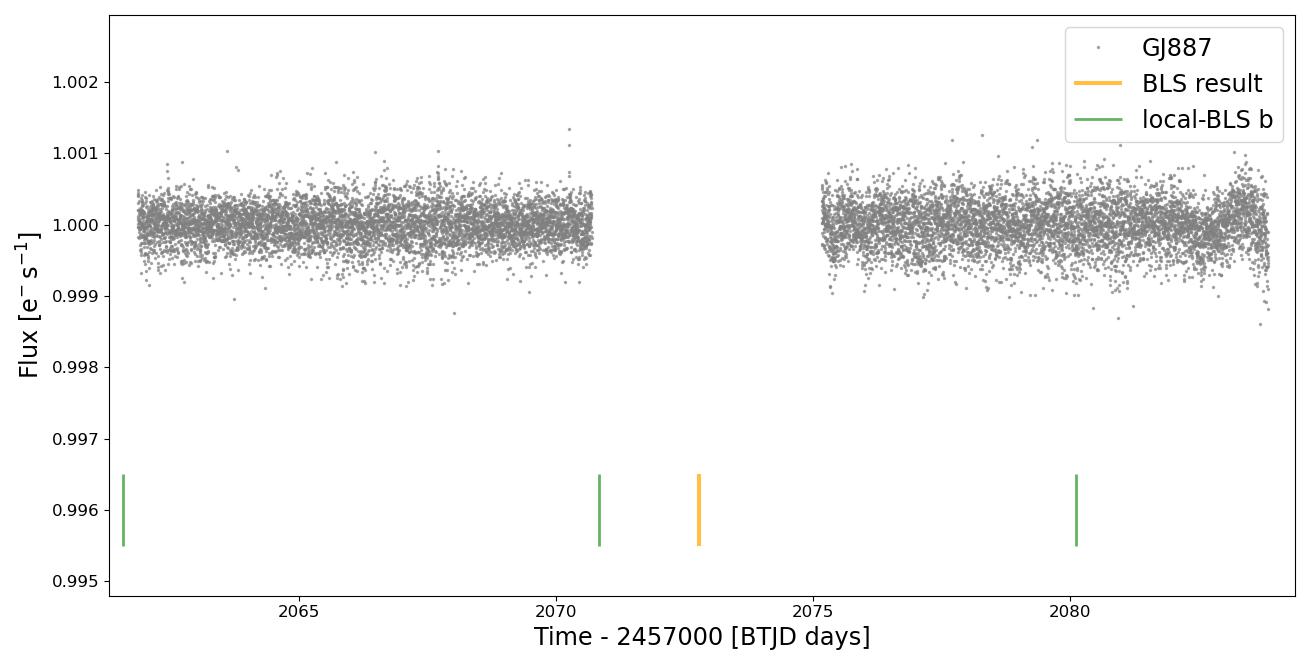}
\\
\caption{\small{Light curve from TESS photometry of GJ~887. The left panel shows data from sector 2 and the right panel shows data from sector 28. Orange lines show the signals detected by global-BLS.  Green lines show the signals detected by local-BLS for planet b.
}}
\label{fig:GJ887lc}
\end{figure*}

We found a peak at a period of 15 days and a SNR~$\sim$~39 in the global-BLS spectrum and a signal of SNR~$\sim$~14 around the period of planet b. The local-BLS spectrum for planet c does not present any noteworthy feature. In Fig.~\ref{fig:GJ887} we show the global-BLS spectrum and corresponding local-BLS spectrum of planet b. The TESS light curves of both sectors are shown in Fig.~\ref{fig:GJ887lc}. The indicated transits correspond to the maximum peak for global-BLS (top panels) and for the local-BLS (bottom panels). There is a flux drop around BTJD = $2\,458\,366$, when cadences are affected by scattered Moon and Earth light. Also in its vicinity there is increased jitter noise caused by momentum dump events (See DRN for sector 2). The affected cadences are responsible for the peak found by the global-BLS analysis. The other transit times coincide with TESS downlink gaps.  

\begin{figure}[!]
\begin{minipage}{\linewidth}
\begin{center}

\includegraphics[angle=0,trim = 0mm 00mm 0mm 00mm, clip, width=\linewidth]{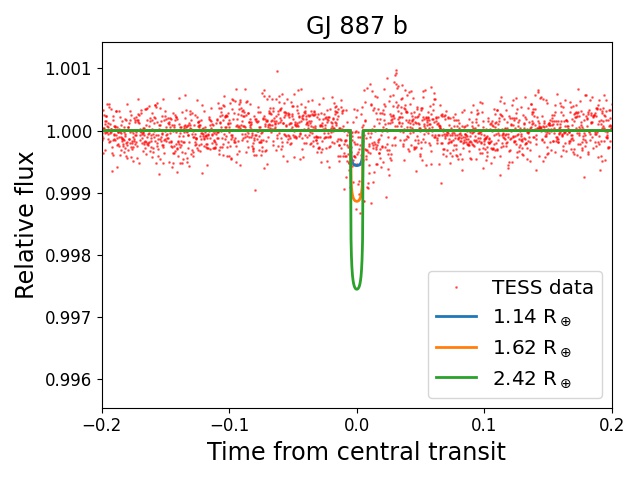}
\end{center}
\caption{\small{Same as Fig.~\ref{fig:61virb} but for planet GJ~887~b except that the parameters are obtained from the local-BLS analysis.}}
\label{fig:GJ887fold}
 \end{minipage}
\end{figure}

As we could not obtain full orbital parameters from the literature, we instead folded the curve using the parameters obtained from the local-BLS analysis and compared it with the synthetic models superimposed. This is shown in Fig.~\ref{fig:GJ887fold}, where we can see a central drop with some offset from phase zero; this corresponds to cadences closer to $2\,458\,366$ BTJD. We therefore cannot claim a transit detection because the data involved are not reliable, and no other flux drops of that type were observed in the data.

\subsubsection{HD~136352} 

This star hosts three RV planets detected by \citet{Udry2019} using the High Accuracy Radial velocity Planet Searcher (HARPS) spectrograph \citep{Mayor2003}. The two inner planets were discovered to transit using TESS photometry \citep{Kane2020}, and a follow up with the CHaracterising ExOPlanets Satellite (CHEOPS) telescope showed that the third outer planet ($P_{d} = 107$ days) also transits its star \citep{delrez2021}. The analysis presented here was performed independently.
The two innermost planets b and c have masses of 4.81 and 10.8 M$_{\oplus}$ and orbital periods of 11 days and 27 days, respectively, and were therefore included in our sample.

\begin{figure}[!]
\begin{minipage}{\linewidth}
\begin{center}

\includegraphics[angle=0,trim = 0mm 00mm 0mm 00mm, clip, width=\linewidth]{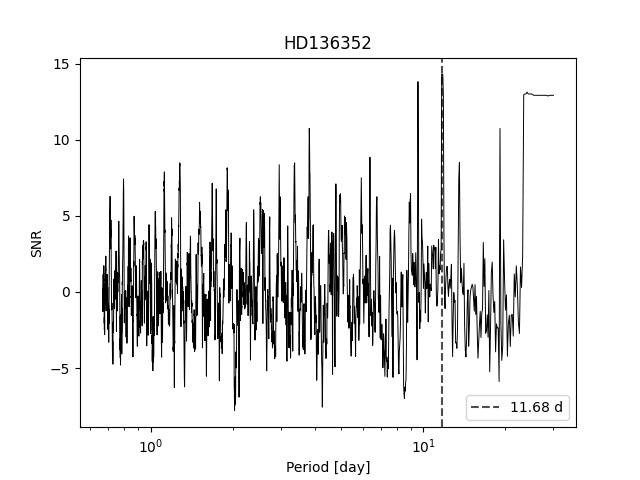}
\includegraphics[angle=0,trim = 0mm 00mm 0mm 00mm, clip, width=\linewidth]{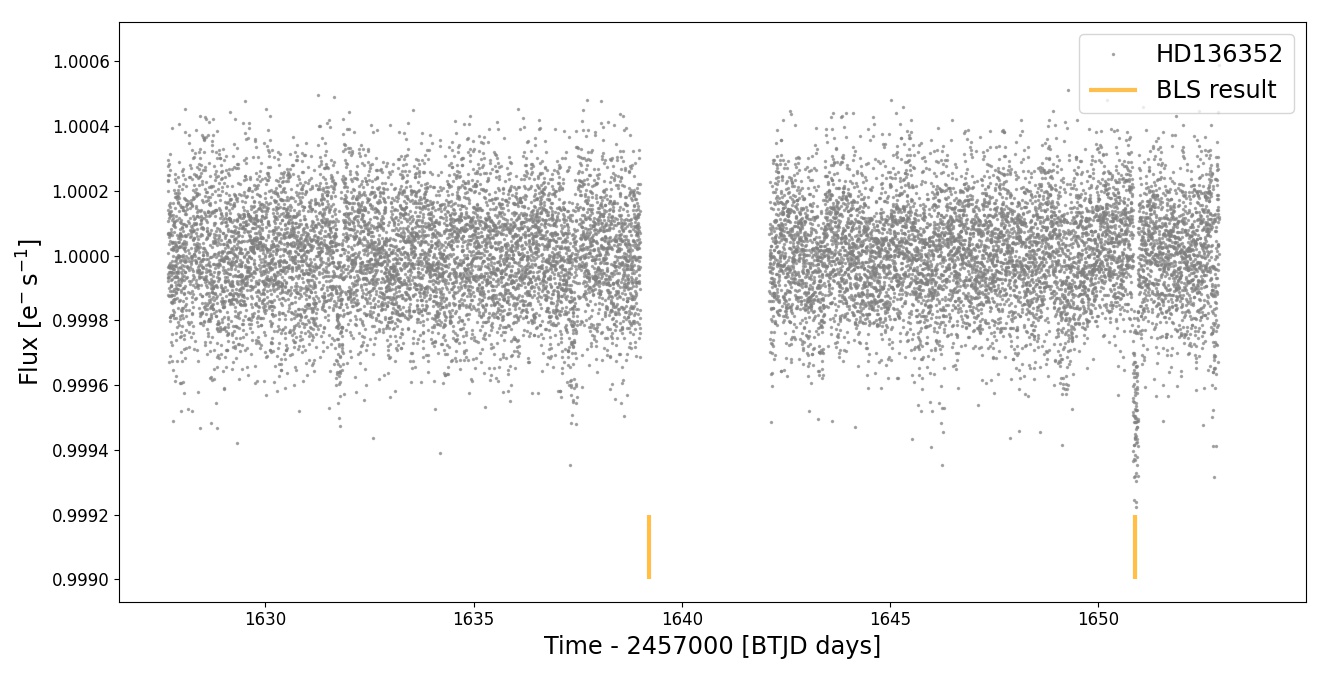}

\end{center}
\caption{\small{Same as Fig.\ref{fig:61vir} but for HD~136352. Top panel only shows the Global-BLS periodogram for this target.}}
\label{fig:HD136352}
 \end{minipage}
\end{figure}

A blind search using the global-BLS analysis shows a significant peak close to the period of planet b (SNR~$\sim$~14 at 11.68 days, see also Table~\ref{apptab3:bls}). We show in Fig.~\ref{fig:HD136352} the respective periodogram and the TESS light curve with the detected transits indicated.
We see that the algorithm detected the deepest transit (produced by planet c; around BTJD= $2\,458\,651$), which for the peak period means a second transit should have occurred during the data gap between TESS orbits around BTJD = $2\,458\,639$. When we search for these planets blindly, we miss them. One of the reasons is that we have a full baseline of 25 days. Only one transit of planet c is therefore present in the TESS light curve. Another reason is that the PDCSAP light curve exhibits residual systematic errors from the reduction and correction process. To solve this, \cite{Kane2020} produced their own custom light curve from TPFs of this bright star.

\begin{figure}[!]
\begin{minipage}{\linewidth}
\begin{center}

\includegraphics[angle=0,trim = 0mm 00mm 0mm 00mm, clip, width=\linewidth]{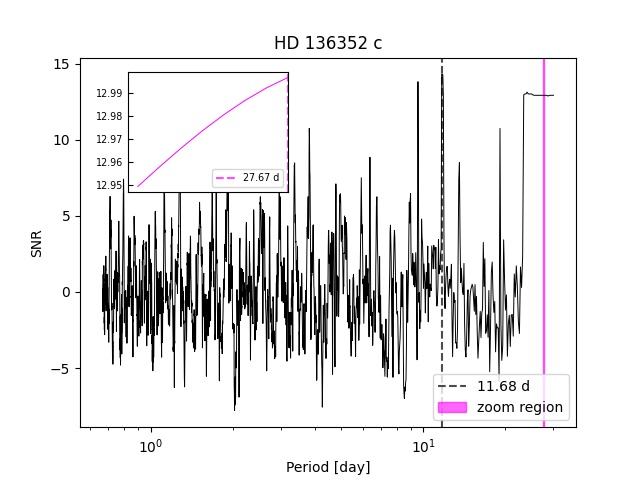}
\includegraphics[angle=0,trim = 0mm 00mm 0mm 00mm, clip, width=\linewidth]{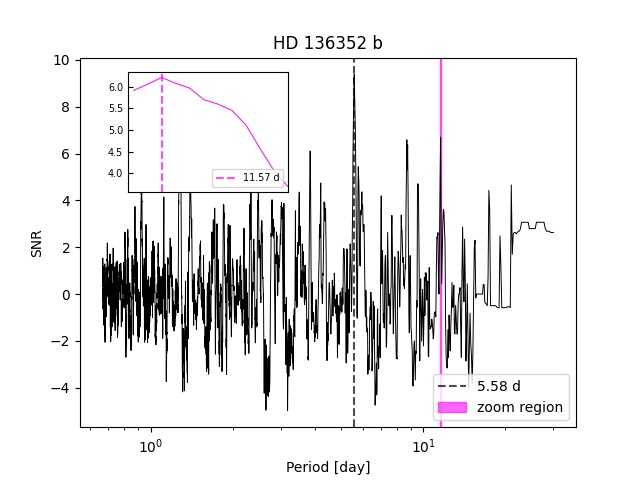}
\includegraphics[angle=0,trim = 0mm 00mm 0mm 00mm, clip, width=\linewidth]{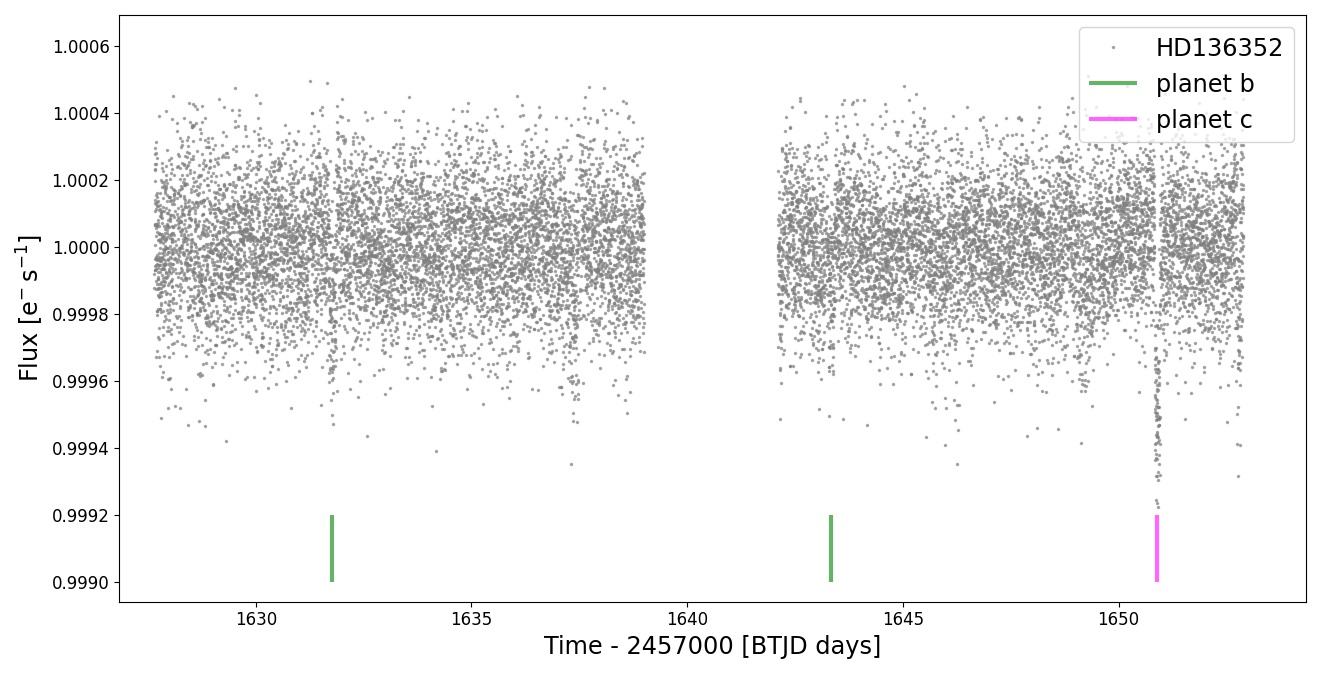}

\end{center}
\caption{\small{HD~136352 b and c. Top panel: Global-BLS periodogram plus inset showing local-BLS spectrum for planet c. Middle panel: Global-BLS periodogram computed masking the signal of planet c plus inset showing local-BLS spectrum for planet b. Dashed vertical black lines indicate the maximum peaks and pink zones point out the regions presented in the insets. Bottom panel: Light curve from TESS photometry with green and magenta lines indicating the signal detected by the local-BLS analysis for planets b and c, respectively.}}
\label{fig:HD136352bc}
 \end{minipage}
\end{figure}

On the other hand, restricting the periods to be close to the two inner planets reveals one deep transit for planet c and two transits of planet b (see Fig.~\ref{fig:HD136352bc}), and shows that the period detected in the global analysis is incorrect.
In Fig.~\ref{fig:HD136352bc} we show global-BLSs with zoomed regions in the vicinity of the period of each planet (SNR$_{c}$~$\sim$~13 in Per$_{c}$~$\sim$~27.67 days and SNR$_{b}$~$\sim$~6.22 in Per$_{b}$~$\sim$~11.57 days, see Table~\ref{apptab3:bls}). The transit of planet c was masked before computing the local-BLS spectrum for planet b. In the same figure we present the TESS light curve with the corresponding planet transits (found by the local-BLS) indicated with colors.

\begin{figure}[!]
\begin{minipage}{\linewidth}
\begin{center}
\includegraphics[angle=0,trim = 0mm 00mm 0mm 00mm, clip, width=\linewidth]{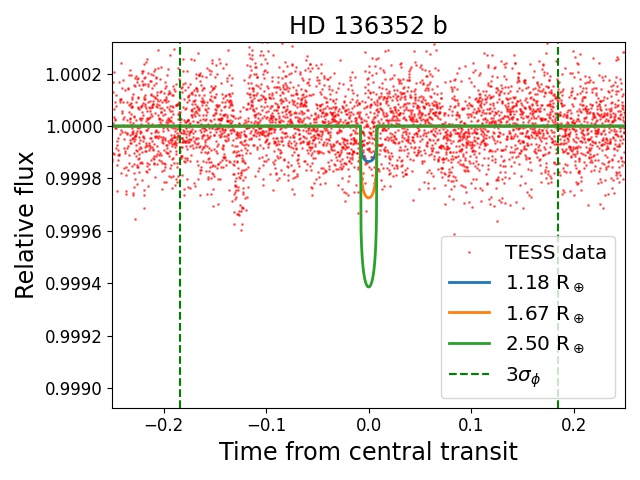}
\includegraphics[angle=0,trim = 0mm 00mm 0mm 00mm, clip, width=\linewidth]{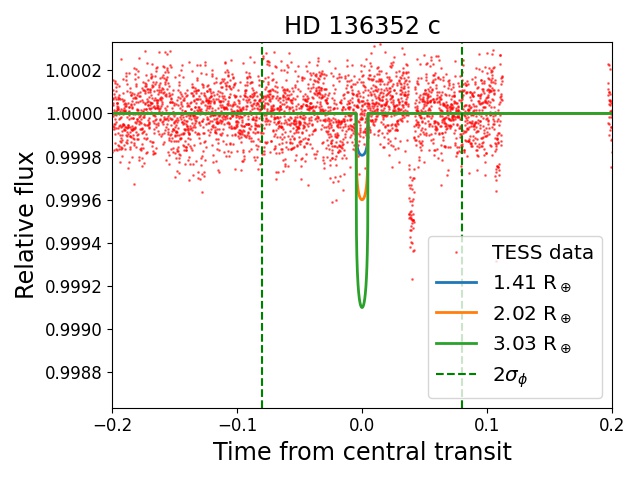}
\end{center}
\caption{\small{Same as Fig~\ref{fig:61virb} but for planets HD~136352~b and c. Except that here dashed green lines indicate $\pm$ 2 or 3~$\sigma_\phi$ (as indicated in each figure) as the orbital phase uncertainty.}}%
\label{fig:HD136352fold}
 \end{minipage}
\end{figure}

In two panels of Fig.~\ref{fig:HD136352fold}, we present the phase-folded light curves for each planet from the literature parameters, together with the corresponding models. Both transits appear clearly in the data but with an offset from phase zero, which may be a consequence of the low precision in the determination of T$_{C}$ from RV measurements. Another possible cause could be a timing difference in transit times between the epochs of the HARPS and TESS observations.

\subsubsection{HD~158259} 

\begin{figure}[!]
\begin{minipage}{\linewidth}
\begin{center}
\includegraphics[angle=0,trim = 0mm 00mm 0mm 00mm, clip, width=\linewidth]{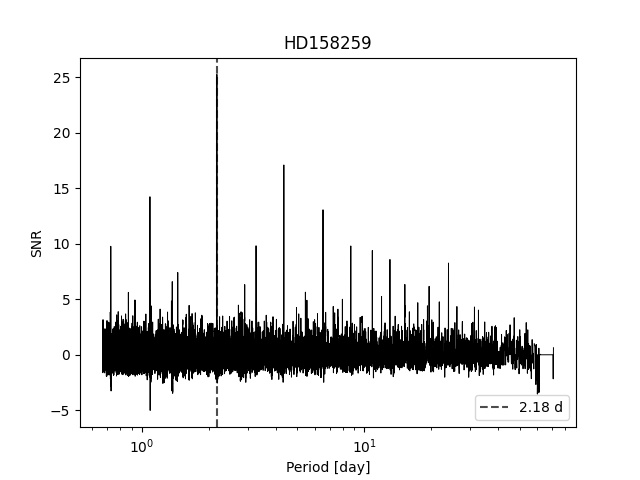}
\includegraphics[angle=0,trim = 0mm 00mm 0mm 00mm, clip, width=\linewidth]{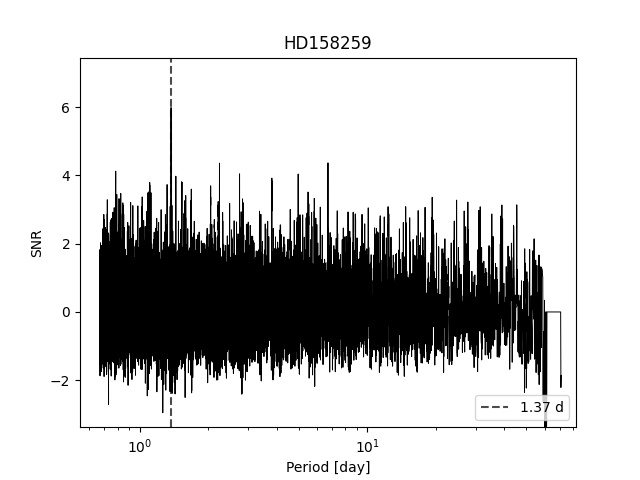}

\end{center}
\caption{\small{HD~158259. Top panel: Global-BLS periodogram showing a high S/N of transiting known planet b. Bottom panel: Global-BLS masking planet b. Dashed vertical black lines indicate the maximum peak and period respective.}}
\label{fig:HD158259}
 \end{minipage}
\end{figure}

\begin{figure*}

\centering
\includegraphics[width=0.5\linewidth]{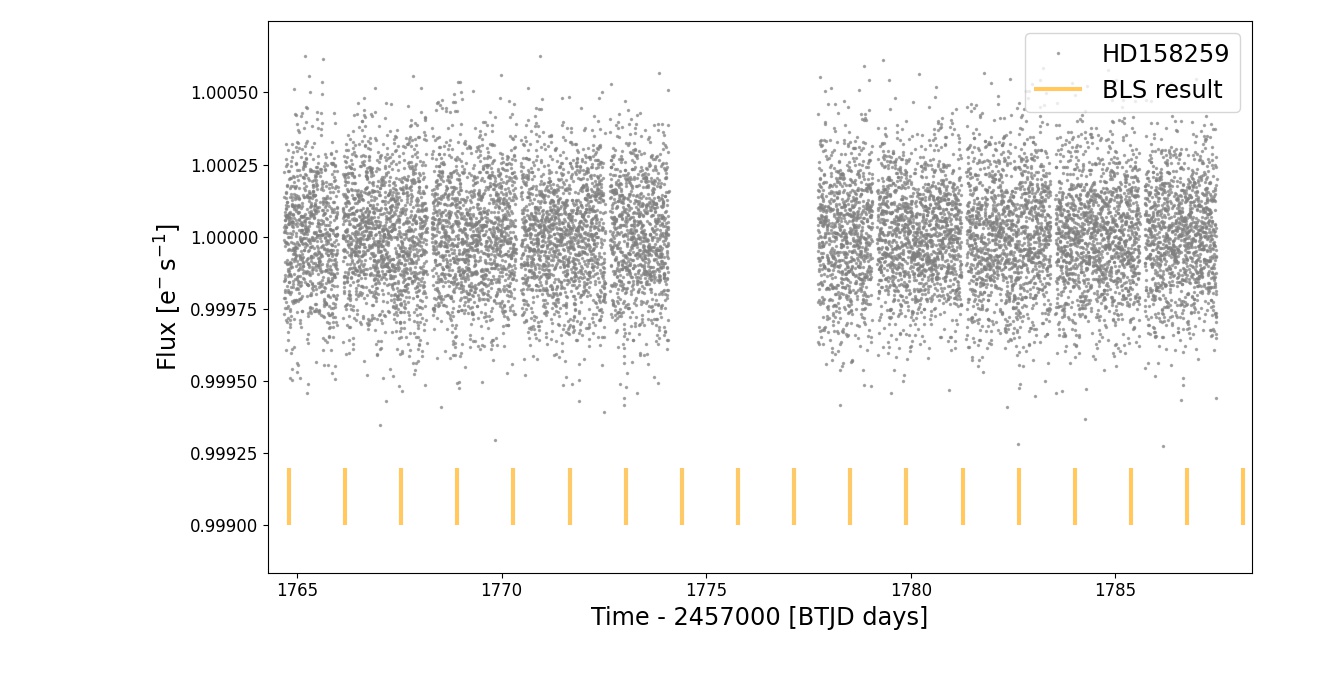}
\includegraphics[width=0.5\linewidth]{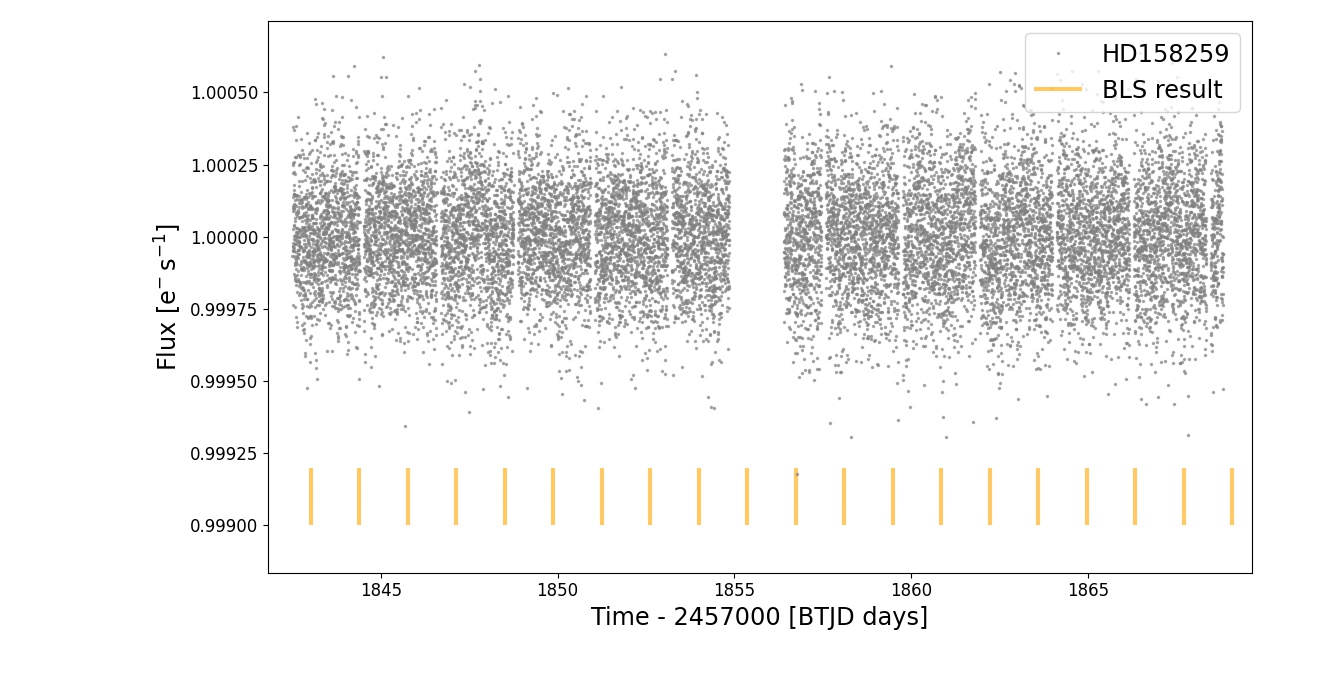}
\\
\centering
\includegraphics[width=0.5\linewidth]{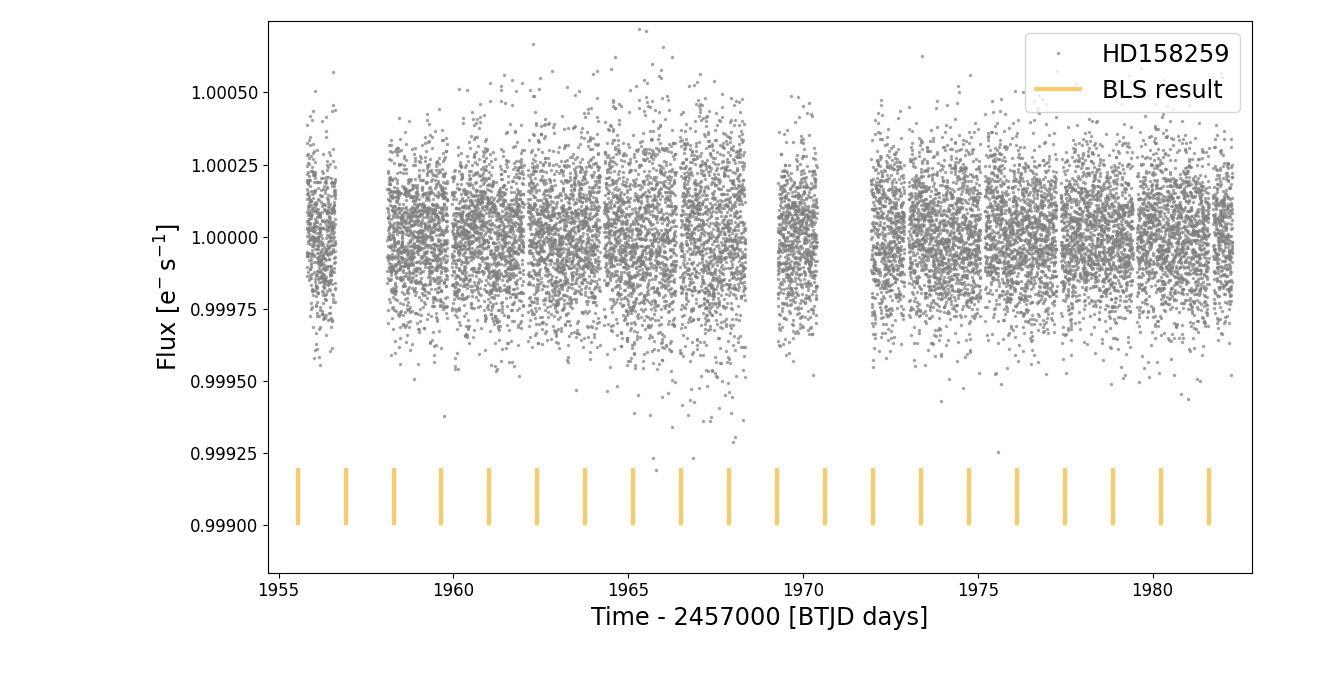}
\includegraphics[width=0.5\linewidth]{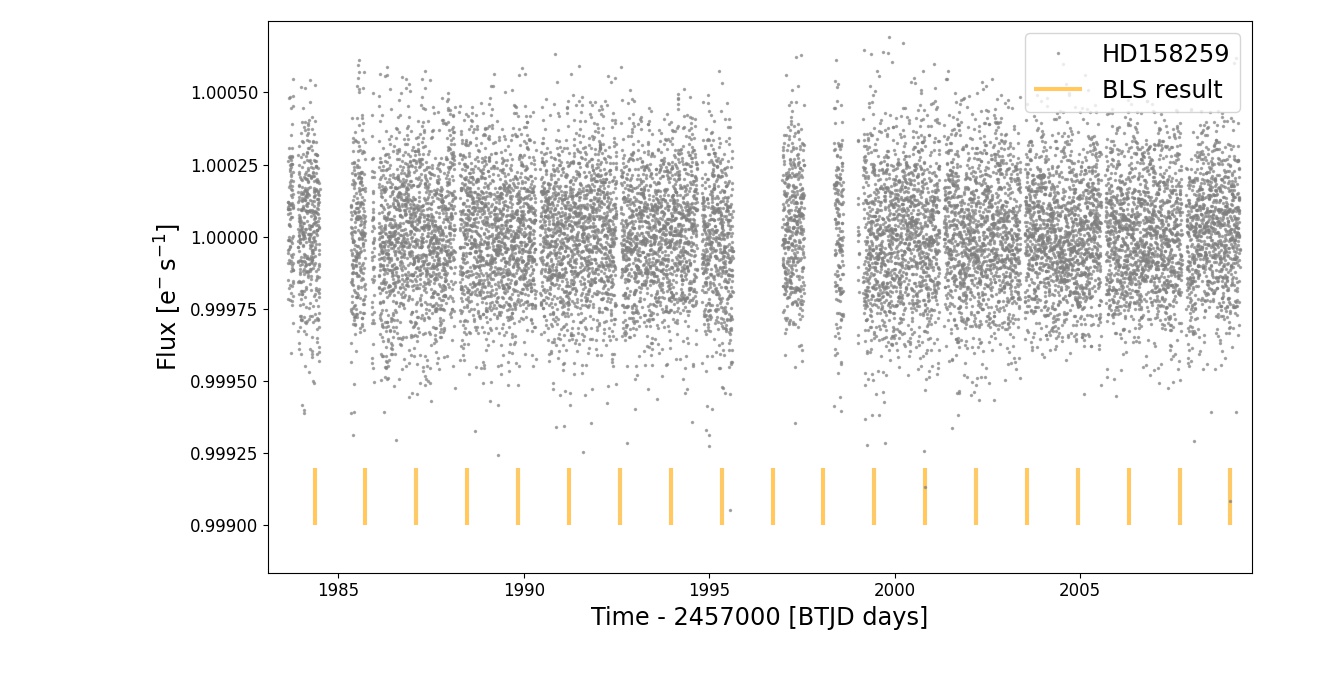} 
\\
\includegraphics[width=0.5\linewidth]{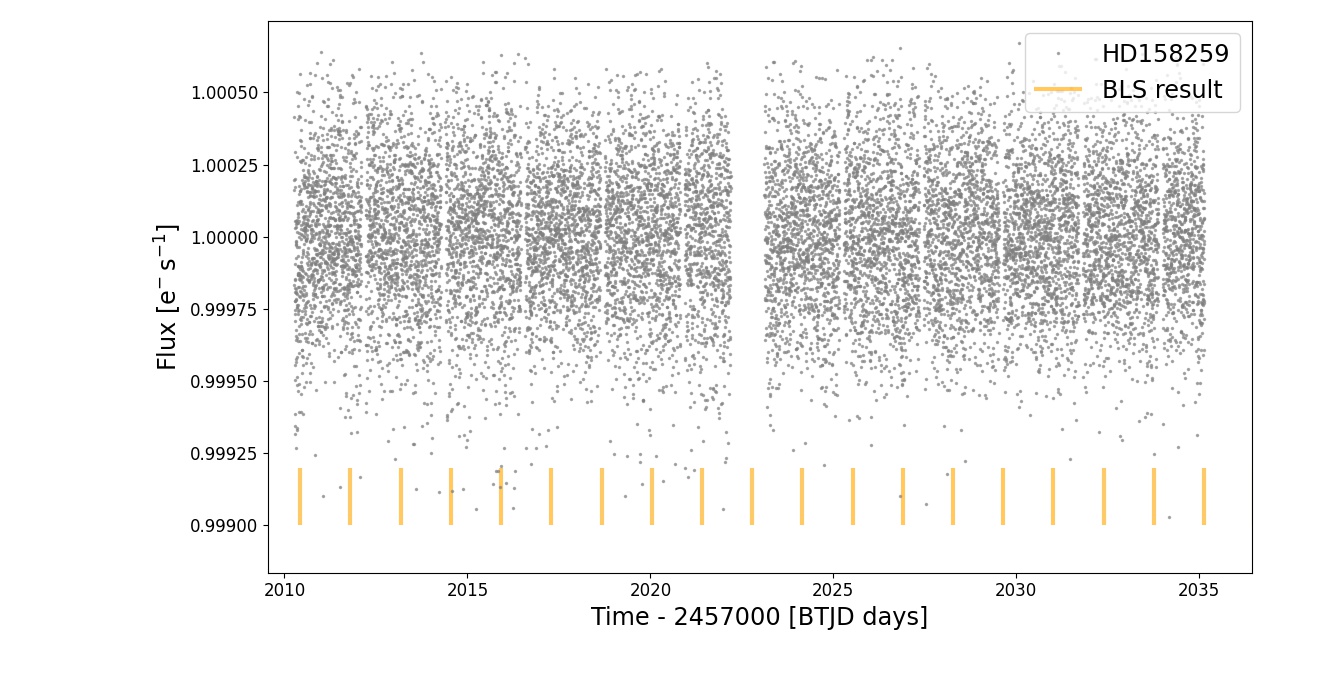} 
\caption{\small{Light curve from TESS photometry of HD~158259. Panels present, from left to right and top to bottom, data from sectors 17, 20, 24, 25, and 26. Orange lines show the signals detected by global-BLS. The light curve is masked to remove the transit of planet b, and the thin gaps in the data are the consequences of this. 
}}
\label{fig:HD158259lc}
\end{figure*}

\begin{figure}[!]
\begin{minipage}{\linewidth}
\begin{center}
\includegraphics[angle=0,trim = 0mm 00mm 0mm 00mm, clip, width=\linewidth]{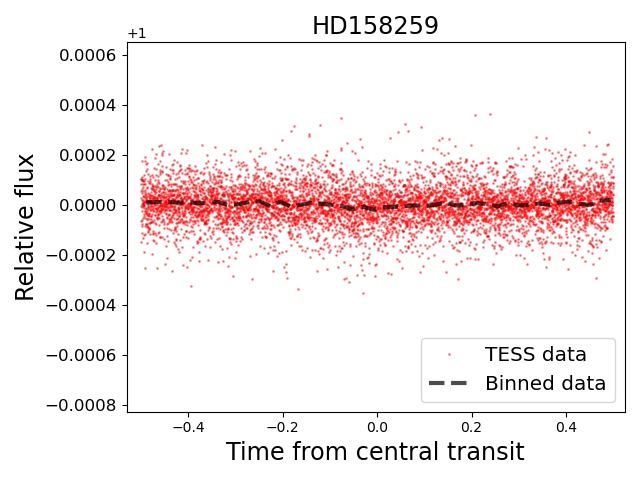}
\end{center}
\caption{\small{HD~158259. Folded light curve in a period of 1.37 days from the BLS-spectrum obtained masking cadences corresponding to the transit of planet b. Red points are the TESS data and the dashed line corresponds to the binned flux.}}
\label{fig:HD158259_n}
 \end{minipage}
\end{figure}

\begin{figure}[!]
\begin{minipage}{\linewidth}
\begin{center}
\includegraphics[angle=0,trim = 0mm 00mm 0mm 00mm, clip, width=\linewidth]{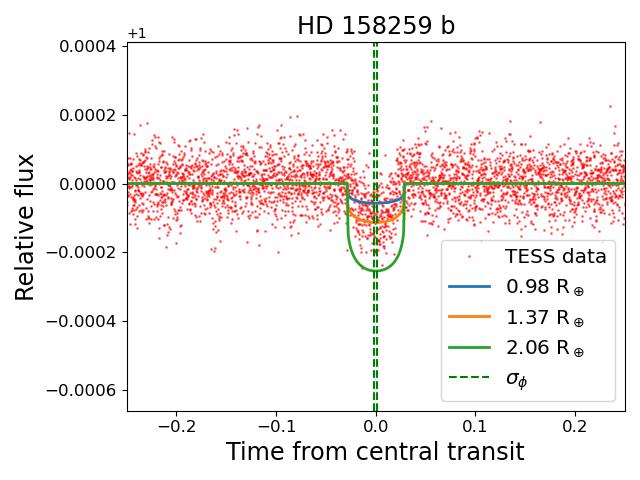}
\includegraphics[angle=0,trim = 0mm 00mm 0mm 00mm, clip, width=\linewidth]{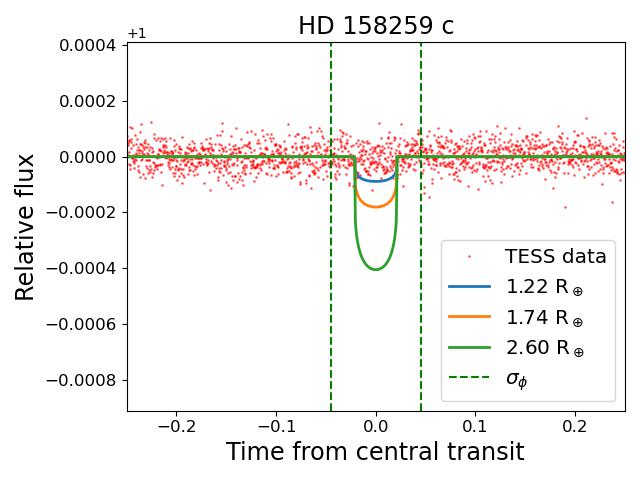}
\includegraphics[angle=0,trim = 0mm 00mm 0mm 00mm, clip, width=\linewidth]{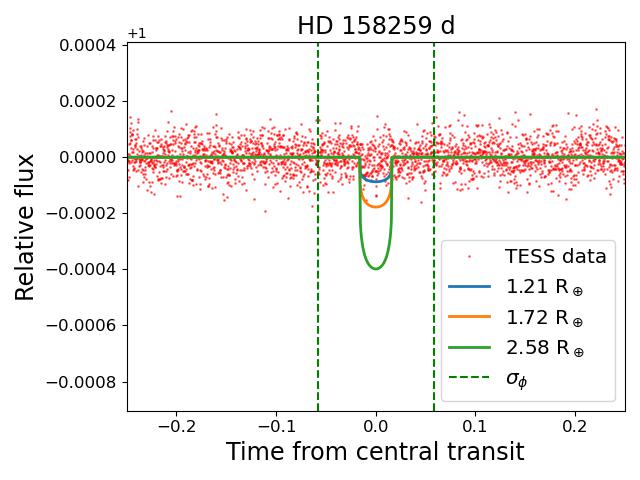}
\end{center}
\caption{\small{Same as Fig.~\ref{fig:61virb} but for planets HD~158259~b,~c and~d.}}%
\label{fig:HD158259fold1}
 \end{minipage}
\end{figure}

\begin{figure}[!]
\begin{minipage}{\linewidth}
\begin{center}

\includegraphics[angle=0,trim = 0mm 00mm 0mm 00mm, clip, width=\linewidth]{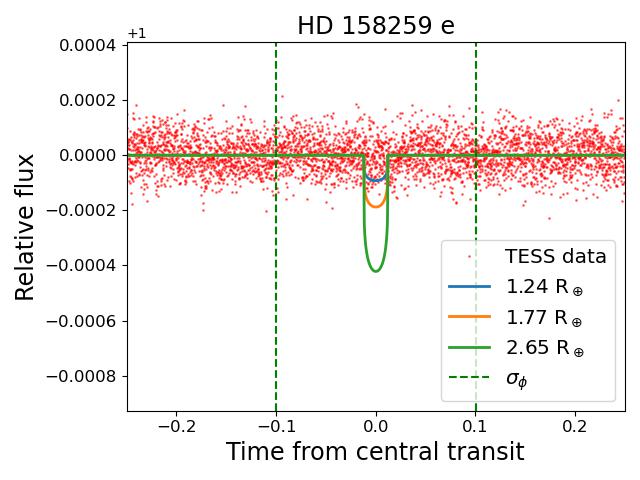}
\includegraphics[angle=0,trim = 0mm 00mm 0mm 00mm, clip, width=\linewidth]{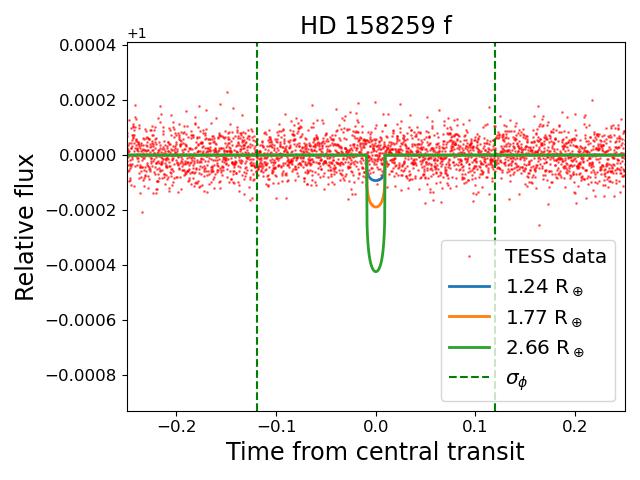}
\end{center}
\caption{\small{Same as Fig.~\ref{fig:61virb} but for planets HD~158259~e and~f.}}
\label{fig:HD158259fold2}
 \end{minipage}
\end{figure}

HD~158259 is a system with five planets detected by RV data from the SOPHIE spectrograph. The planets have masses between 2 and 6~M$_{\oplus}$, and at least one of them transits its host star (planet b with a period of 2.16 days), and was detected using TESS data by \citet{Hara2020}. With the exception of the innermost planet b, all of them are included in our sample. We used data from October 2019 to July 2020 (sectors 17, 20, and 24 through 26). As shown in Table~\ref{apptab3:bls} and Fig.~\ref{fig:HD158259}, the global-BLS analysis shows a high peak corresponding to the period of planet b. After masking this planet, the BLS spectrum exhibits a moderate peak of the order of the SNR threshold we considered ($\sim$6) at a period of 1.37 days, which does not correspond to any of the known planets. This last apparent signal is not clearly visible over the data noise. In Fig.~\ref{fig:HD158259lc} we present the light curves from each TESS sector, with vertical lines indicating the positions of the transits. Also, in Fig.~\ref{fig:HD158259_n} we present this light curve folded in orbital phase using the same parameters; this supports our previous inference that no transit is observed.
Figures~\ref{fig:HD158259fold1} and~\ref{fig:HD158259fold2} present the phase-folded light curves using the parameters of Table~\ref{apptab2:planet} for planets b to f. Planet b exhibits a clear transit, while the others do not.

\subsubsection{HD~20794} 

\begin{figure}[!]
\begin{minipage}{\linewidth}
\begin{center}
\includegraphics[angle=0,trim = 0mm 00mm 0mm 00mm, clip, width=\linewidth]{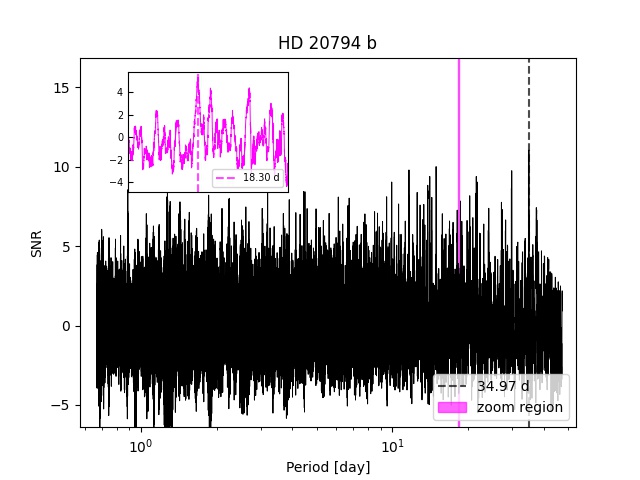}

\end{center}
\caption{\small{Same as Fig.~\ref{fig:GJ887} but for HD~20794~b.}}

\label{fig:HD20794}
 \end{minipage}
\end{figure}

\begin{figure*}
%
\centering
\includegraphics[width=0.5\linewidth]{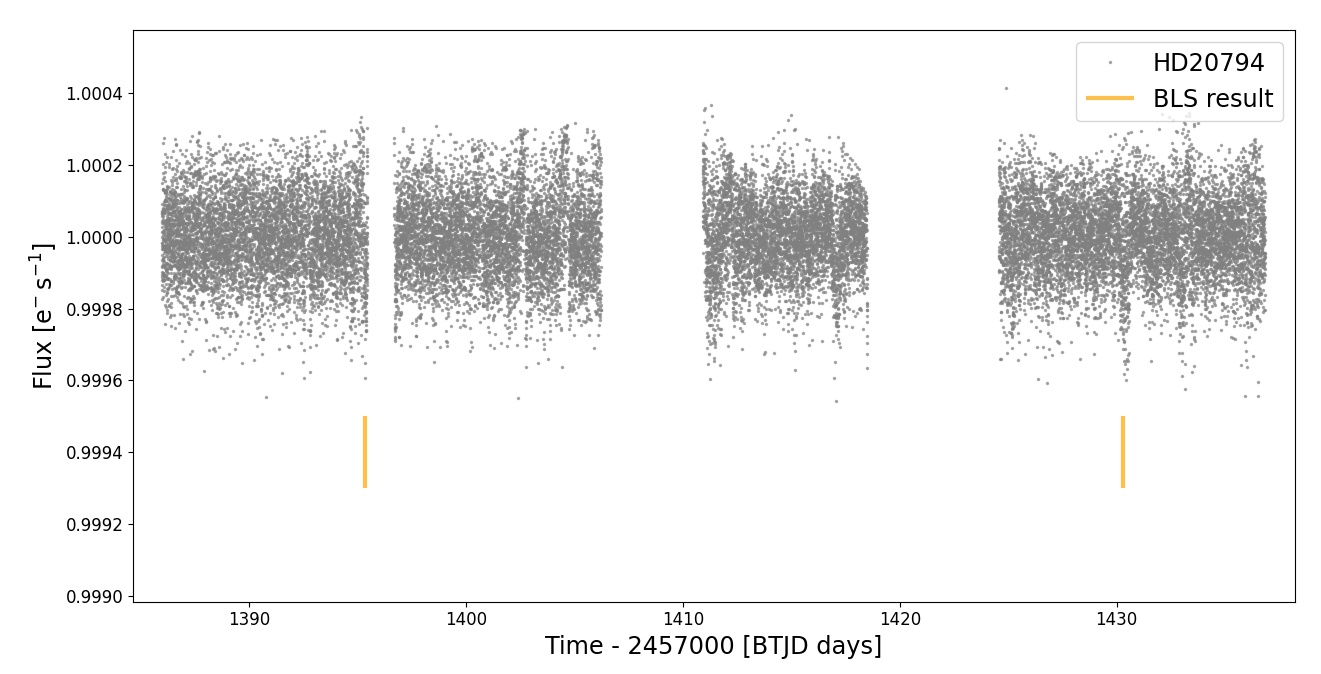}
\includegraphics[width=0.5\linewidth]{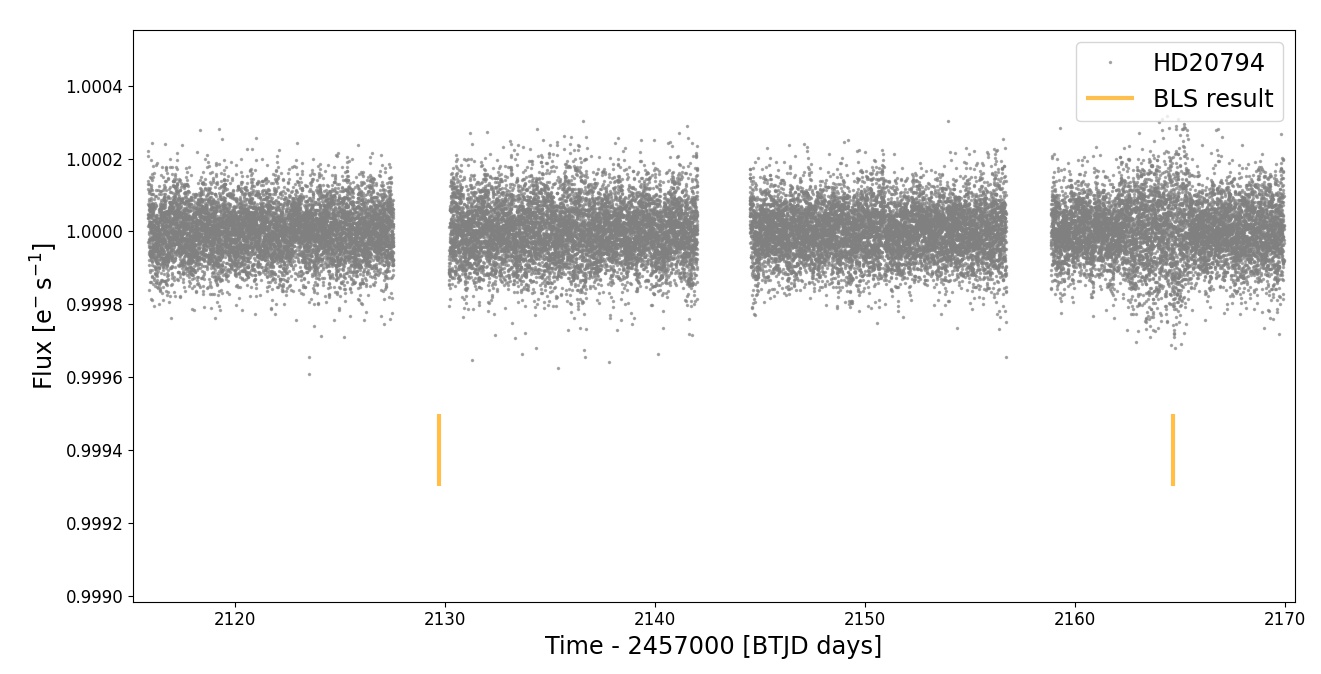}

\caption{\small{Light curve from TESS photometry of HD~20794. Left panel shows data from sectors 3 and 4 and right panel shows data from sectors 30 and 31. Orange lines show the signals detected by global-BLS.
}}
\label{fig:HD20794lc}

\end{figure*}

\begin{figure}[!]
\begin{minipage}{\linewidth}
\begin{center}
\includegraphics[angle=0,trim = 0mm 00mm 0mm 00mm, clip, width=\linewidth]{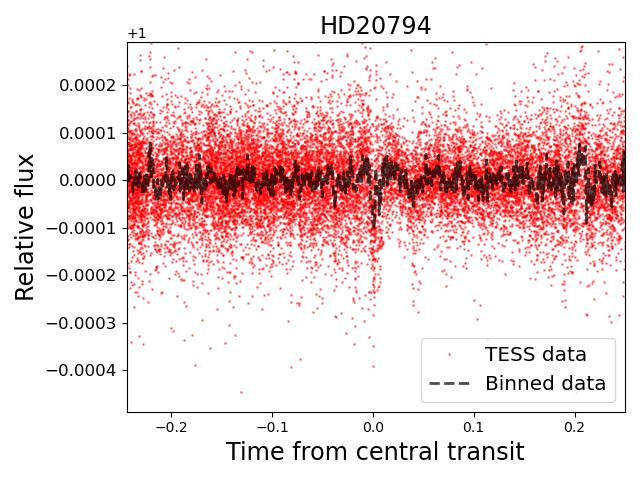} 
\end{center}
\caption{\small{HD~20794. Folded light curve in a period of 34.97 days from the BLS spectrum. Red points are the TESS data and the dashed line corresponds to the binned flux.}}
\label{fig:HD20794bls}
 \end{minipage}
\end{figure}

HD~20794 hosts four RV planets but only one with an orbital period shorter than 30 days. HD~20794~b has a period of 18.32 days \citep{pepe2011}. TESS observed this star in sectors 3, 4, 30, and 31. In the global-BLS analysis we observed a peak with SNR $\sim$ 11 at around 34 days. On the other hand, the local-BLS periodogram centered at the period of the planet does not exhibit any significant peak (see Fig.~\ref{fig:HD20794}). Checking the light curve visually (Fig.~\ref{fig:HD20794lc}) we noted that the flux drops detected by the global-BLS are often associated with unstable pointing at cadences near momentum dumps (one can be seen near $2\,458\,430$ BTJD) and gaps in the light curves (See Fig. 7 in TESS Data Release Notes of Sector 4\footnote{https://archive.stsci.edu/missions/tess/doc/tess$_{-}$drn/tess$_{-}$sector$_{-}$04$_{-}$
drn05$_{-}$v04.pdf}). In Fig.~\ref{fig:HD20794bls} we present the folded light curve using the BLS results. The measured 0.007\% (70 ppm) depth signal is not reliable. The central cadences of folded curve have a sharp shape that does not resemble a transit. This is mainly due to noisy cadences. In addition, similar features are seen across the orbital phase.

\begin{figure}[!]
\begin{minipage}{\linewidth}
\begin{center}
\includegraphics[angle=0,trim = 0mm 00mm 0mm 00mm, clip, width=\linewidth]{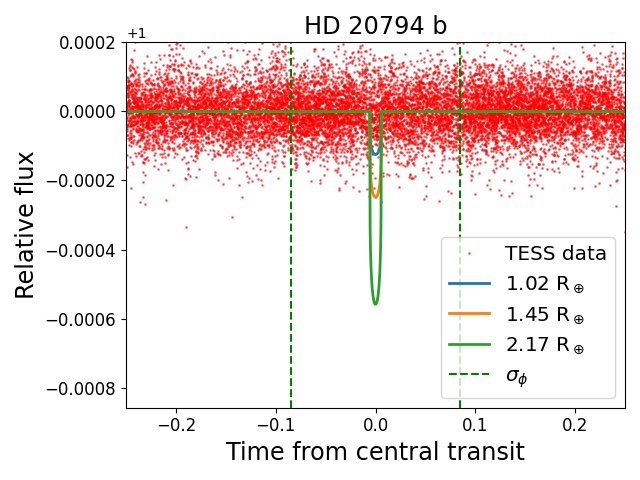}
\end{center}
\caption{\small{Same as Fig.~\ref{fig:61virb} but for HD~20794~b.}}
\label{fig:HD20794b}
 \end{minipage}
\end{figure}

In Fig.~\ref{fig:HD20794b} we show the phase-folded light curve with literature parameters of planet b (see Table~\ref{apptab2:planet}). We can conclude that there is no transit signals of this planet in the TESS data.

\subsubsection{HD~219134} 

HD~219134 is a multiplanetary system composed of six planets all detected using the RV technique \citep{vogt2015,motalebi2015}; three of them have periods shorter than 30 days (b, c, and f), but only planet f with a period of 22 days and a minimum mass of 7.3~M$_{\oplus}$ is in our sample because it has not been detected in transit so far \citep{guillon2017b}.

\begin{figure*}
%
\centering
\includegraphics[width=0.5\textwidth]{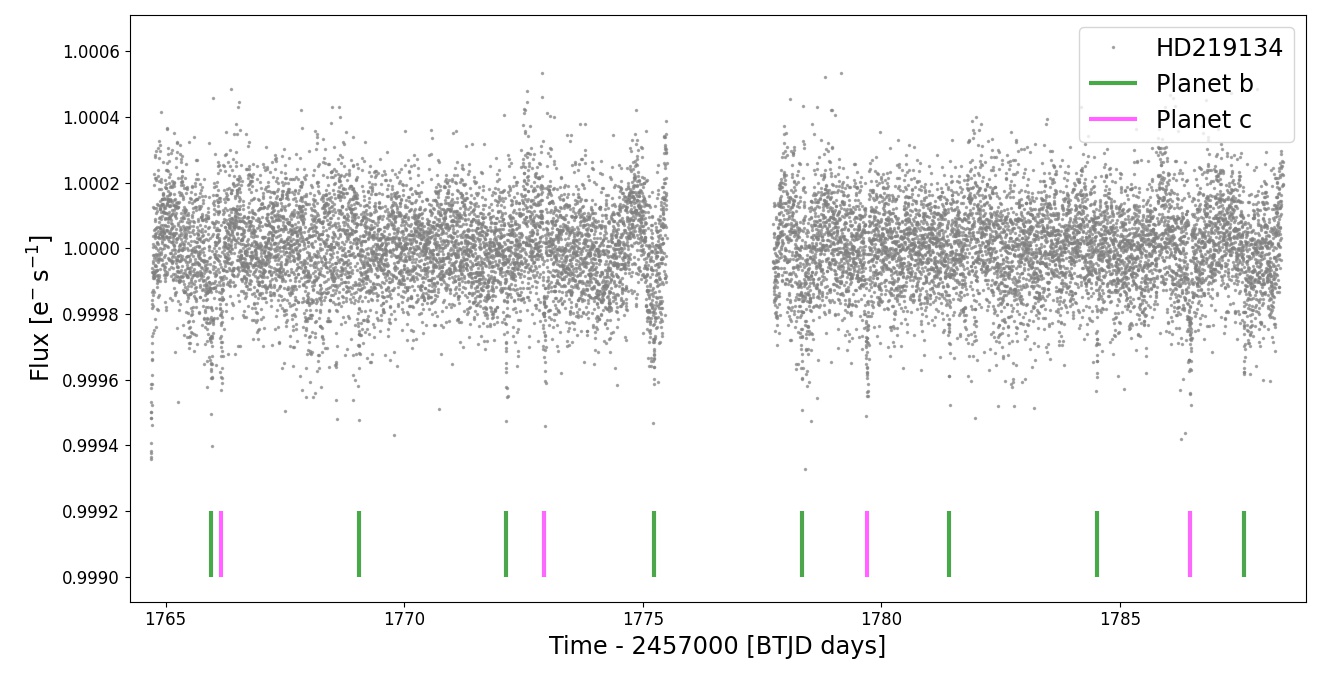}
\includegraphics[width=0.5\textwidth]{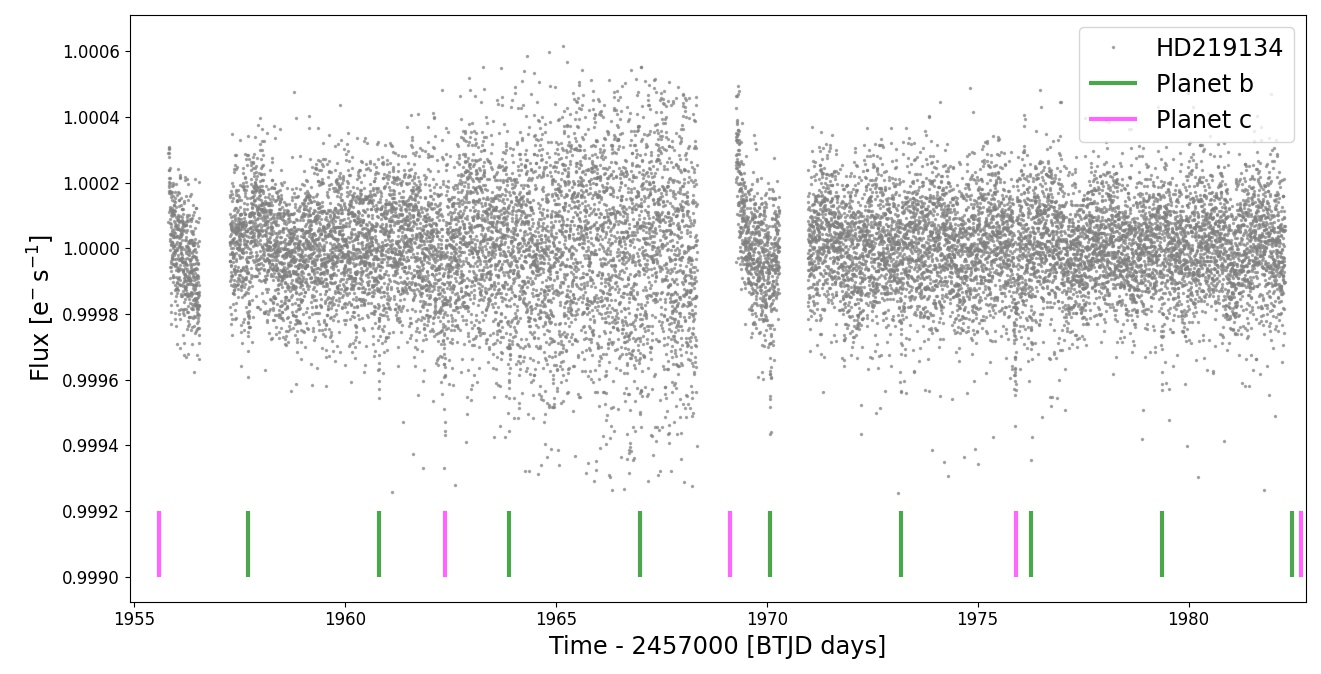}
\\
\includegraphics[width=0.5\textwidth]{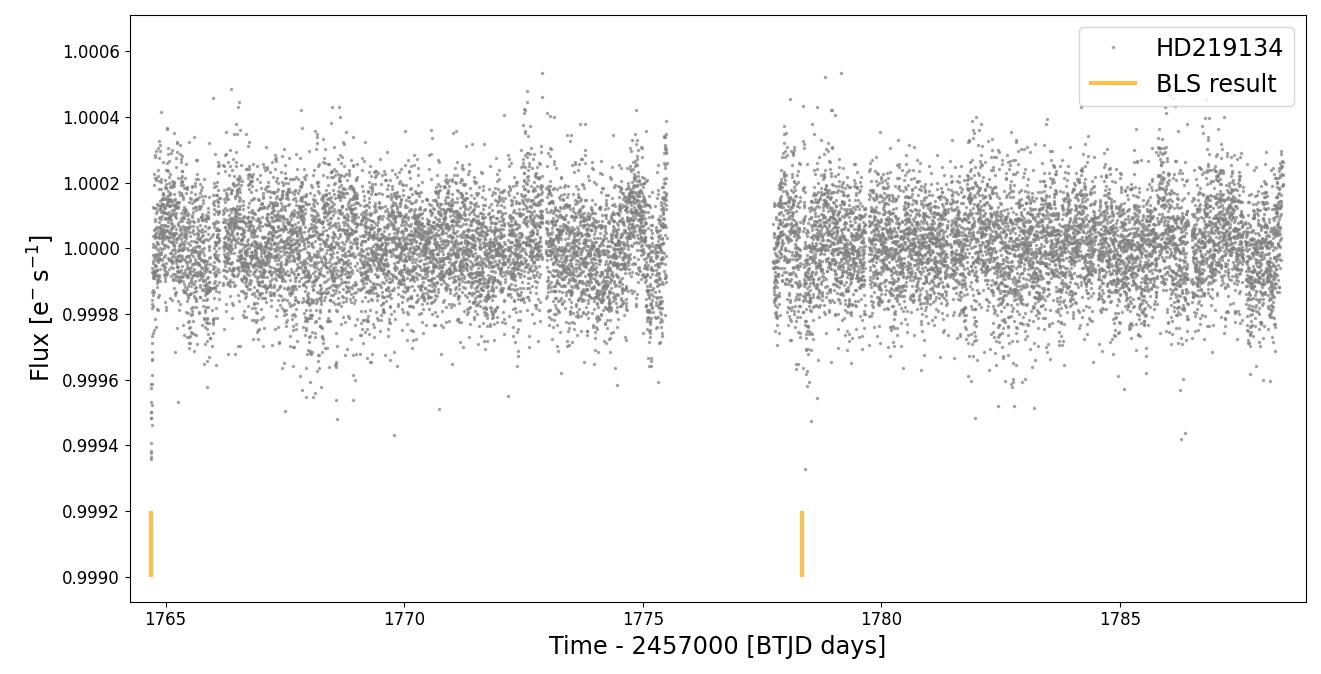}
\includegraphics[width=0.5\textwidth]{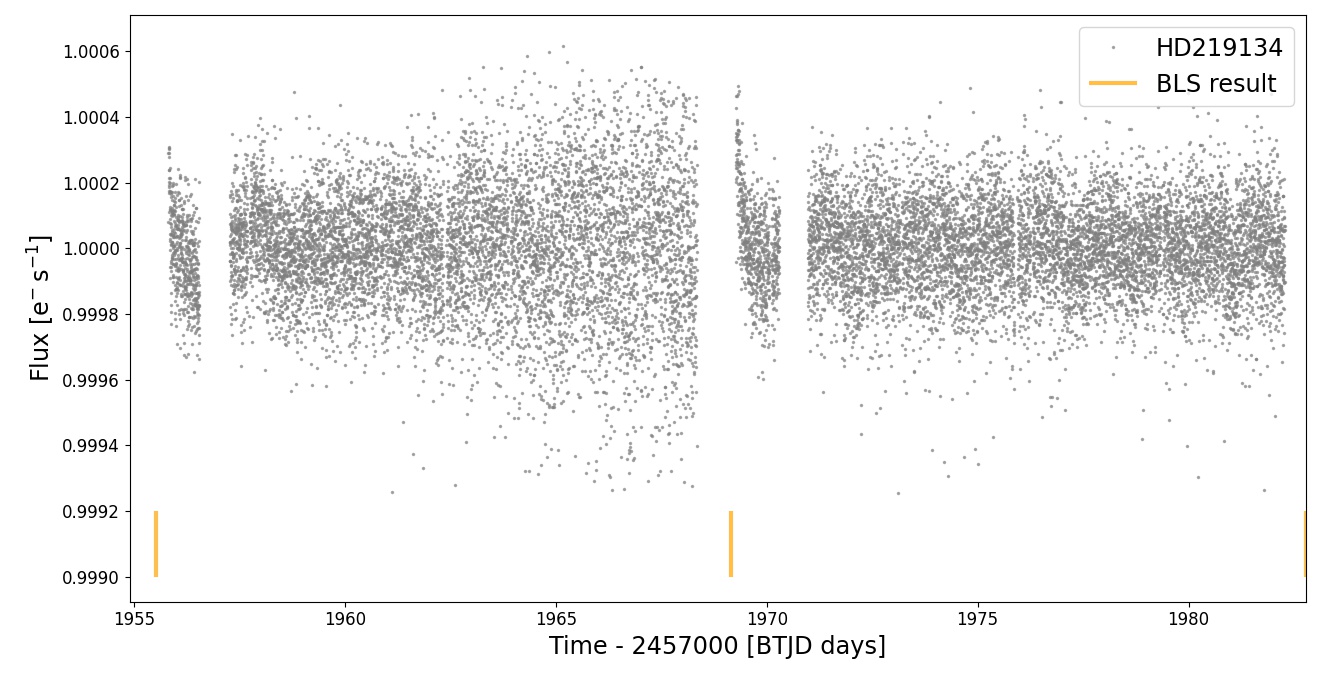}
\\
\includegraphics[width=0.5\textwidth]{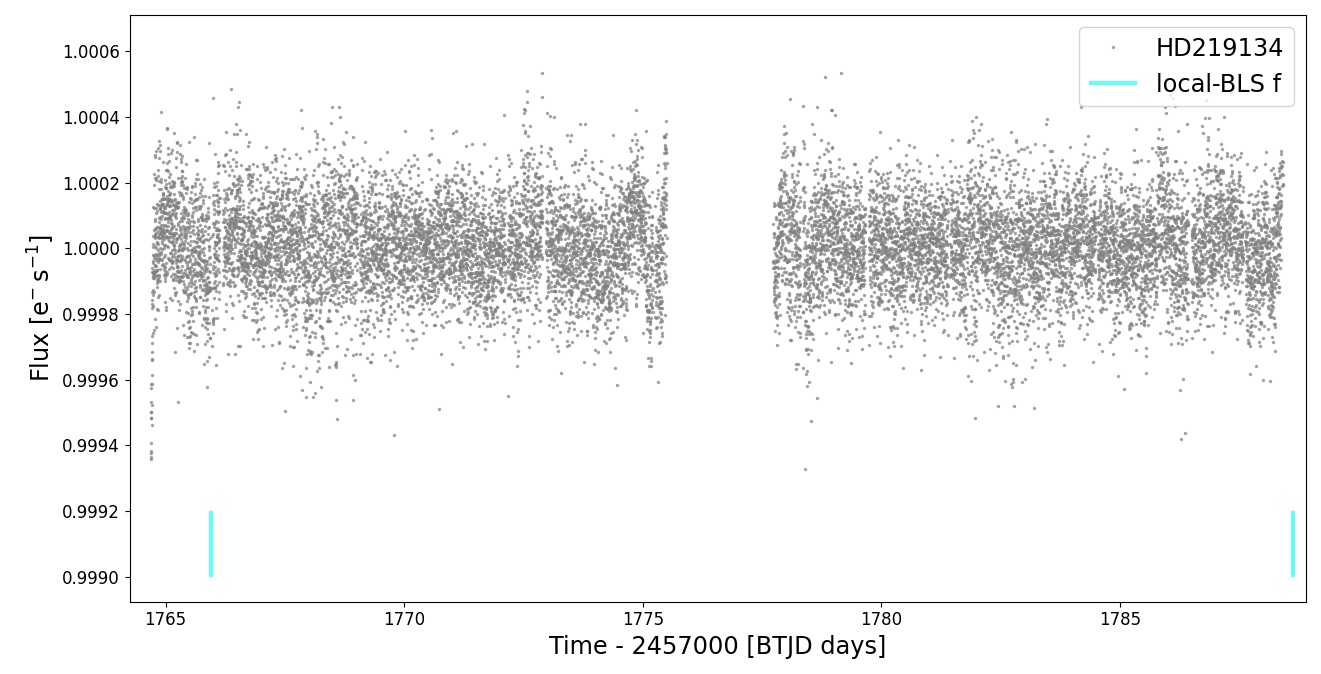}
\includegraphics[width=0.5\textwidth]{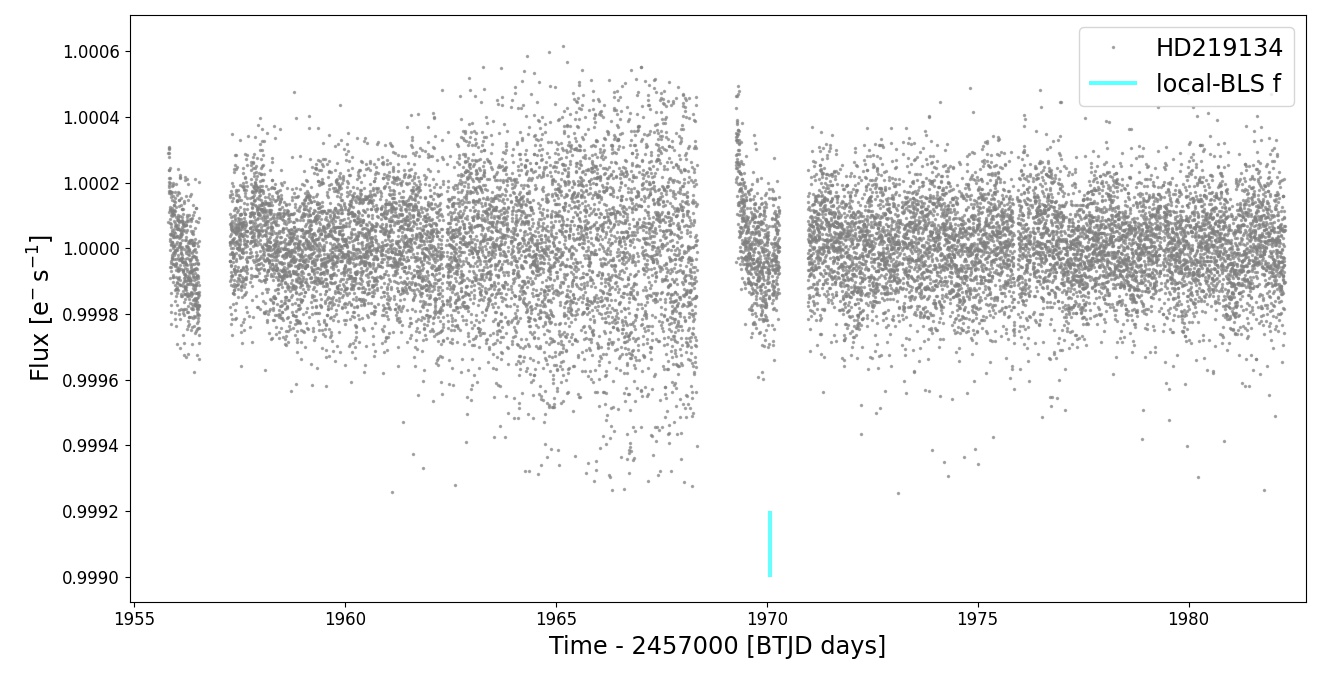}

\caption{\small{Light curve from TESS photometry of HD~219134. Left panels show data from sector 17 and right panels show data from sector 24. In  the top panels, the green and magenta lines show transits of planet b and c, respectively. In the middle panels, orange lines show the signals detected by global-BLS. In the bottom panels, cyan lines point out the signals detected by local-BLS. In the middle and bottom panels,light curves are shown masking the cadences corresponding to the transits of planets b and c.
}}
\label{fig:HD219134lc}
\end{figure*}

\begin{figure}[!]
\begin{minipage}{\linewidth}
\begin{center}
\includegraphics[angle=0,trim = 0mm 00mm 0mm 00mm, clip, width=\linewidth]{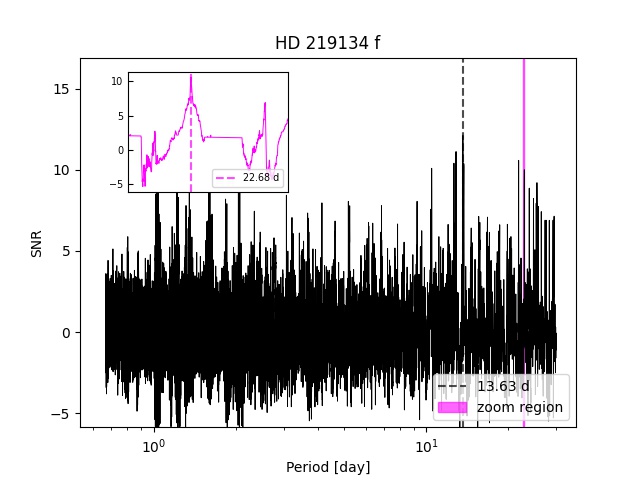}

\end{center}
\caption{\small{Same as Fig.~\ref{fig:GJ887} but for HD~219134~f.}}

\label{fig:HD219134}
 \end{minipage}
\end{figure}

\begin{figure}[!]
\begin{minipage}{\linewidth}
\begin{center}
\includegraphics[angle=0,trim = 0mm 00mm 0mm 00mm, clip, width=\linewidth]{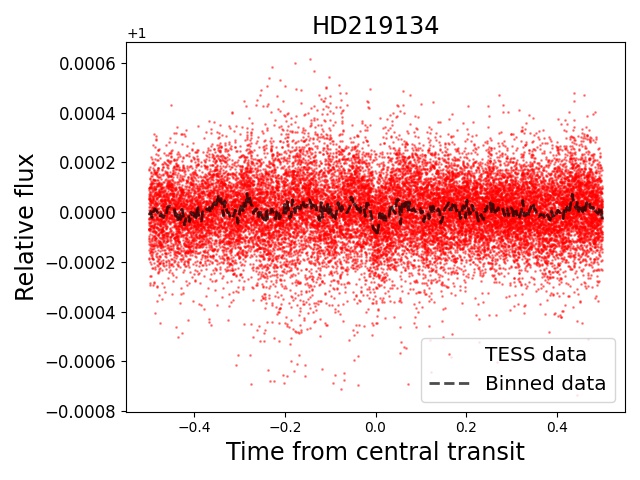}
\end{center}
\caption{\small{HD~219134. Folded light curve in a period of 10.45 days from the BLS-spectrum obtained discarding first cadences of sector 17. Red points are the TESS data and the dashed line corresponds to the binned flux.}}
\label{fig:HD219134new}
 \end{minipage}
\end{figure}

Before computing the global- and local-BLS power spectra, we masked the TESS cadences corresponding to transits of planets b and c (see top panel of Fig.~\ref{fig:HD219134lc}). Both BLS periodograms are shown in Fig.~\ref{fig:HD219134} and the results are listed in Table~\ref{apptab3:bls}. In the global-BLS spectrum we found a peak with SNR$\sim$12 at a period of 13.6 days. In the middle panel of Fig.~\ref{fig:HD219134lc}, the times of transit for this signal are indicated in the TESS light curve. We can see that one of the detected transits would correspond to a deep drop at the beginning of the observations of sector 17. The signals at the beginning of the observation are not reliable and are usually caused by systematic errors. The  second transit is close to noisy cadences just after the downlink data gap (BTJD $\sim$ $2\,458\,778.4$ ) and the two transits of sector 24 are located in data gaps. We recomputed the BLS algorithm discarding the cadences at the beginning of sector 17 ($<$ 1764.75) and obtained a new peak at 10.45 days. Figure \ref{fig:HD219134new} presents the light curve folded at that period. The transit has a duration of almost 3.8 hours, which is long for the measured period of 10 days. 
This, in addition to its irregular shape, leads us to the assumption that the signal is not due to a planetary transit. On the other hand, if we take into account the unknown eccentricity, we obtain that depth (0.008\%), period, and duration would be compatible with an eccentricity of $e$~$\gtrsim$~0.126. In any case, this system warrants the gathering of further data from other high-precision telescopes such as CHEOPS \citep{benz2021} to help confirm or rule out this transit.

\begin{figure}[!]
\begin{minipage}{\linewidth}
\begin{center}
\includegraphics[angle=0,trim = 0mm 00mm 0mm 00mm, clip, width=\linewidth]{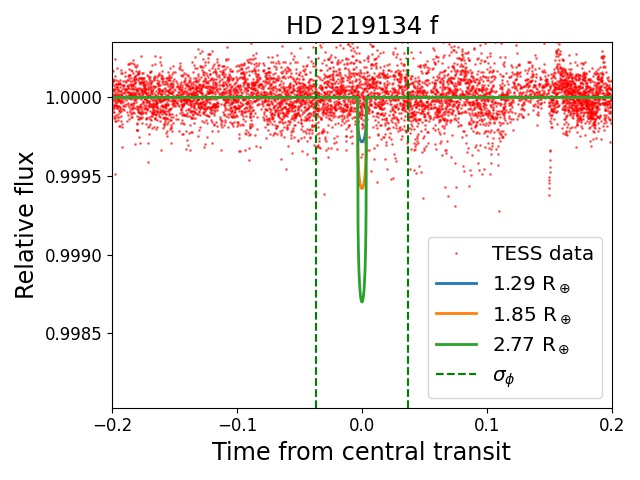}
\end{center}
\caption{\small{Same as Fig.~\ref{fig:61virb} but for HD~219134~f.}}

\label{fig:HD219134f}
 \end{minipage}
\end{figure}

Considering the peak seen in the local-BLS spectrum around the period of planet f, the bottom panel of Fig.~\ref{fig:HD219134lc} shows that the cadences that produced the power in the BLS spectrum are located in gaps caused by the previous masked planets. Furthermore, in Fig.~\ref{fig:HD219134f} we present the transit model superimposed on the phase-folded data using the literature parameters listed in Table~\ref{apptab2:planet}. Clearly, the data discard central transits of planets larger than around 1.3 $R_\oplus$.

\subsubsection{HD~31527} 

Announced by \cite{Udry2019}, HD~31527 is a system with three Neptune-mass planets with periods of 16.6, 51.2, and 271.7 days detected by RV measurements from the HARPS spectrograph. Our sample only includes the innermost planet, HD~31527~b, with a mass of 10.47~M$_{\oplus}$. The host star was observed by TESS in sectors 5 and 32 with a total baseline of 726 days, in two intervals of $\sim$ 26 days per sector. 

\begin{figure}[!]
\begin{minipage}{\linewidth}
\begin{center}
\includegraphics[angle=0,trim = 0mm 00mm 0mm 00mm, clip, width=\linewidth]{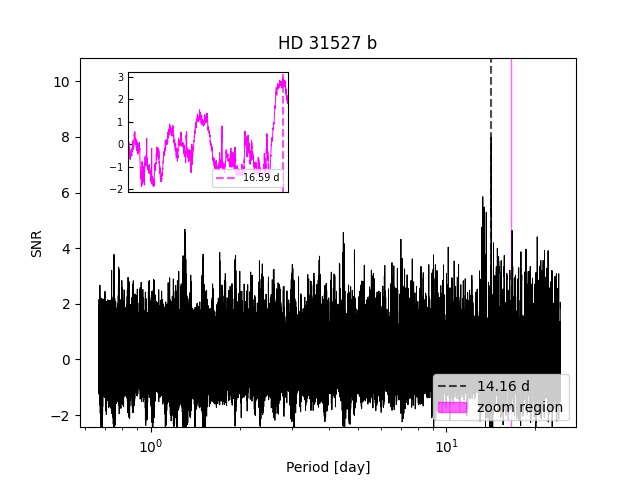}
\end{center}
\caption{\small{Same as Fig.~\ref{fig:GJ887} but for HD~31527~b.}}

\label{fig:HD31527}
 \end{minipage}
\end{figure}

\begin{figure*}
%
\centering
\includegraphics[width=0.5\textwidth]{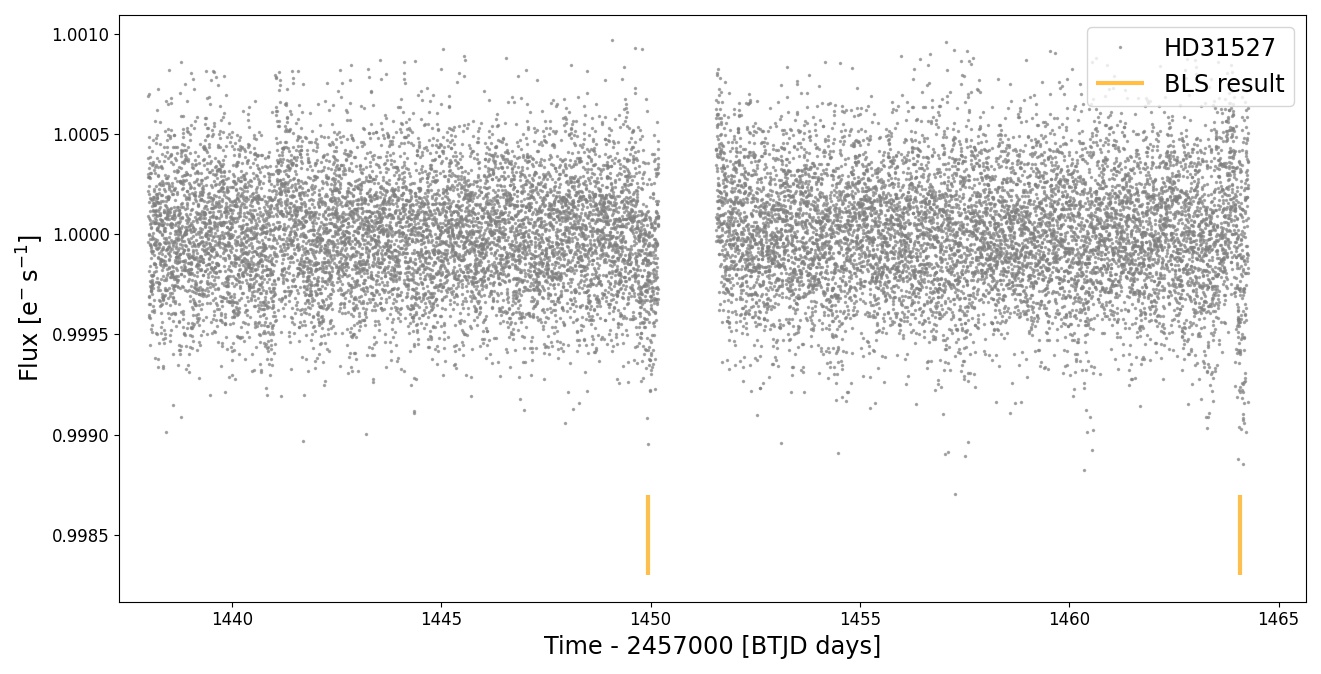}
\includegraphics[width=0.5\textwidth]{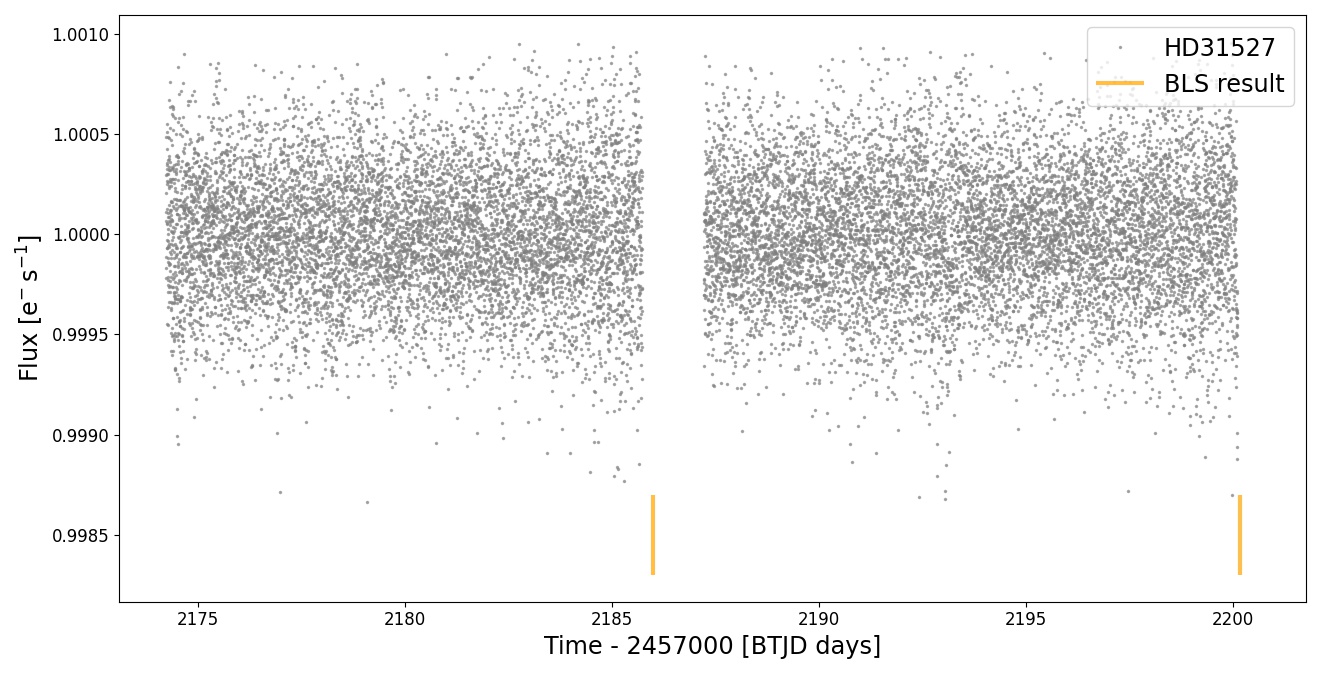}

\caption{\small{Light curve from TESS photometry of HD~31527. The left panel shows data from sector 5 and the right panel shows data from sector 35. Orange lines show the signals detected by global-BLS.
}}
\label{fig:HD31527lc}

\end{figure*}

\begin{figure}[!]
\begin{minipage}{\linewidth}
\begin{center}
\includegraphics[angle=0,trim = 0mm 00mm 0mm 00mm, clip, width=\linewidth]{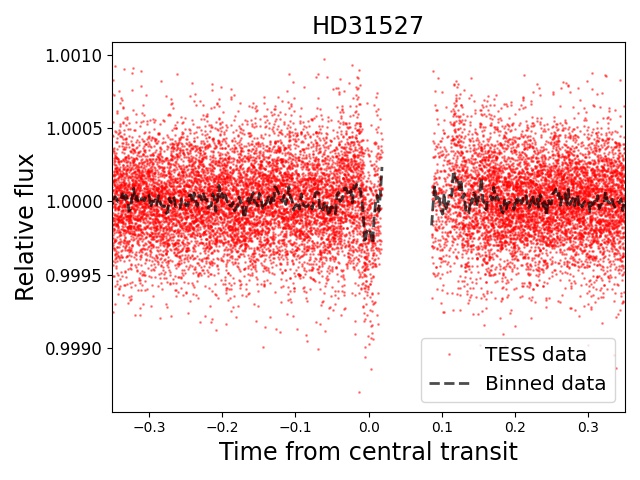}
\end{center}
\caption{\small{HD~31527. Folded light curve using parameters obtained from the global-BLS with a period of 14.16 days. Red points are the TESS data and the dashed line corresponds to the binned flux.}}
\label{fig:HD31527F}
 \end{minipage}
\end{figure}

\begin{figure}[!]
\begin{minipage}{\linewidth}
\begin{center}
\includegraphics[angle=0,trim = 0mm 00mm 0mm 00mm, clip, width=\linewidth]{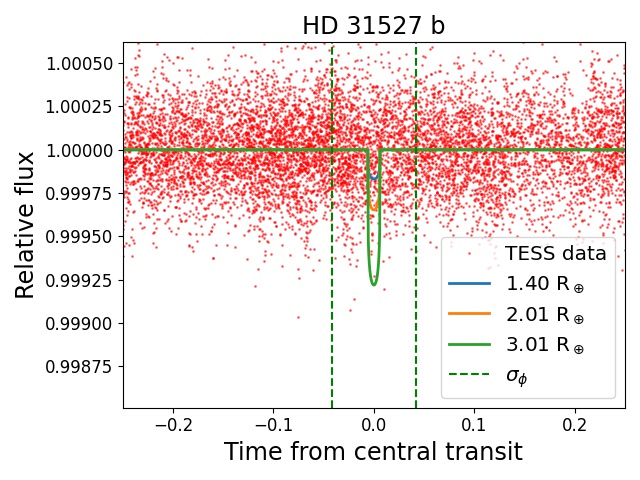}
\end{center}
\caption{\small{Same as Fig.~\ref{fig:61virb} but for HD~31527~b.}}

\label{fig:HD31527fold}
 \end{minipage}
\end{figure}

The global-BLS spectrum exhibits a peak with SNR$\sim$8 at a period of 14.15 days (Fig.~\ref{fig:HD31527}, Table~\ref{apptab3:bls}). The TESS light curve of this star is shown in Fig.~\ref{fig:HD31527lc} and the folded light curve is shown in Fig.~\ref{fig:HD31527F}. All of the putative transits occur close to data gaps (specifically both transits in sector 32 are located in the gaps). In sector 5, the cadences at the end of each orbit are affected by scattered light from Earth. In particular, the cadences at the end of this sector show  pointing jitter noise that is caused by a guiding problem (we recommend interested users see Figs. 1 and 3 in the TESS Data Release Notes of the Sector 5\footnote{https://archive.stsci.edu/missions/tess/doc/tess$_{-}$drn/tess$_{-}$sector$_{-}$05$_{-}$
drn07$_{-}$v02.pdf}). If we apply the BLS-periodogram discarding the latest cadences, the peak disappears and no other significant peak is found. Therefore, we do not consider this a reliable detection.
Additionally, the models shown with the TESS light curve phase folded at the period of planet $b$ (Fig.~\ref{fig:HD31527fold}) indicate that the data discard transits of planets larger than around 2 $R_\oplus$.

\subsubsection{HD~40307} 

HD~40307 is a multiplanetary system detected by RV from the HARPS spectrograph. It has four confirmed planets \citep{mayor2009, diaz2016a}, and additional potential planets \citep{tuomi2013}. Three confirmed planets are included in our sample. These are the super-Earth planets with orbital periods P$_{\rm b}$ = 4.3, P$_{\rm c}$ = 9.6, and P$_{\rm d}$ = 20.5 days. HD~40307 was observed by TESS over 22 sectors (see Table~\ref{tab1:star}) with a full baseline of 980 days approximately.

\begin{figure}[!]
\begin{minipage}{\linewidth}
\begin{center}
\includegraphics[angle=0,trim = 0mm 00mm 0mm 00mm, clip, width=\linewidth]{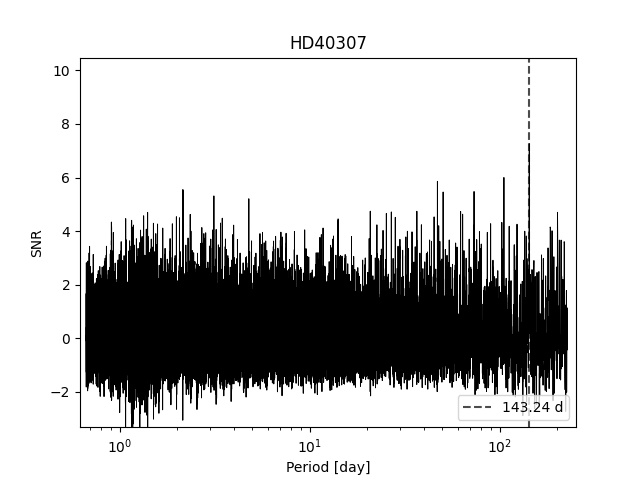}
\end{center}
\caption{\small{HD~40307. Global-BLS periodogram. Vertical dashed line indicates the maximum peak detected.}}
\label{fig:HD40307}
 \end{minipage}
\end{figure}

The global-BLS analysis shows a peak with SNR$\sim$7 at a period of 143 days (Fig.~\ref{fig:HD40307}). The BLS-algorithm placed putative transits in the data gaps between sectors and in cadences with increased instrumental noise in the sectors 30 (momentum dump event near $2\,459\,122.7$ BTJD) and 35 (near $2\,459\,265.9$ BTJD, just before the instrument was turned off because of an eclipse).

\begin{figure}[!]
\begin{minipage}{\linewidth}
\begin{center}
\includegraphics[angle=0,trim = 0mm 00mm 0mm 00mm, clip, width=\linewidth]{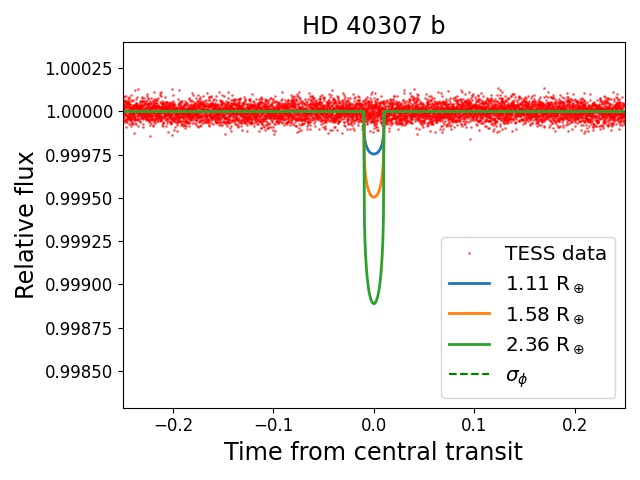}
\includegraphics[angle=0,trim = 0mm 00mm 0mm 00mm, clip, width=\linewidth]{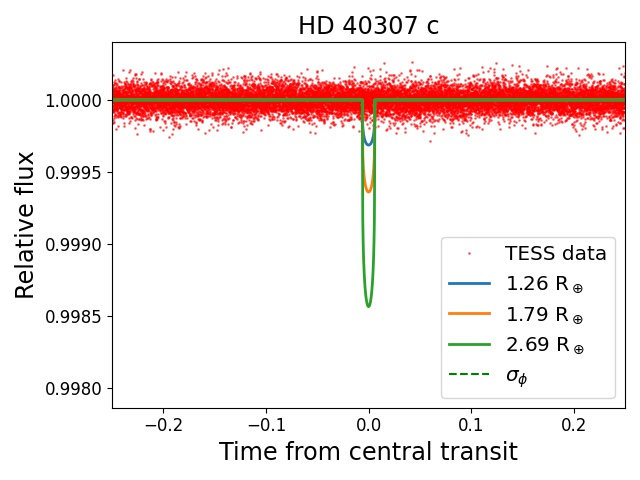}
\includegraphics[angle=0,trim = 0mm 00mm 0mm 00mm, clip, width=\linewidth]{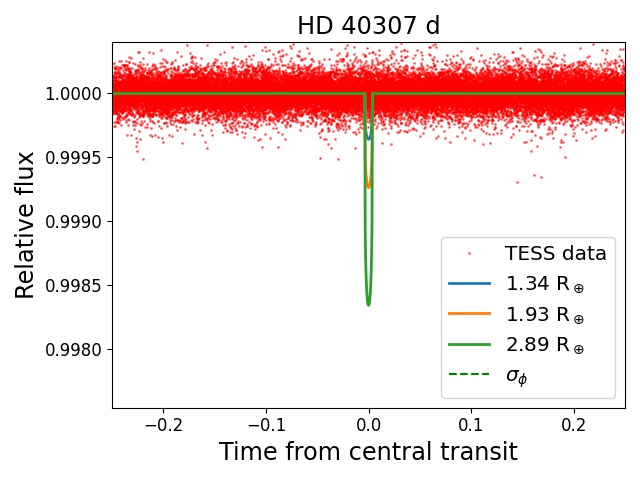}
\end{center}
\caption{\small{Same as Fig.~\ref{fig:61virb} but for planets HD~40307~b, c, and d. In these cases, the uncertainties of the orbital phases are outside the limits of the plot.}}

\label{fig:HD40307fold}
 \end{minipage}
\end{figure}

On the other hand, the local-BLS analyses for individual planets did not yield any significant results (Table~\ref{apptab3:bls}).
Finally, contrasting with the models calculated using the literature parameters (see Fig.~\ref{fig:HD40307fold}), we see that the TESS data clearly discards transits larger than 0.02\% for b and c and deeper than 0.03\% for d. Previous work on HD 40307 by \citet{Kane2021} also concludes the planets of the system are not transiting.

\subsubsection{HD~7924}

\begin{figure}[!]
\begin{minipage}{\linewidth}
\begin{center}
\includegraphics[angle=0,trim = 0mm 00mm 0mm 00mm, clip, width=\linewidth]{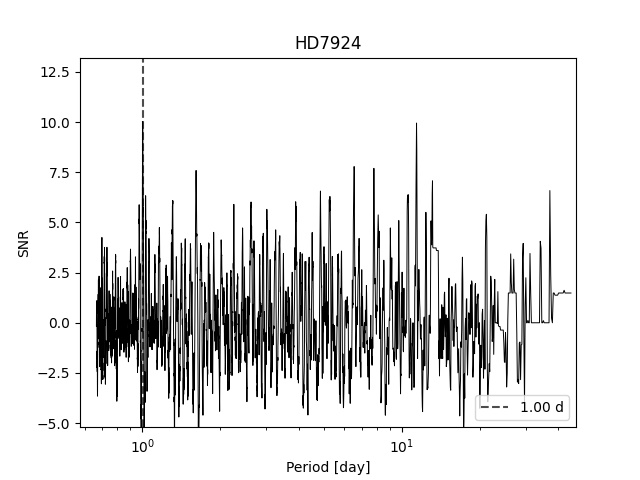}
\end{center}
\caption{\small{HD~7924. Global-BLS periodogram. The vertical dashed line indicates the maximum peak detected.}}
\label{fig:HD7924}
 \end{minipage}
\end{figure}

HD~7924 is a K-type star that hosts three super-Earth planets with periods of 5.39, 15.29, and 24.45 days \citep{howard2009,fulton2015}. All of them are included in our sample. This target was observed by TESS in sectors 18 and 19, making a full baseline of approximately 44 days. In the global-BLS analysis, we obtained a SNR of slightly larger than 10 at a period of 1.004 days (see Fig.~\ref{fig:HD7924}). By visual inspection of the signals detected by the BLS in the light curve, we noticed a transit with a depth of 40 ppm, which we attributed to instrumental systematic error, because some of the cadences correspond with momentum dumps over the spacecraft. We removed those cadences coinciding with momentum dump events, which are shown in the respective Figs. 7 in the Data Release Notes of sector 18 \footnote{https://archive.stsci.edu/missions/tess/doc/tess$_{-}$drn/tess$_{-}$sector$_{-}$18$_{-}$
drn25$_{-}$v02.pdf} and sector 19 \footnote{https://archive.stsci.edu/missions/tess/doc/tess$_{-}$drn/tess$_{-}$sector$_{-}$19$_{-}$
drn26$_{-}$v02.pdf}, and recomputed the BLS periodogram. We obtained a new period of 3.92 days with a SNR at the limit of our threshold (SNR$\sim$6), the duration of the respective transit is almost 6 hours with a depth of 0.007\% (70 ppm). These parameters are not consistent with a circular orbit, because the duration should be close to 3 hours. Nevertheless, the transit with the measured depth and period would be consistent with a duration of 6 hours if the orbit were to have an eccentricity of $e \gtrsim$ 0.683.

\begin{figure}[!]
\begin{minipage}{\linewidth}
\begin{center}
\includegraphics[angle=0,trim = 0mm 00mm 0mm 00mm, clip, width=\linewidth]{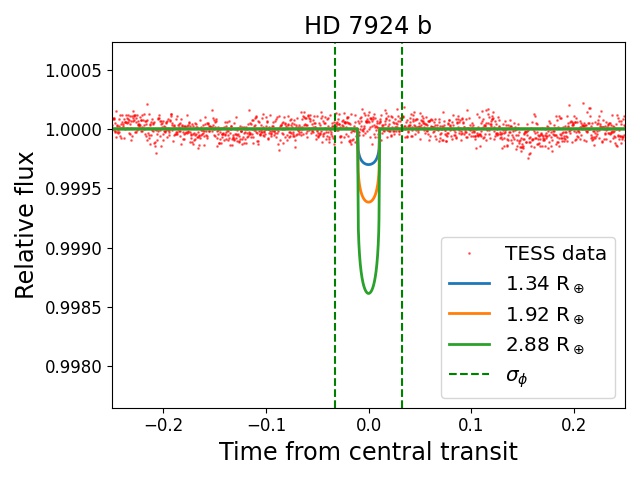}
\includegraphics[angle=0,trim = 0mm 00mm 0mm 00mm, clip, width=\linewidth]{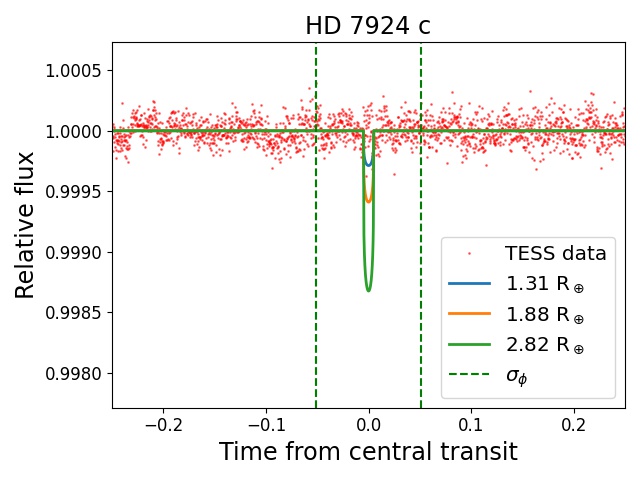}
\includegraphics[angle=0,trim = 0mm 00mm 0mm 00mm, clip, width=\linewidth]{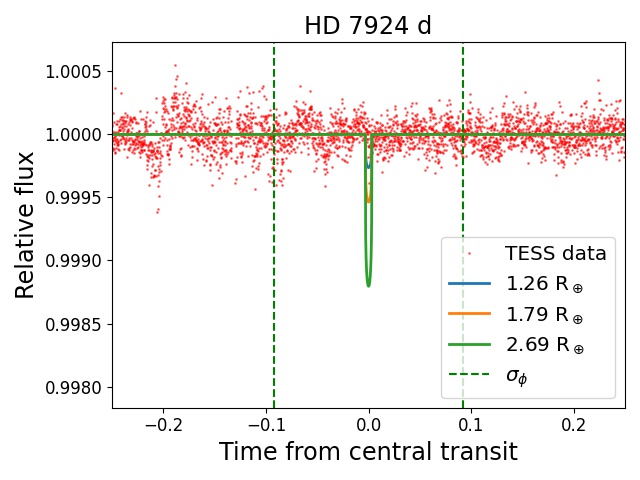}
\end{center}
\caption{\small{Same as Fig.~\ref{fig:61virb} but for planets HD~7924~b, c and d.}}

\label{fig:HD7924fold}
 \end{minipage}
\end{figure}

The models for each planet are shown in Fig.~\ref{fig:HD7924fold}. In all cases, the precision would allow the detection of transits  of planets as small as around 1.2 $R_\oplus$, both in terms of depth and sampling. We conclude that no planetary transits are detected in this target.

\begin{figure}[!]
\begin{minipage}{\linewidth}
\begin{center}
\includegraphics[angle=0,trim = 0mm 00mm 0mm 00mm, clip, width=\linewidth]{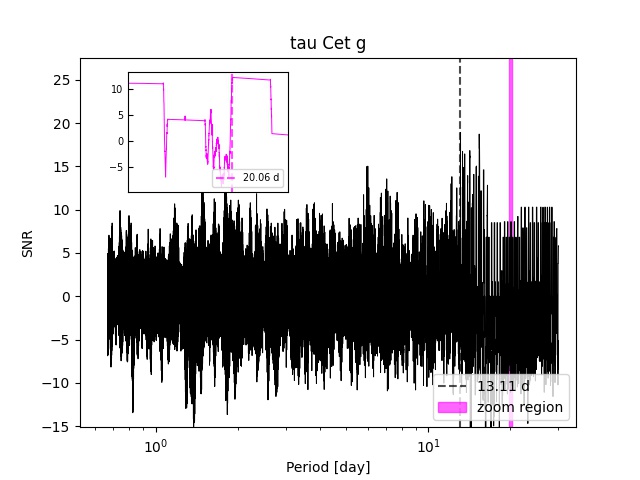}
\end{center}
\caption{\small{Same as Fig.~\ref{fig:GJ887} but for tau~Cet~g.}}

\label{fig:taucet}
 \end{minipage}
\end{figure}

\subsubsection{tau~Cet} 

tau~Cet has been reported to host four planets with masses of less than 4~M$_{\oplus}$ and periods of P$_{g}$= 20 , P$_{h}$= 49.41, P$_{e}$= 162.87, and P$_{f}$= 636.13 days \citep{feng2017}. None of them have  been detected in transit, but only planet g was included in our sample because of its period. The BLS spectra exhibit peaks with large SNR $\sim$18.9 in 13 days for the global analysis and $\sim$12 in 20 days for the local one (see Table~\ref{apptab3:bls} and Fig.~\ref{fig:taucet}).

Checking the data release notes for sectors 3 and 30 (specifically seeing
Fig.~6\footnote{https://archive.stsci.edu/tess/tess$_{-}$drn.html}), we note that the light curve produced by the SPOC pipeline for this object is dominated by residual systematic errors. This is seen in the large scatter present in the baseline BLS power. In this way, the signals detected correspond to instrumental effects.

\begin{figure}[!]
\begin{minipage}{\linewidth}
\begin{center}
\includegraphics[angle=0,trim = 0mm 00mm 0mm 00mm, clip, width=\linewidth]{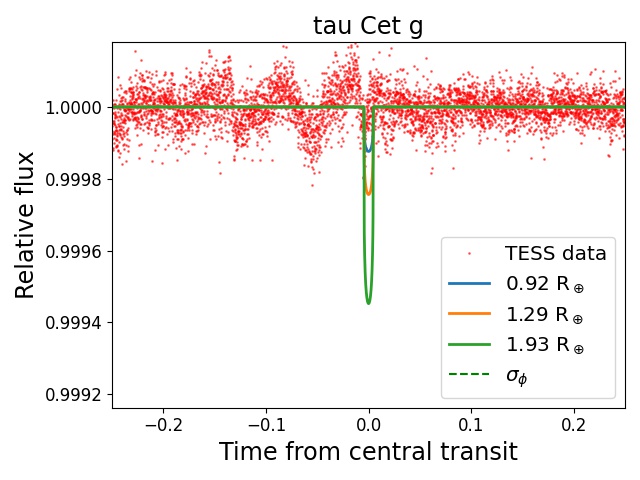}
\end{center}

\caption{\small{Same as Fig.~\ref{fig:61virb} but for tau~Cet~g. In this case, the orbital phase uncertainties are outside the limits of the figure.}}

\label{fig:taucetg}
 \end{minipage}
\end{figure}

Finally, in Fig.~\ref{fig:taucetg} we show the light curve folded with parameters from the literature and the models of possible transits for the measured planetary masses. It is clear that, in spite of the increased systematic errors, the current data would allow the detection of the transits of these planets, at least for the two composition assumptions producing the largest planets.

\subsection{Constraints on planetary masses\label{sect.mass}}

 From the absence of transits, one can put upper limits on the orbital inclination of each planet. Based on the parameters of the systems, and assuming the planets do not transit, we computed the maximum orbital inclination. In Table~\ref{apptab2:planet} we show the inclination angles calculated in two ways: Column 9 is the  maximum possible inclination assuming no transits of any kind ---not even grazing--- are present ($i_{\rm max}$) and column 10 is the inclination constraint if only full transits are assumed to be discarded ($i_{\rm full}$).

\section{Summary and Discussion \label{sect.summ}}

TESS data give us the opportunity to detect transits of Earth-size planets. Here we present the results of a search for transits of a sample of 68 planets detected by RV surveys in 43 systems. We used TESS photometric
data reduced using the processing pipeline developed by the SPOC. We ruled out transits for 66 out of the surveyed 68 planets.

In this work, we wanted to carry out a general exploration using the whole automatically processed data (PDCSAP$_{-}$FLUX light curves). In the first part, we performed a blind planetary search and a local search for the known planets.

A BLS spectrum was computed for each system using all available sectors. The frequency grid spacing was given by the maximum baseline of observations. In cases such as 61~Vir, GJ~887, and HD~20794, the downlink gaps, together with some photometric systematic effects, introduced false positives in the periodograms. Similarly, the gaps between nonconsecutive sectors produced spurious peaks in the BLS spectrum, such as in the HD~40307 system.
To refine this process, it would be useful to produce custom light curves and proceed to explore them sector by sector or simply consider scanning them in consecutive sectors.

For HD~136352, we missed the detection of the two transiting planets in the global-BLS search, although visual inspection revealed the clear transit of planet c. One of the possible reasons for this is the systematic error in the light curve, another is the observational baseline available for this target. 
On the other hand, targets such as YZ Cet are active cool stars and present flares in their light curves that could hide or weaken the transit signals should they occur and therefore affect the determination of the transit parameters and planet characterization \citep{Bentley2009,Giacobbe2012,Loyd2014}. To improve the study of this kind of object, it would be necessary to account for the presence of flares when looking for transits. This topic is treated in a recent work about Proxima Centauri; the authors obtain better results when identifying and modeling the flares using a template, which they then subtract from the data, and perform the transit exploration \citep{gilbert2021}; similar processes were performed for the AU Mic system by \cite{gilbert2021b} and \cite{Szab2021}.
 
In the second part of this work, we modeled synthetic transits for every planet in the sample, considering three different radii from compositions following Fortney et al. \citeyearpar{Fortney_2007,Fortney_Err2007}:  pure ice, pure rock, and pure iron composition. These models were used to visually evaluate whether or not such transits could be present in the TESS data. We used these models to check detection limits of each transit, given a maximum and minimum depth and duration for pure ice and pure iron planets, respectively, comparing with the corresponding upper limit radius. This procedure allows us to establish that none of the studied exoplanets are transiting their host star, except for HD~136352 b and c. However, for these two planets, we found a displacement from the literature midpoint. These offsets could be caused by the gravitational interaction between these planets or by the third outer planet "d", producing a transit timing variation (TTV) signal. In the work of \citep{delrez2021}, a TTV analysis was performed, but the authors did not measure any significant TTV signal. Therefore, we attribute this discrepancy to the low precision of the Doppler method in determining the transit midpoint.

The presence of TTVs in any of the studied systems would certainly affect our ability to detect the transit signals, as they would no longer occur with strict periodicity, and the BLS signal would be degraded. A detailed analysis of the TTVs of these systems is outside the scope of this paper, and we simply assume that there are no timing variations in the studied planets.

Overall, our work allows us to rule out transits for 28 of the 68 planets with radii larger than a pure iron composition radius in our sample, and 51 out of 68 considering a terrestrial planet composition (under the assumption of central transits).

\begin{acknowledgements}

This research has made use of the NASA {Exoplanet Archive}, which is operated by the California Institute of Technology, under contract with the National Aeronautics and Space Administration under the Exoplanet Exploration Program.
This paper includes data collected by the TESS mission. Funding for the TESS mission is provided by the NASA's Science Mission Directorate.
We acknowledge the use of public TESS data from
pipelines at the TESS Science Office and at the TESS Science
Processing Operations Center.
We also thank the referee for the careful review of the manuscript and for the constructive suggestions and comments, which improved the content and quality of the article.

\end{acknowledgements}

\bibliographystyle{aa}
\bibliography{grandbiblio.bib}

\clearpage
\onecolumn

\begin{appendix}

\section{Tables }

The following tables describe the characteristics of the planets studied in this work. In Table \ref{apptab2:planet} we list the literature orbital parameters and planetary radii for pure ice, pure rock, and pure iron compositions, in addition to the orbital inclinations obtained in Sects. \ref{sect.model} and \ref{sect.mass}, respectively. In Table \ref{apptab3:bls} we list the results of the BLS search performed in Sect. \ref{sect.transitsearch}. Finally, Table \ref{apptab4:upplim} presents the upper detection limits for each target.

\longtab{
\centering
\begin{longtable}{lccccccccc}	

\caption{\label{apptab2:planet} Sample of planets studied in this work.}\\
\hline
\hline
Planet	&	Orbital period			&	M$_{P} \sin i$			&	References 	&	T$_{C}$			&	R$_{ice}$	&	R$_{rock}$	&	R$_{iron}$	&	$i_{\rm max}$	&	$i_{\rm full}$	\\
	&	[d]			&	[M$_{\oplus}$]			&		&	[BJD]			&	[R$_{\oplus}$]	&	[R$_{\oplus}$]	&	[R$_{\oplus}$]	&	[deg]	&	[deg]	\\
\hline
\endfirsthead
\caption{continued.}\\
\hline
\hline
Planet	&	Orbital period			&	M$_{P} \sin i$			&	References 	&	T$_{C}$			&	R$_{ice}$	&	R$_{rock}$	&	R$_{iron}$	&	$i_{\rm max}$	&	$i_{\rm full}$	\\
	&	[d]			&	[M$_{\oplus}$]			&		&	[BJD]			&	[R$_{\oplus}$]	&	[R$_{\oplus}$]	&	[R$_{\oplus}$]	&	[deg]	&	[deg]	\\
\hline
\endhead
\hline
\endfoot
\hline
\endlastfoot

61 Vir b	&	4.215	$\pm$	0.001	&	5.10	$\pm$	0.50	&	 1 	&	2453367.071	$\pm$ 0.574	&	2.54	&	1.70	&	1.19	&	84.04	&	84.33	\\
BD-06 1339 b	&	3.873	$\pm$	0.000	&	8.50	$\pm$	1.30	&	 2 	&	2455221.463	$\pm$ 0.102	&	2.87	&	1.91	&	1.34	&	86.09	&	86.42	\\
BD-08 2823 b	&	5.600	$\pm$	0.020	&	13.00	$\pm$	3.00	&	 3 	&	2454638.452	$\pm$ 1.574	&	3.16	&	2.11	&	1.46	&	86.12	&	86.42	\\
DMPP-1 c	&	6.584	$\pm$	0.003	&	9.60	$\pm$	0.53	&	 4 	&	-- $\pm$ --	&	2.95	&	1.97	&	1.37	&	--	&	--	\\
DMPP-1 d	&	2.882	$\pm$	0.001	&	3.35	$\pm$	0.38	&	 4 	& -- $\pm$ --	&	2.29	&	1.53	&	1.08	&	--	&	--	\\
DMPP-1 e	&	5.516	$\pm$	0.002	&	4.13	$\pm$	0.66	&	 4 	&	--	$\pm$ --	&	2.41	&	1.61	&	1.13	&	--	&	--	\\
GJ 1061 b	&	3.204	$\pm$	0.001	&	1.37	$\pm$	0.15	&	 5 	&	2458300.280	$\pm$ 0.972	&	1.81	&	1.21	&	0.86	&	87.14	&	87.69	\\
GJ 1061 c	&	6.689	$\pm$	0.005	&	1.74	$\pm$	0.23	&	 5 	&	2458300.317	$\pm$ 2.201	&	1.93	&	1.29	&	0.92	&	88.10	&	88.49	\\
GJ 1061 d	&	13.031	$\pm$	0.030	&	1.64	$\pm$	0.23	&	 5 	&	2458305.283	$\pm$ 4.216	&	1.90	&	1.27	&	0.90	&	88.55	&	88.84	\\
GJ 1132 c	&	8.929	$\pm$	0.010	&	2.64	$\pm$	0.44	&	 6 	&	2457506.020	$\pm$ 0.340	&	2.15	&	1.44	&	1.02	&	88.63	&	88.87	\\
GJ 15 A b	&	11.441	$\pm$	0.002	&	3.03	$\pm$	0.45	&	 7 	&	2456995.860	$\pm$ 0.310	&	2.23	&	1.49	&	1.05	&	88.39	&	88.55	\\
GJ 163 b	&	8.632	$\pm$	0.002	&	10.60	$\pm$	0.60	&	 8 	&	2452936.121	$\pm$ 0.404	&	3.02	&	2.01	&	1.40	&	87.83	&	88.09	\\
GJ 163 c	&	25.631	$\pm$	0.026	&	6.80	$\pm$	0.90	&	 8 	&	2452922.230	$\pm$ 2.295	&	2.72	&	1.82	&	1.27	&	89.09	&	89.19	\\
GJ 180 b	&	17.133	$\pm$	0.003	&	6.49	$\pm$	0.68	&	 9 	&	2451567.018	$\pm$ 7.334	&	2.69	&	1.80	&	1.26	&	88.83	&	88.96	\\
GJ 180 c	&	24.329	$\pm$	0.052	&	6.40	$\pm$	3.70	&	 10 	&	-- $\pm$ --	&	2.68	&	1.79	&	1.25	&	89.11	&	89.21	\\
GJ 273 b	&	18.650	$\pm$	0.006	&	2.89	$\pm$	0.26	&	 11 	&	2456249.295	$\pm$ 0.489	&	2.20	&	1.47	&	1.04	&	89.07	&	89.19	\\
GJ 273 c	&	4.723	$\pm$	0.000	&	1.18	$\pm$	0.16	&	 11 	&	2456238.580	$\pm$ 0.260	&	1.74	&	1.16	&	0.83	&	87.68	&	87.91	\\
GJ 3138 b	&	1.220	$\pm$	0.000	&	1.78	$\pm$	0.34	&	 11 	&	2456685.470	$\pm$ 0.092	&	1.94	&	1.30	&	0.92	&	82.70	&	83.20	\\
GJ 3138 c	&	5.974	$\pm$	0.001	&	4.18	$\pm$	0.60	&	 11 	&	2456685.407	$\pm$ 0.250	&	2.42	&	1.62	&	1.14	&	87.51	&	87.73	\\
GJ 3293 e	&	13.254	$\pm$	0.008	&	3.28	$\pm$	0.64	&	 11 	&	2456426.847	$\pm$ 0.538	&	2.28	&	1.52	&	1.07	&	88.56	&	88.70	\\
GJ 3323 b	&	5.364	$\pm$	0.001	&	2.02	$\pm$	0.25	&	 11 	&	2456803.797	$\pm$ 0.213	&	2.01	&	1.34	&	0.95	&	88.82	&	89.14	\\
GJ 3473 c	&	15.509	$\pm$	0.033	&	7.41	$\pm$	0.91	&	 12 	&	2458575.620	$\pm$ 0.420	&	2.78	&	1.85	&	1.30	&	88.81	&	88.97	\\
GJ 357 c	&	9.125	$\pm$	0.001	&	3.40	$\pm$	0.46	&	 13 	&	2458314.300	$\pm$ 0.420	&	2.30	&	1.53	&	1.08	&	88.40	&	88.59	\\
GJ 433 b	&	7.371	$\pm$	0.001	&	5.79	$\pm$	--	&	 14 	&	2454285.651	$\pm$ 2.114	&	2.62	&	1.75	&	1.23	&	87.66	&	87.87	\\
GJ 676 A d	&	3.601	$\pm$	0.000	&	4.40	$\pm$	0.30	&	 15 	&	2455500.024	$\pm$ 0.219	&	2.45	&	1.64	&	1.15	&	86.04	&	86.29	\\
GJ 682 b	&	17.478	$\pm$	0.062	&	4.40	$\pm$	3.70	&	 10 	&	-- $\pm$ --	&	2.45	&	1.64	&	1.15	&	88.82	&	88.97	\\
GJ 876 d	&	1.938	$\pm$	0.000	&	6.83	$\pm$	0.40	&	 16 	&	2450601.406	$\pm$ 0.027	&	2.73	&	1.82	&	1.27	&	86.45	&	86.99	\\
GJ 887 b	&	9.262	$\pm$	0.001	&	4.20	$\pm$	0.60	&	 17 	&	-- $\pm$ --		&	2.42	&	1.62	&	1.14	&	87.93	&	88.12	\\
GJ 887 c	&	21.789	$\pm$	0.005	&	7.60	$\pm$	1.20	&	 17 	&	-- $\pm$ --	&	2.80	&	1.87	&	1.30	&	88.93	&	89.04	\\
HD 10180 c	&	5.760	$\pm$	0.000	&	13.20	$\pm$	0.40	&	 18 	&	2454003.318	$\pm$ 0.722	&	3.17	&	2.12	&	1.47	&	85.43	&	85.66	\\
HD 10180 d	&	16.357	$\pm$	0.004	&	12.00	$\pm$	0.70	&	 18 	&	2454027.412	$\pm$ 1.871	&	3.11	&	2.07	&	1.44	&	87.78	&	87.89	\\
HD 109271 b	&	7.854	$\pm$	0.001	&	17.16	$\pm$	1.27	&	 2 	&	2455721.655	$\pm$ 4.104	&	3.36	&	2.24	&	1.55	&	86.38	&	86.55	\\
HD 136352 b	&	11.582	$\pm$	0.002	&	4.81	$\pm$	0.57	&	 19 	&	2455494.391	$\pm$ 0.353	&	2.50	&	1.67	&	1.18	&	86.95	&	87.09	\\
HD 136352 c	&	27.582	$\pm$	0.009	&	10.80	$\pm$	1.05	&	 19 	&	2455477.862	$\pm$ 0.411	&	3.03	&	2.02	&	1.41	&	88.40	&	88.49	\\
HD 1461 b	&	5.772	$\pm$	0.000	&	6.44	$\pm$	0.61	&	 20 	&	2455152.144	$\pm$ 2.213	&	2.69	&	1.79	&	1.26	&	85.20	&	85.41	\\
HD 1461 c	&	13.505	$\pm$	0.002	&	5.59	$\pm$	0.73	&	 20 	&	2455147.235	$\pm$ 5.293	&	2.60	&	1.73	&	1.22	&	87.21	&	87.32	\\
HD 158259 c	&	3.432	$\pm$	0.000	&	5.60	$\pm$	0.60	&	 21 	&	2457496.682	$\pm$ 0.102	&	2.60	&	1.74	&	1.22	&	82.47	&	82.75	\\
HD 158259 d	&	5.198	$\pm$	0.001	&	5.41	$\pm$	0.71	&	 21 	&	2457497.705	$\pm$ 0.213	&	2.58	&	1.72	&	1.21	&	84.32	&	84.53	\\
HD 158259 e	&	7.951	$\pm$	0.002	&	6.08	$\pm$	0.94	&	 21 	&	2457491.522	$\pm$ 0.683	&	2.65	&	1.77	&	1.24	&	85.70	&	85.86	\\
HD 158259 f	&	12.028	$\pm$	0.009	&	6.14	$\pm$	1.37	&	 21 	&	2457496.957	$\pm$ 1.104	&	2.66	&	1.77	&	1.24	&	86.74	&	86.87	\\
HD 181433 b	&	9.375	$\pm$	0.002	&	7.40	$\pm$	--	&	 22 	&	2452786.205	$\pm$ 0.411	&	2.78	&	1.85	&	1.30	&	87.13	&	87.30	\\
HD 20003 b	&	11.848	$\pm$	0.002	&	11.66	$\pm$	1.04	&	 19 	&	2455483.762	$\pm$ 0.377	&	3.09	&	2.06	&	1.43	&	88.09	&	88.20	\\
HD 20781 b	&	5.314	$\pm$	0.001	&	1.93	$\pm$	0.35	&	 19 	&	2455503.289	$\pm$ 0.236	&	1.98	&	1.32	&	0.94	&	85.23	&	85.44	\\
HD 20781 c	&	13.891	$\pm$	0.003	&	5.33	$\pm$	0.67	&	 19 	&	2455506.439	$\pm$ 0.355	&	2.57	&	1.71	&	1.20	&	87.68	&	87.81	\\
HD 20781 b	&	29.158	$\pm$	0.010	&	10.61	$\pm$	1.20	&	 19 	&	2455513.243	$\pm$ 0.425	&	3.02	&	2.01	&	1.40	&	88.45	&	88.55	\\
HD 20794 b	&	18.315	$\pm$	0.008	&	2.70	$\pm$	0.30	&	 23 	&	2454779.850	$\pm$ 0.002	&	2.17	&	1.45	&	1.02	&	88.06	&	88.14	\\
HD 213885 c	&	4.785	$\pm$	0.001	&	19.95	$\pm$	1.36	&	 24 	&	2458396.635	$\pm$ 0.054	&	3.48	&	2.31	&	1.60	&	84.83	&	85.13	\\
HD 215497 b	&	3.934	$\pm$	0.001	&	6.60	$\pm$	3.00	&	 25 	&	2454858.922	$\pm$ 0.476	&	2.70	&	1.80	&	1.26	&	84.01	&	84.35	\\
HD 21693 b	&	22.679	$\pm$	0.009	&	8.23	$\pm$	1.05	&	 19 	&	2455480.755	$\pm$ 0.745	&	2.85	&	1.90	&	1.33	&	88.46	&	88.55	\\
HD 219134 f	&	22.714	$\pm$	0.015	&	7.30	$\pm$	0.40	&	 26 	&	2457716.310	$\pm$ 0.500	&	2.77	&	1.85	&	1.29	&	88.26	&	88.37	\\
HD 31527 b	&	16.554	$\pm$	0.003	&	10.47	$\pm$	0.88	&	 19 	&	2455501.459	$\pm$ 0.264	&	3.01	&	2.01	&	1.40	&	87.48	&	87.61	\\
HD 40307 b	&	4.311	$\pm$	0.000	&	3.81	$\pm$	0.30	&	 20 	&	2454520.397	$\pm$ 1.560	&	2.36	&	1.58	&	1.11	&	85.58	&	85.84	\\
HD 40307 c	&	9.621	$\pm$	0.000	&	6.43	$\pm$	0.44	&	 20 	&	2454514.501	$\pm$ 3.913	&	2.69	&	1.79	&	1.26	&	87.48	&	87.65	\\
HD 40307 d	&	20.429	$\pm$	0.002	&	8.74	$\pm$	0.58	&	 20 	&	2454514.877	$\pm$ 6.127	&	2.89	&	1.93	&	1.34	&	88.46	&	88.57	\\
HD 45184 b	&	5.885	$\pm$	0.000	&	12.19	$\pm$	1.03	&	 19 	&	2455499.415	$\pm$ 0.105	&	3.12	&	2.08	&	1.44	&	85.20	&	85.45	\\
HD 45184 c	&	13.135	$\pm$	0.003	&	8.81	$\pm$	1.02	&	 19 	&	2455494.882	$\pm$ 0.336	&	2.89	&	1.93	&	1.35	&	87.25	&	87.38	\\
HD 51608 b	&	14.073	$\pm$	0.002	&	12.77	$\pm$	1.20	&	 19 	&	2455493.604	$\pm$ 0.168	&	3.15	&	2.10	&	1.46	&	87.41	&	87.57	\\
HD 69830 b	&	8.667	$\pm$	0.003	&	10.20	$\pm$	--	&	 27 	&	2453499.281	$\pm$ 0.971	&	2.99	&	2.00	&	1.39	&	87.14	&	87.32	\\
HD 7924 b	&	5.398	$\pm$	0.000	&	8.68	$\pm$	0.52	&	 28 	&	2455586.380	$\pm$ 0.086	&	2.88	&	1.92	&	1.34	&	85.98	&	86.24	\\
HD 7924 c	&	15.299	$\pm$	0.003	&	7.86	$\pm$	0.72	&	 28 	&	2455586.290	$\pm$ 0.400	&	2.82	&	1.88	&	1.31	&	87.83	&	87.97	\\
HD 7924 d	&	24.451	$\pm$	0.015	&	6.44	$\pm$	0.78	&	 28 	&	2455579.100	$\pm$ 1.000	&	2.69	&	1.79	&	1.26	&	88.14	&	88.25	\\
HIP 54373 b	&	7.760	$\pm$	0.003	&	8.62	$\pm$	1.84	&	 29 	&	2455199.109	$\pm$ 0.706	&	2.88	&	1.92	&	1.34	&	87.26	&	87.54	\\
HIP 54373 c	&	15.144	$\pm$	0.008	&	12.44	$\pm$	2.11	&	 29 	&	2455189.728	$\pm$ 2.062	&	3.13	&	2.09	&	1.45	&	88.27	&	88.46	\\
HIP 57274 b	&	8.135	$\pm$	0.004	&	11.60	$\pm$	1.30	&	 30 	&	2455801.779	$\pm$ 0.271	&	3.08	&	2.06	&	1.43	&	84.95	&	85.30	\\
YZ Cet b	&	1.969	$\pm$	0.000	&	0.75	$\pm$	0.13	&	 31 	&	2456846.911	$\pm$ 0.095	&	1.53	&	1.01	&	0.73	&	86.88	&	87.36	\\
YZ Cet c	&	3.060	$\pm$	0.000	&	0.98	$\pm$	0.14	&	 31 	&	2456847.050	$\pm$ 0.140	&	1.65	&	1.10	&	0.79	&	87.66	&	88.05	\\
YZ Cet d	&	4.656	$\pm$	0.000	&	1.14	$\pm$	0.17	&	 31 	&	2456847.440	$\pm$ 0.160	&	1.72	&	1.15	&	0.82	&	88.20	&	88.51	\\
$\tau$~Cet g	&	20.000	$\pm$	0.020	&	1.75	$\pm$	0.25	&	 32 	&	2451755.426	$\pm$ 1.910	&	1.93	&	1.29	&	0.92	&	88.25	&	88.32	\\
55 Cnc e	&	0.737	$\pm$	0.000	&	7.99	$\pm$	0.25	&	 33 	&	2457063.210	$\pm$ 0.001	&	2.83	&	1.89	&	1.32	&	73.03	&	73.96	\\
																									
\hline																									
\hline																									
\end{longtable}		
\tablebib{
(1) \citet{Vogt2010}; (2) \citet{LoCurto2013}; (3) \citet{Hebrard2010} ; (4) \citet{Staab2020}; (5)\citet{Dreizler2020}; (6) \citet{bonfils2018}; (7) \citet{Pinamonti2018}; (8) \citet{bonfils2013}; (9) \citet{feng2020}; (10) \citet{tuomi2014}; (11) \citet{AstuDefru2017a}; (12) \citet{Kemmer2020}; (13) \citet{Luque2019}; (14) \citet{delfosse2013}; (15) \citet{Sahlmann2016}; (16) \citet{rivera2010b}; (17) \citet{Jeffers2020}; (18) \citet{Kane2014}; (19) \citet{Udry2019}; (20) \citet{diaz2016a}; (21) \citet{Hara2020}; (22) \citet{campanella2011}; (23) \citet{pepe2011}; (24) \citet{Espinoza2020}; (25) \citet{locurto2010}; (26) \citet{guillon2017b}; (27) \citet{lovis2006}; (28) \citet{fulton2015}; (29) \citet{feng2019}; (30) \citet{fischer2012}; (31) \citet{AstuDefru2017b}; (32) \citet{feng2017}; (33)  \citet{bou2018}.}																	}

\longtab{
\centering
\begin{longtable}{lcc|lcc} 											

\caption{\label{apptab3:bls} BLS parameters.}\\

\hline											
\hline											
Host star	&	Per	&	SNR	&	Planet	&	Per$_{10\sigma}$	&	SNR$_{10\sigma}$	\\
	&	[d]	&		&		&	[d]	&		\\
\hline											
\endfirsthead											
\caption{continued.}\\											
\hline											
\hline											
Host star	&	Per	&	SNR	&	Planet	&	Per$_{10\sigma}$	&	SNR$_{10\sigma}$	\\
	&	[d]	&		&		&	[d]	&		\\
\hline											
\endhead											
\hline											
\endfoot
\hline
\endlastfoot

61 Vir	&	7.451	&	10.47	&	61 Vir b	&	4.221	&	1.09	\\
BD-06 1339	&	0.793	&	2.39	&	BD-06 1339 b	&	3.869	&	0.77	\\
BD-08 2823	&	10.640	&	4.66	&	BD-08 2823 b	&	5.638	&	2.81	\\
DMPP-1	&	4.848	&	2.95	&	DMPP-1 c	&	6.580	&	1.67	\\
	&		&		&	DMPP-1 d	&	2.890	&	1.53	\\
	&		&		&	DMPP-1 e	&	5.521	&	2.00	\\
GJ 1061	&	16.618	&	3.25	&	GJ 1061 b	&	3.209	&	1.61	\\
	&		&		&	GJ 1061 c	&	6.709	&	1.59	\\
	&		&		&	GJ 1061 d	&	12.986	&	2.12	\\
GJ 1132	&	1.629	&	18.17	&	GJ 1132 c	&	8.917	&	2.48	\\
	&	19.118	&	4.39	&		&		&		\\
GJ 15 A	&	4.339	&	9.67	&	GJ 15 A b	&	11.458	&	1.83	\\
GJ 163	&	13.326	&	3.16	&	GJ 163 b	&	8.635	&	1.56	\\
	&		&		&	GJ 163 c	&	25.483	&	1.80	\\
GJ 180 	&	25.814	&	4.87	&	GJ 180 b	&	17.119	&	2.21	\\
	&		&		&	GJ 180 c	&	23.994	&	3.14	\\
GJ 273	&	1.393	&	2.97	&	GJ 273 b	&	18.600	&	0.80	\\
	&		&		&	GJ 273 c	&	4.727	&	0.51	\\
GJ 3138	&	1.417	&	2.23	&	GJ 3138 b	&	1.220	&	1.70	\\
	&		&		&	GJ 3138 c	&	5.978	&	1.05	\\
GJ 3293	&	39.520	&	3.30	&	GJ 3293 e	&	13.294	&	2.22	\\
GJ 3323	&	1.336	&	4.18	&	GJ 3323 b	&	5.365	&	1.89	\\
GJ 3473 	&	1.198	&	7.19	&	GJ 3473 c	&	15.828	&	1.99	\\
        &	2.017	&	2.03	&		&		&		\\
GJ 357	&	3.931	&	21.31	&	GJ 357 c	&	9.123	&	2.69	\\
	&	15.215	&	4.49	&		&		&		\\
GJ 433 	&	0.796	&	2.66	&	GJ 433 b	&	7.364	&	1.60	\\
GJ 676 A	&	8.628	&	3.68	&	GJ 676 A d	&	3.600	&	1.09	\\
GJ 682 	&	2.766	&	2.73	&	GJ 682 b	&	17.795	&	1.62	\\
GJ 876	&	13.791	&	5.20	&	GJ 876 d	&	1.938	&	1.18	\\
GJ 887 	&	15.358	&	39.33	&	GJ 887 b	&	9.270	&	14.34	\\
	&		&		&	GJ 887 c	&	21.775	&	3.20	\\
HD 10180	&	34.947	&	3.25	&	HD 10180 c	&	5.760	&	1.18	\\
	&		&		&	HD 10180 d	&	16.360	&	2.17	\\
HD 109271	&	5.037	&	2.12	&	HD 109271 b	&	7.855	&	0.63	\\
HD 136352	&	11.680	&	14.29	&	HD 136352 b	&	11.568	&	6.22	\\
	&		&		&	HD 136352 c	&	27.671	&	13.00	\\
HD 1461 	&	2.490	&	4.90	&	HD 1461 b	&	5.771	&	0.18	\\
	&		&		&	HD 1461 c	&	13.523	&	0.01	\\
HD 158259 	&	2.178	&	25.26	&	HD 158259 c	&	3.435	&	1.19	\\
        	&	1.372	&	5.99	&	HD 158259 d	&	5.196	&	2.59	\\
        	&	    	&	    	&	HD 158259 e	&	7.944	&	2.56	\\
        	&   		&	    	&	HD 158259 f	&	11.954	&	2.54	\\
HD 181433 	&	2.549	&	2.12	&	HD 181433 b	&	9.357	&	0.20	\\
HD 20003	&	43.185	&	3.45	&	HD 20003 b	&	11.835	&	1.88	\\
HD 20781	&	11.499	&	3.59	&	HD 20781 b	&	5.311	&	1.68	\\
	&		&		&	HD 20781 c	&	13.920	&	2.15	\\
	&		&		&	HD 20781 d	&	29.065	&	1.14	\\
HD 20794	&	34.971	&	11.06	&	HD 20794 b	&	18.305	&	5.32	\\
HD 213885	&	1.008	&	17.30	&	HD 213885 c	&	4.781	&	1.62	\\
	&	26.054	&	2.85	&		&		&		\\
HD 215497	&	1.331	&	2.78	&	HD 215497 b	&	3.931	&	1.68	\\
HD 21693	&	1.576	&	3.56	&	HD 21693 b	&	22.626	&	2.86	\\
HD 219134 	&	3.093	&	39.04	&	HD 219134 f	&	22.682	&	10.58	\\
	&	6.766	&	32.80	&		&		&		\\
	&	13.710	&	12.17	&		&		&		\\
HD 31527	&	14.155	&	7.99	&	HD 31527 b	&	16.585	&	2.98	\\
HD 40307 	&	143.236	&	7.26	&	HD 40307 b	&	4.312	&	1.20	\\
	&		&		&	HD 40307 c	&	9.625	&	0.38	\\
	&		&		&	HD 40307 d	&	20.445	&	1.76	\\
HD 45184	&	13.760	&	5.14	&	HD 45184 b	&	5.888	&	1.97	\\
	&		&		&	HD 45184 c	&	13.141	&	0.99	\\
HD 51608	&	4.238	&	5.11	&	HD 51608 b	&	14.088	&	1.41	\\
HD 69830	&	5.850	&	5.44	&	HD 69830 b	&	8.666	&	3.64	\\
HD 7924	&	1.005	&	10.02	&	HD 7924 b	&	5.396	&	2.92	\\
	&		&		&	HD 7924 c	&	15.267	&	0.23	\\
	&		&		&	HD 7924 d	&	24.558	&	0.21	\\
HIP 54373 	&	1.405	&	3.76	&	HIP 54373 b	&	7.733	&	2.88	\\
	&		&		&	HIP 54373 c	&	15.144	&	1.80	\\
HIP 57274 	&	0.958	&	1.84	&	HIP 57274 b	&	8.120	&	0.86	\\
YZ Cet	&	4.533	&	5.21	&	YZ Cet b	&	1.970	&	3.16	\\
	&		&		&	YZ Cet c	&	3.060	&	4.49	\\
	&		&		&	YZ Cet d	&	4.659	&	1.29	\\
$\tau$~Cet 	&	13.115	&	18.93	&	$\tau$ Cet g	&	20.060	&	12.27	\\
55 Cnc	&	0.736	&	105.41	&	55 Cnc e	&	0.737	&	113.38	\\
											
\hline											
\hline											
\end{longtable} 
}

\longtab{
\centering
\begin{longtable}{lcc} 					
\caption{\label{apptab4:upplim} Upper limits.}\\					

\hline					
\hline	

Planet	&	Upper limit $(R_{\rm p}/R_{\star})^2$	&	R$_{\rm p}$	\\
	&	[ppm]	&	[R$_{\oplus}$]	\\
\hline					
\endfirsthead					
\caption{continued.}\\					
\hline					
\hline					
Planet	&	Upper limit $(R_{\rm p}/R_{\star})^2$	&	R$_{\rm p}$	\\
	&	[ppm]	&	[R$_{\oplus}$]	\\
\hline					
\endhead					
\hline					
\endfoot
\hline
\endlastfoot

61 Vir b	&	116	&	1.133	\\
BD-06 1339 b	&	422	&	1.350	\\
BD-08 2823 b	&	581	&	1.866	\\
DMPP-1 c	&	--	&	--	\\
DMPP-1 d	&	--	&	--	\\
DMPP-1 e	&	--	&	--	\\
GJ 1061 b	&	809	&	0.484	\\
GJ 1061 c	&	829	&	0.490	\\
GJ 1061 d	&	790	&	0.478	\\
GJ 1132 c	&	2010	&	1.012	\\
GJ 15 A b	&	467	&	0.896	\\
GJ 163 b	&	784	&	1.298	\\
GJ 163 c	&	760	&	1.278	\\
GJ 180 b	&	576	&	1.100	\\
GJ 180 c	&	--	&	--	\\
GJ 273 b	&	349	&	0.597	\\
GJ 273 c	&	350	&	0.598	\\
GJ 3138 b	&	759	&	1.503	\\
GJ 3138 c	&	786	&	1.529	\\
GJ 3293 e	&	1040	&	1.421	\\
GJ 3323 b	&	1538	&	0.509	\\
GJ 3473 c	&	2221	&	1.850	\\
GJ 357 c	&	575	&	0.881	\\
GJ 433 b	&	349	&	1.019	\\
GJ 676 A d	&	450	&	1.597	\\
GJ 682 b	&	--	&	--	\\
GJ 876 d	&	343	&	0.606	\\
GJ 887 b	&	--	&	--	\\
GJ 887 c	&	--	&	--	\\
HD 10180 c	&	220	&	1.794	\\
HD 10180 d	&	225	&	1.816	\\
HD 109271 b	&	304	&	2.391	\\
HD 136352 b	&	157	&	1.381	\\
HD 136352 c	&	293	&	1.886	\\
HD 1461 b	&	171	&	1.596	\\
HD 1461 c	&	153	&	1.513	\\
HD 158259 c	&	189	&	1.935	\\
HD 158259 d	&	200	&	1.989	\\
HD 158259 e	&	199	&	1.984	\\
HD 158259 f	&	192	&	1.950	\\
HD 181433 b	&	290	&	1.540	\\
HD 20003 b	&	332	&	1.832	\\
HD 20781 b	&	576	&	2.182	\\
HD 20781 c	&	615	&	2.254	\\
HD 20781 d	&	331	&	1.653	\\
HD 20794 b	&	95	&	0.978	\\
HD 213885 c	&	305	&	2.034	\\
HD 215497 b	&	410	&	1.878	\\
HD 21693 b	&	296	&	1.717	\\
HD 219134 f	&	168	&	1.100	\\
HD 31527 b	&	319	&	2.092	\\
HD 40307 b	&	194	&	1.095	\\
HD 40307 c	&	199	&	1.109	\\
HD 40307 d	&	199	&	1.107	\\
HD 45184 b	&	147	&	1.427	\\
HD 45184 c	&	147	&	1.430	\\
HD 51608 b	&	300	&	1.731	\\
HD 69830 b	&	143	&	1.110	\\
HD 7924 b	&	218	&	1.257	\\
HD 7924 c	&	277	&	1.415	\\
HD 7924 d	&	232	&	1.295	\\
HIP 54373 b	&	565	&	1.296	\\
HIP 54373 c	&	505	&	1.225	\\
HIP 57274 b	&	384	&	1.668	\\
YZ Cet b	&	725	&	0.493	\\
YZ Cet c	&	792	&	0.516	\\
YZ Cet d	&	744	&	0.500	\\
tau Cet g	&	93	&	0.872	\\
55 Cnc e	&	197	&	1.443	\\

\hline					
\hline					
\end{longtable}                       					
}

\section{BLS transit search figures}

\begin{figure}[!ht]
\begin{minipage}{\linewidth}
\begin{center}
\includegraphics[angle=0,trim = 0mm 05.5mm 0mm 00mm, clip,width=0.45\linewidth]{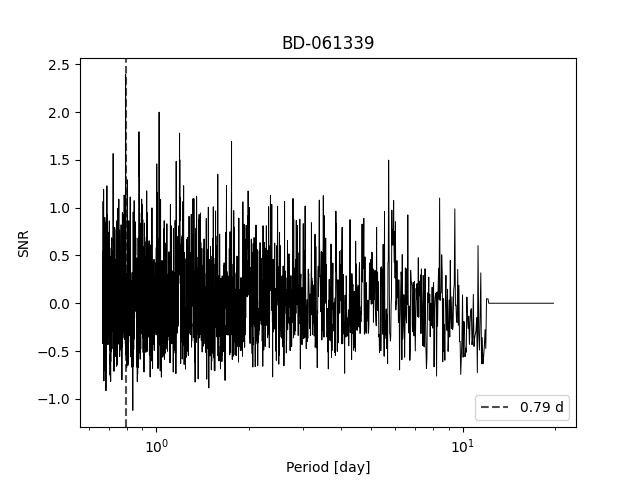}
\includegraphics[angle=0,trim = 0mm 05.5mm 0mm 00mm, clip,width=0.45\linewidth]{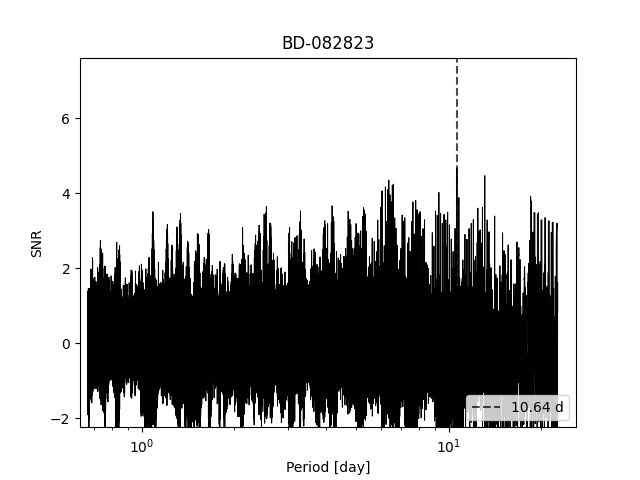}
\includegraphics[angle=0,trim = 0mm 05.5mm 0mm 00mm, clip,width=0.45\linewidth]{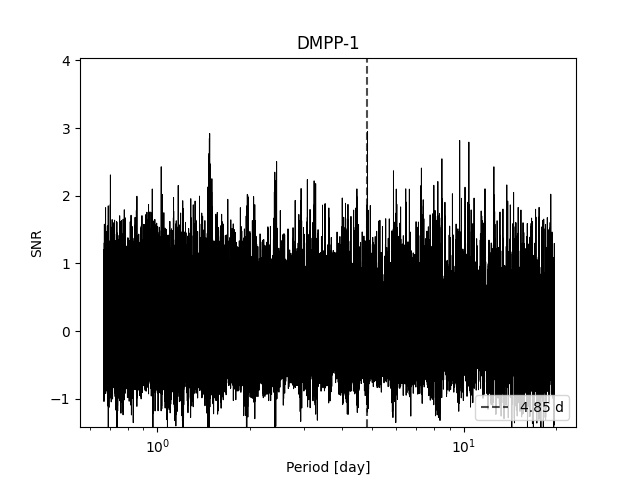}
\includegraphics[angle=0,trim = 0mm 05.5mm 0mm 00mm, clip,width=0.45\linewidth]{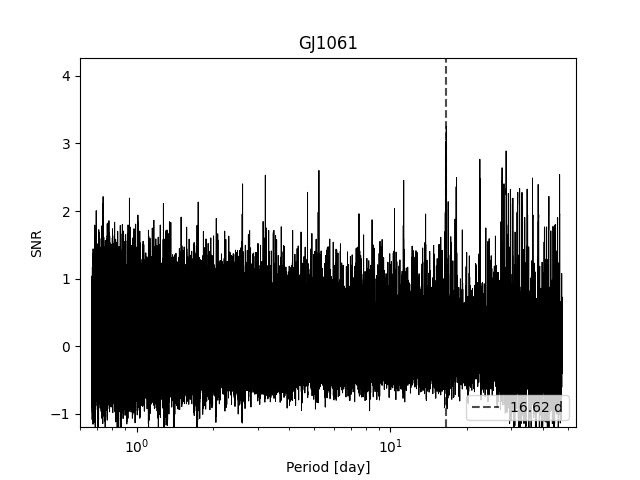}
\includegraphics[angle=0,trim = 0mm 00mm 0mm 00mm, clip,width=0.45\linewidth]{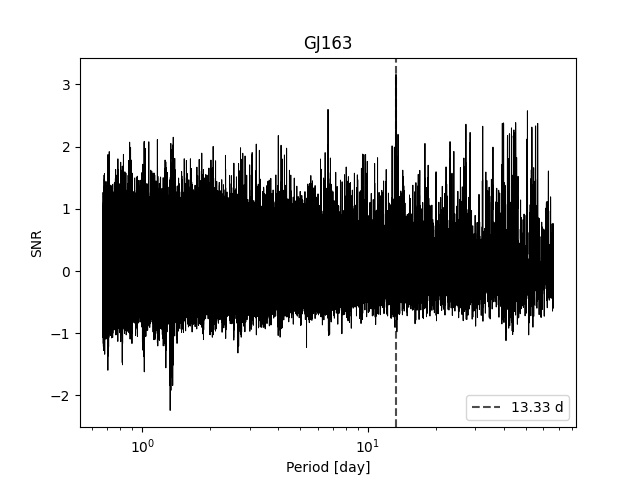}
\includegraphics[angle=0,trim = 0mm 00mm 0mm 00mm, clip,width=0.45\linewidth]{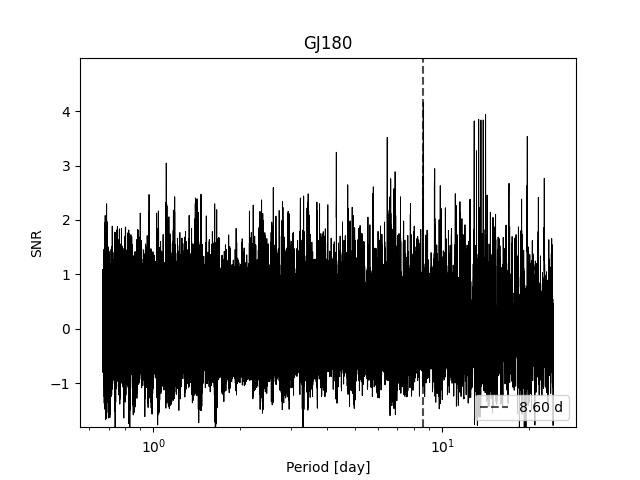}

\end{center}
\caption{\small{BLS power spectrum of  BD-061339, BD-082823, DMPP-1, GJ 1061, GJ 163 and GJ 180.}}
\label{appfig1:bls}
 \end{minipage}
\end{figure}

\begin{figure}[!ht]
\begin{minipage}{\linewidth}
\begin{center}

\includegraphics[angle=0,trim = 0mm 05.5mm 0mm 00mm, clip,width=0.45\linewidth]{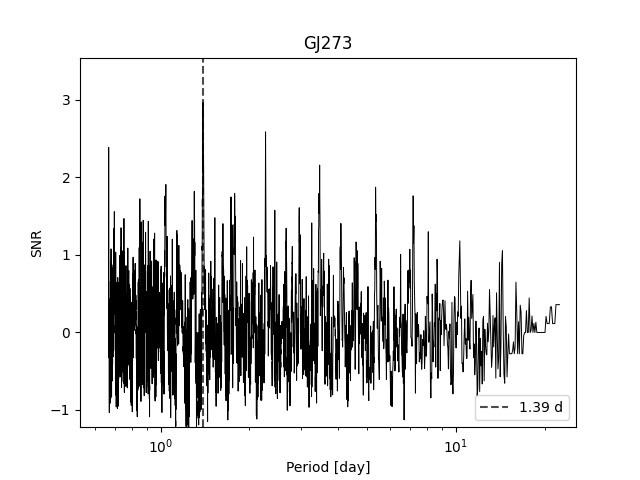}
\includegraphics[angle=0,trim = 0mm 05.5mm 0mm 00mm, clip,width=0.45\linewidth]{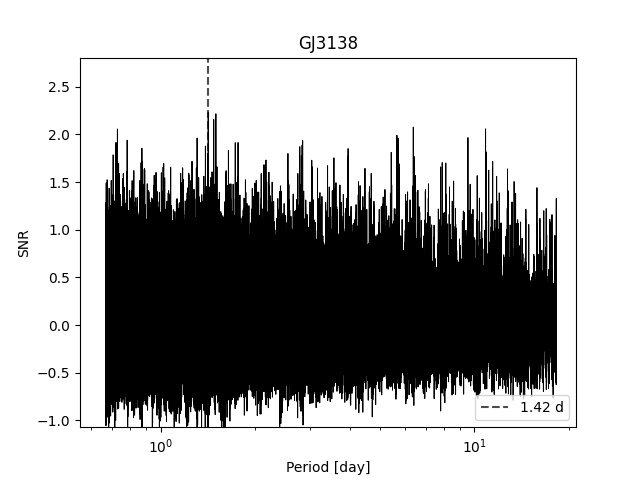}
\includegraphics[angle=0,trim = 0mm 05.5mm 0mm 00mm, clip,width=0.45\linewidth]{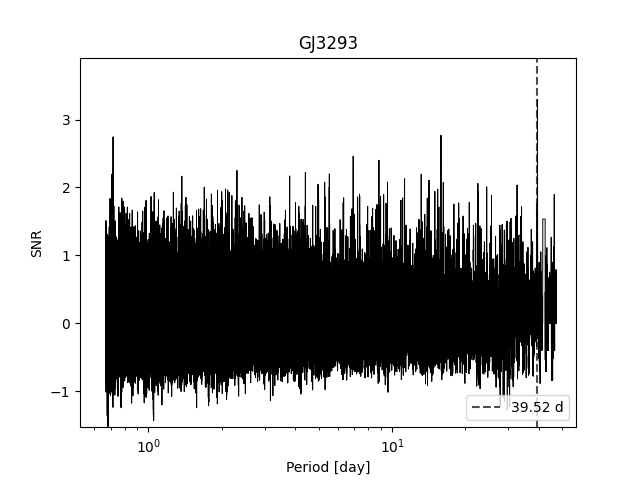}
\includegraphics[angle=0,trim = 0mm 05.5mm 0mm 00mm, clip,width=0.45\linewidth]{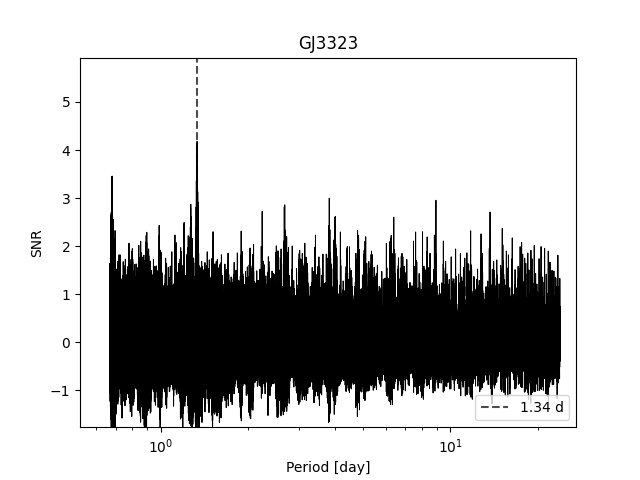}
\includegraphics[angle=0,trim = 0mm 00mm 0mm 00mm, clip,width=0.45\linewidth]{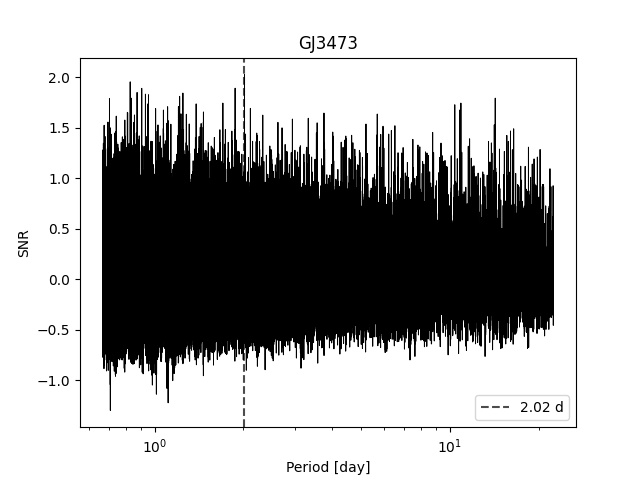}
\includegraphics[angle=0,trim = 0mm 00mm 0mm 00mm, clip,width=0.45\linewidth]{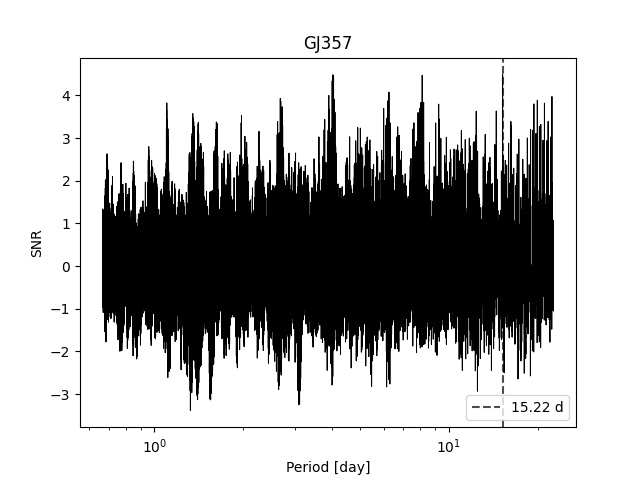}

\end{center}
\caption{\small{BLS power spectrum of  GJ 273 and GJ 3138, GJ 3293, GJ 3323, GJ 3473 and GJ 357. }}
\label{appfig2:bls}
 \end{minipage}
\end{figure}

\begin{figure}[!ht]
\begin{minipage}{\linewidth}
\begin{center}

\includegraphics[angle=0,trim = 0mm 05.5mm 0mm 00mm, clip,width=0.45\linewidth]{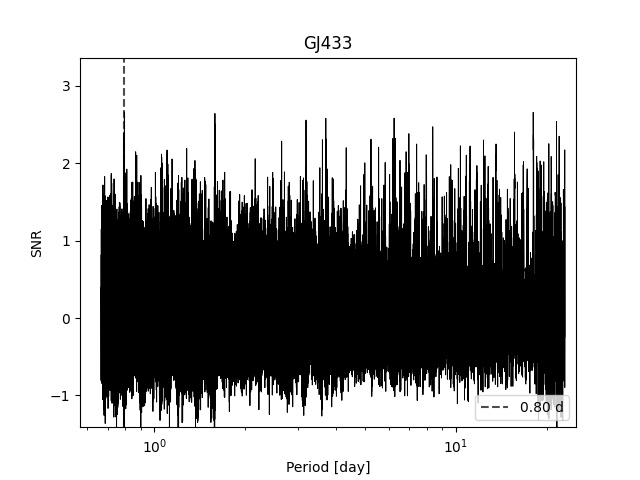}
\includegraphics[angle=0,trim = 0mm 05.5mm 0mm 00mm, clip,width=0.45\linewidth]{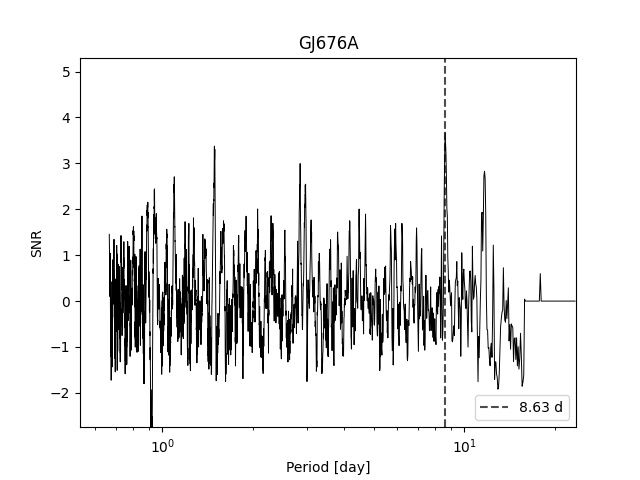}
\includegraphics[angle=0,trim = 0mm 05.5mm 0mm 00mm, clip,width=0.45\linewidth]{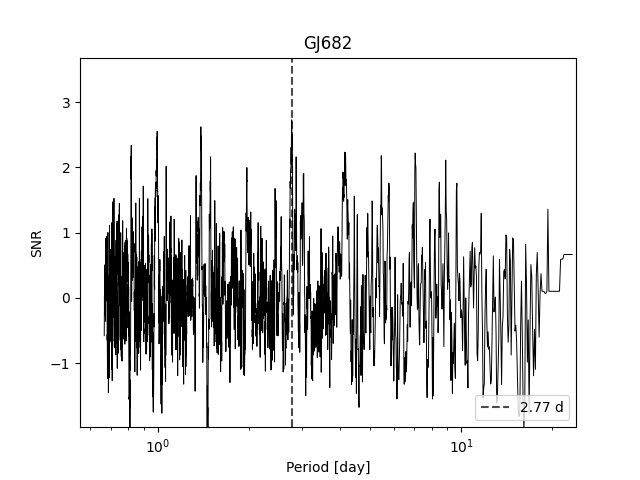}
\includegraphics[angle=0,trim = 0mm 05.5mm 0mm 00mm, clip,width=0.45\linewidth]{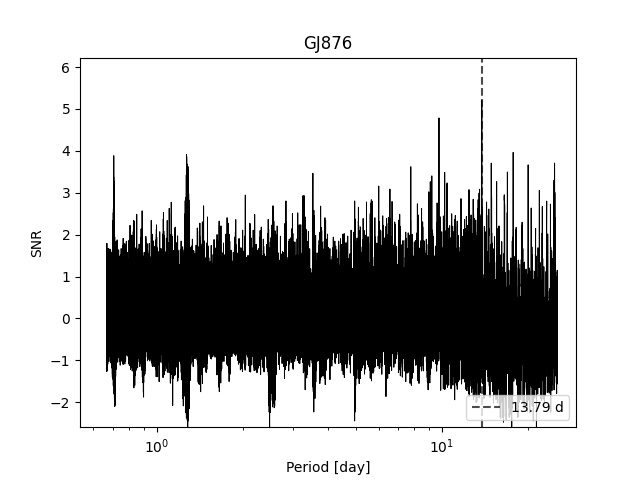}
\includegraphics[angle=0,trim = 0mm 00mm 0mm 00mm, clip,width=0.45\linewidth]{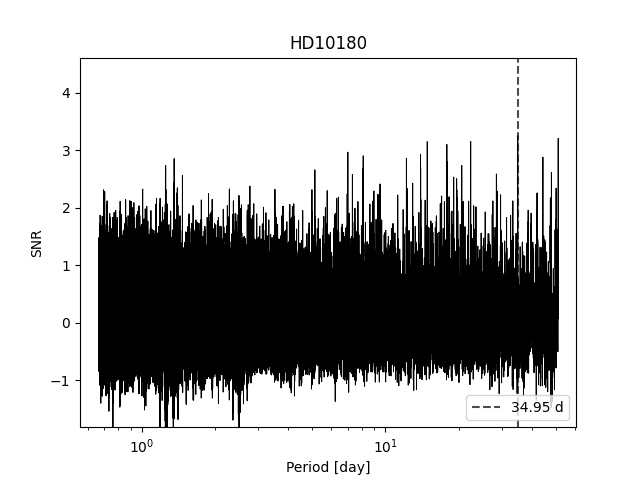}
\includegraphics[angle=0,trim = 0mm 00mm 0mm 00mm, clip,width=0.45\linewidth]{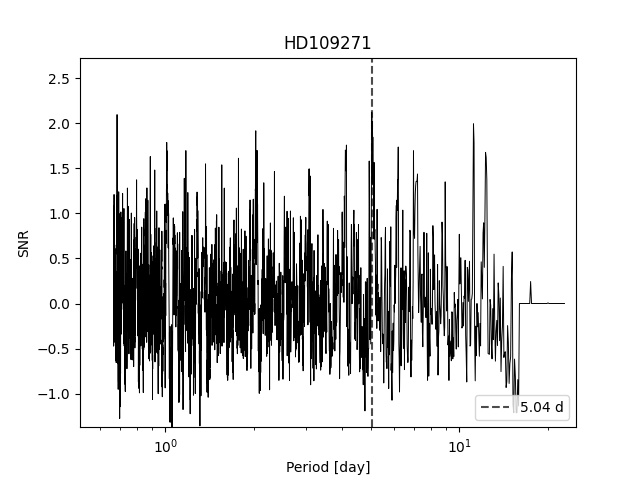}

\end{center}
\caption{\small{BLS power spectrum of  GJ 433, GJ 676 A, GJ 682, GJ 876, HD 10180 and HD 109271. }}
\label{appfig3:bls}
 \end{minipage}
\end{figure}

\begin{figure}[!ht]
\begin{minipage}{\linewidth}
\begin{center}
\includegraphics[angle=0,trim = 0mm 05.5mm 0mm 00mm, clip,width=0.45\linewidth]{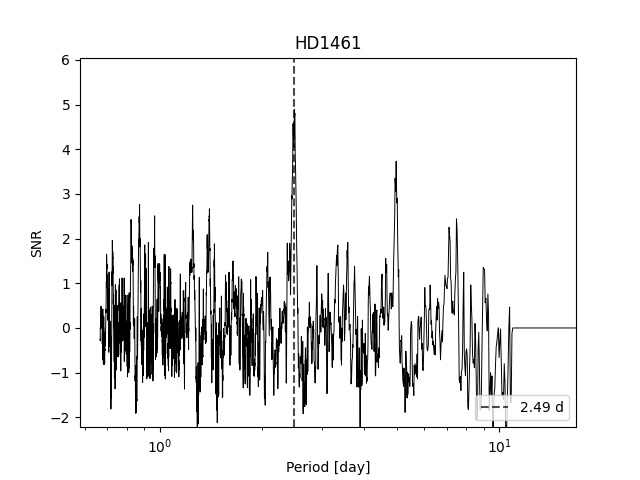}
\includegraphics[angle=0,trim = 0mm 05.5mm 0mm 00mm, clip,width=0.45\linewidth]{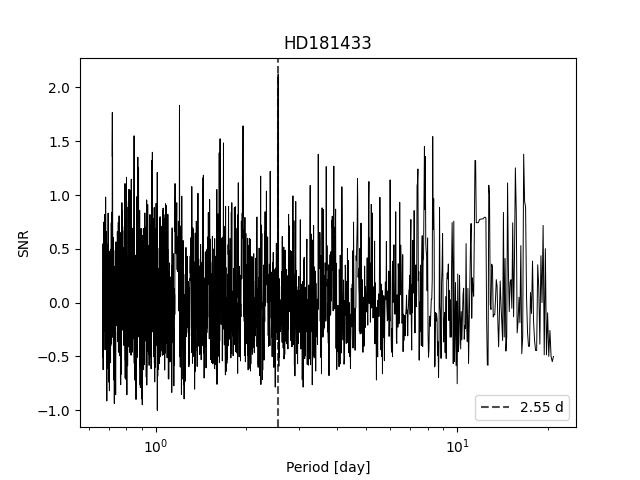}
\includegraphics[angle=0,trim = 0mm 05.5mm 0mm 00mm, clip,width=0.45\linewidth]{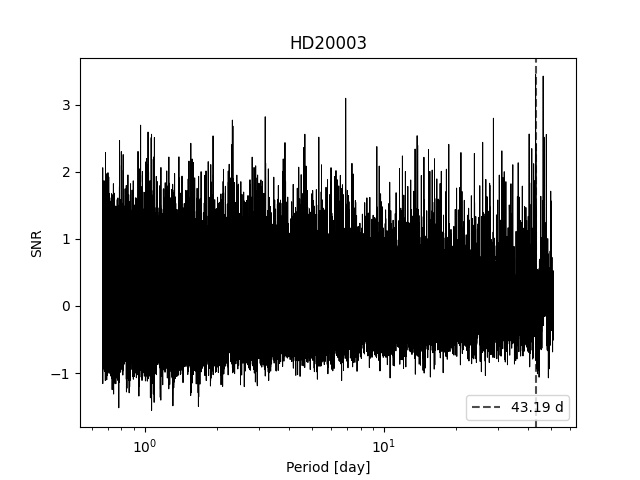}
\includegraphics[angle=0,trim = 0mm 05.5mm 0mm 00mm, clip,width=0.45\linewidth]{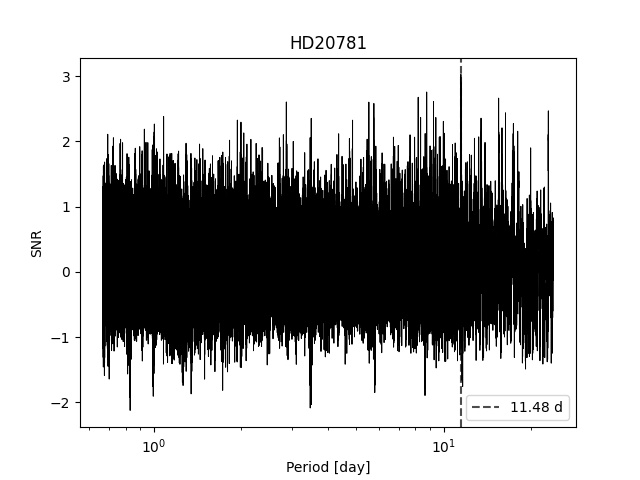}
\includegraphics[angle=0,trim = 0mm 00mm 0mm 00mm, clip,width=0.45\linewidth]{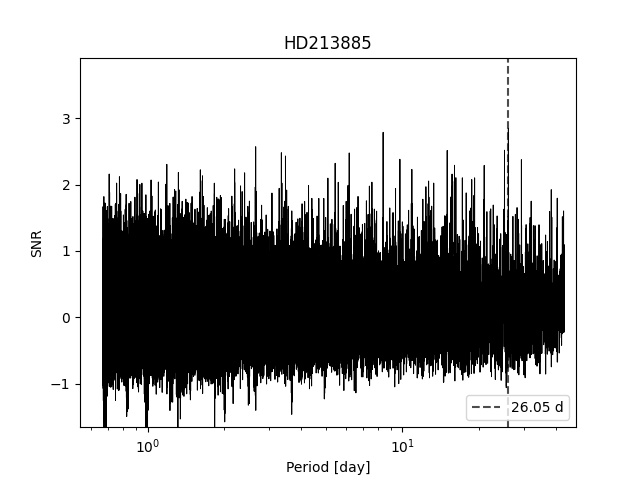}
\includegraphics[angle=0,trim = 0mm 00mm 0mm 00mm, clip,width=0.45\linewidth]{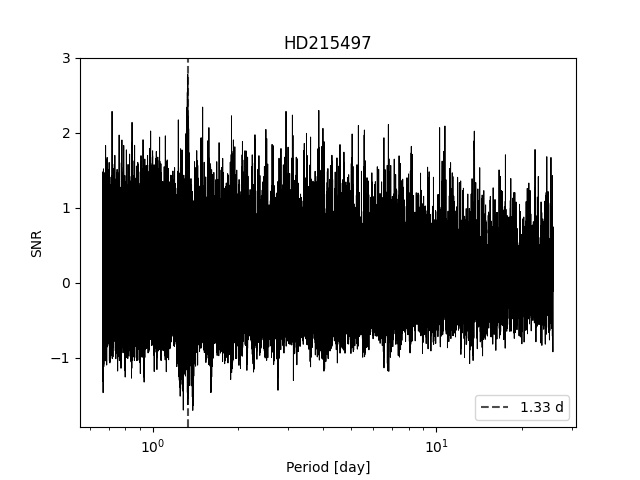}

\end{center}
\caption{\small{BLS power spectrum of HD 1461, HD 181433, HD 20003, HD 20781, HD 213885 and HD 215497.}}
\label{appfig4:bls}
 \end{minipage}
\end{figure}

\begin{figure}[!ht]
\begin{minipage}{\linewidth}
\begin{center}
\includegraphics[angle=0,trim = 0mm 05.5mm 0mm 00mm, clip,width=0.45\linewidth]{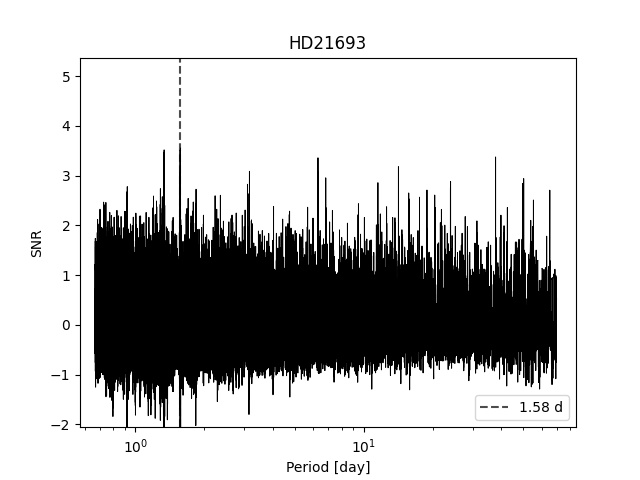}
\includegraphics[angle=0,trim = 0mm 05.5mm 0mm 00mm, clip,width=0.45\linewidth]{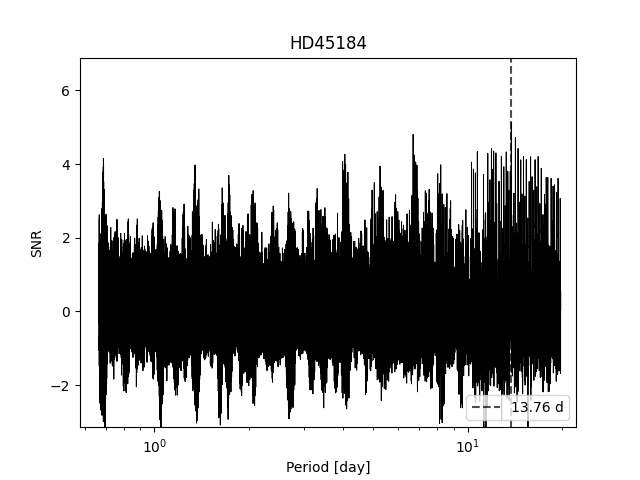}
\includegraphics[angle=0,trim = 0mm 05.5mm 0mm 00mm, clip,width=0.45\linewidth]{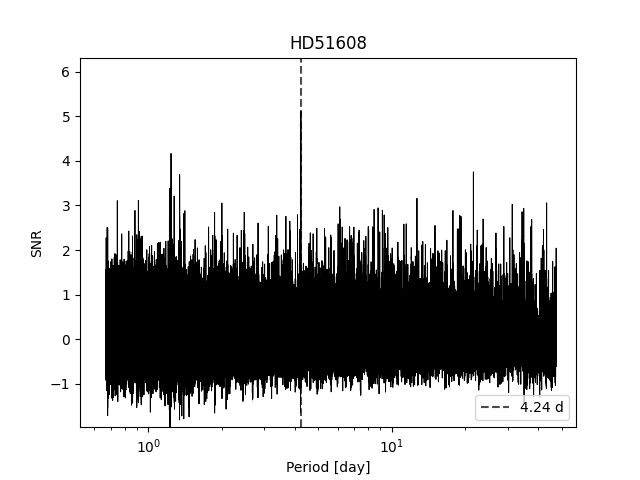}
\includegraphics[angle=0,trim = 0mm 05.5mm 0mm 00mm, clip,width=0.45\linewidth]{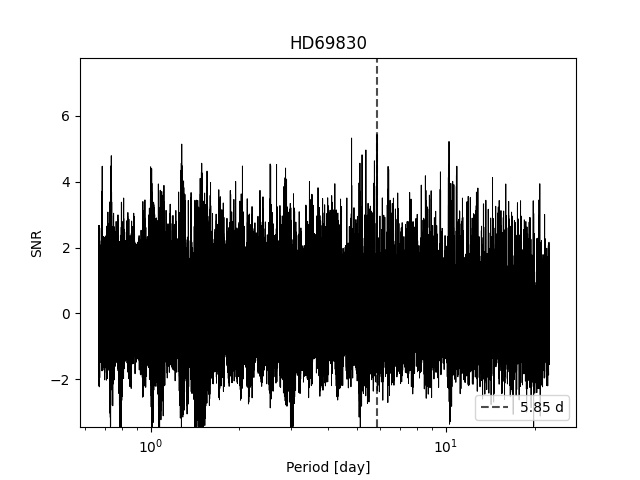}
\includegraphics[angle=0,trim = 0mm 00mm 0mm 00mm, clip,width=0.45\linewidth]{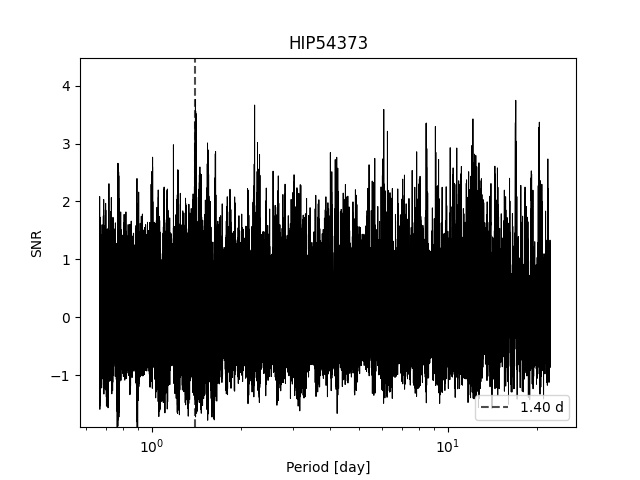}
\includegraphics[angle=0,trim = 0mm 00mm 0mm 00mm, clip,width=0.45\linewidth]{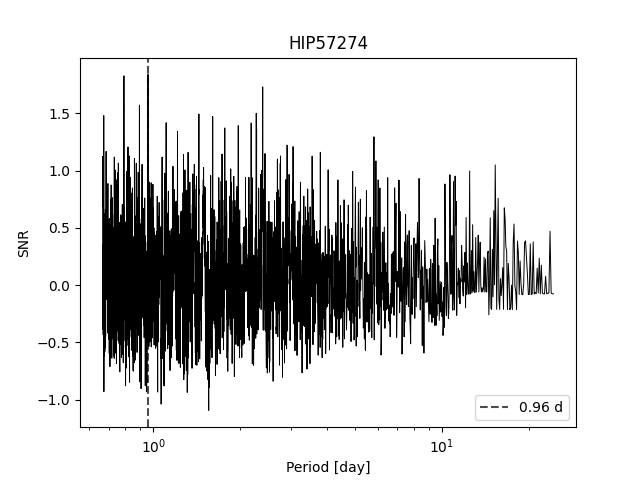}
\includegraphics[angle=0,trim = 0mm 00mm 0mm 00mm, clip,width=0.45\linewidth]{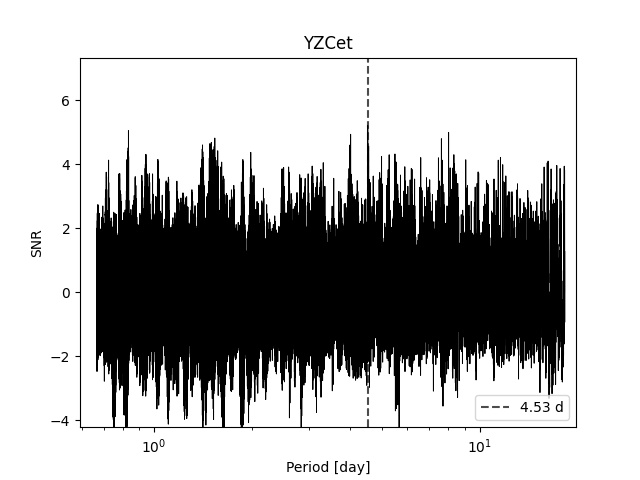}

\end{center}
\caption{\small{BLS power spectrum of HD 21693, HD 45184, HD 51608, HD 69830, HIP 54373, HIP 57274 and YZ Cet.}}
\label{appfig5:bls}
 \end{minipage}
\end{figure}

This appendix presents the BLS power spectrum of the targets without discussion in previous sections, computed as described in Sect.~\ref{sect.transitsearch}. Each figure shows the results of the global-BLSs, as we mentioned earlier, the GJ~3473, GJ~357 and HD~213885 targets have previously detected transiting planets, so we show the global-BLSs of the masked light curves. The mentioned BLS periodograms are in Figs.~\ref{appfig1:bls}, \ref{appfig2:bls}, \ref{appfig3:bls}, \ref{appfig4:bls} and \ref{appfig5:bls}.

 \vspace{1cm}

\vspace{1.5cm}

\vspace{1.5cm}

\vspace{1.5cm}

\vspace{-1.5cm}

\end{appendix}
\end{document}